\documentclass[11pt]{article}
\usepackage[letterpaper,margin=1in]{geometry}

\usepackage{booktabs}
\usepackage{graphicx}
\usepackage{amsmath}
\usepackage{amssymb}
\usepackage{array}
\usepackage{dblfloatfix}
\usepackage{placeins}
\usepackage[nodayofweek]{datetime}

\usepackage[
  pagebackref,
  colorlinks=true,
  linkcolor=blue,
  citecolor=blue,
  urlcolor=blue
]{hyperref}

\usepackage{natbib}

\newcommand{\athresh}{a_{\mathrm{th}}}
\newcommand{\DI}{\textit{DI}} 
\newcommand{\muhat}{\widehat{\mu}} 
\newcommand{\sigmahat}{\widehat{\sigma}}           
\newcommand{\length}{\textit{length}}
\newcommand{\lengthbar}{\overline{\textit{length}}}

\newcommand{\Norm}{\textrm{Norm}}
\newcommand{\norm}{\textrm{norm}}
\newcommand{\POD}{\textrm{POD}}
\newcommand{\response}{\textit{response}}
\newcommand{\Rsoftware}{\texttt{R}}
\newcommand{\xbar}{\bar{x}}

\newcommand{\footremember}[2]{%
  \footnote{#2}
  \newcounter{#1}
  \setcounter{#1}{\value{footnote}}%
}

\newif\ifempty

\renewcommand*{\backrefalt}[4]{%
  \ifcase #1
  \or
    [#2]%
  \else
    [#2]%
  \fi
}

\newdateformat{mydate}{%
  \twodigit{\THEDAY}\ \monthname[\THEMONTH]\ \THEYEAR}

\title{Statistical Methods for Estimating Probability of Detection in Structural Health Monitoring}

\author{Qizheng Xia
    \footremember{isu1}{
    Department of Industrial and Manufacturing Systems Engineering,
    Iowa State University,
    Ames, IA 50010, USA
    }
\and
William Q. Meeker
    \footremember{isu2}{
    Department of Statistics,
    Iowa State University,
    Ames, IA 50010, USA
    }
\and
Dennis Roach
    \footremember{sandia}{
    FAA Airworthiness Assurance Center,
    Sandia National Laboratories,
    Box 5800, Albuquerque, NM 87185, USA
    }
    \and
    Seth S. Kessler
    \footremember{metis}{
    Metis Design Corporation,
    205 Portland St, 4th Floor,
    Boston, MA 02114, USA
    }
\and
Qing Li
    \footremember{isu3}{
    Department of Industrial and Manufacturing Systems Engineering,
    Center for Nondestructive Evaluation,
    Iowa State University,
    Ames, IA 50010, USA,
    qlijane@iastate.edu
    (Corresponding author)
    }
}

\date{\mydate\today}

\begin{document}

\maketitle

\begin{abstract}
There is much interest in the potential to use structural
  health monitoring (SHM) technology to augment traditional
  nondestructive inspection (NDI) methods to improve safety,
  increase asset availability, and reduce maintenance and inspection
  costs. SHM has the potential to be used in many applications,
  including critical components in aircraft and
  pipelines. Probability of detection (POD) plays a critical role in
  aircraft structural integrity programs, leading to increased
  interest in developing methods to assess POD in SHM
  applications. In contrast to traditional NDI laboratory
  experiments involving specimens with cracks, SHM sensors are
  fixed, and SHM data are acquired over time as cracks grow or
  otherwise evolve. Thus, traditional statistical methods for
  assessing POD \citep[e.g., as described in][]{milhdbk1823a2009} must be
  replaced or extended to properly handle repeated-measures
  data. The purpose of this paper is to review the basic statistical
  concepts of POD and show how these concepts can be extended or
  adapted for SHM-POD applications. The paper presents statistical
  methods for modifying and extending existing POD methods,
  including a simple size-of-damage-at-detection (SoDaD) method and a
  random-parameter (RP) method for repeated-measures data. The methods are
  compared using three case studies involving Piezoelectric
  Transducer (PZT), Carbon Nanotube (CNT), and Comparative Vacuum
  Monitoring (CVM) sensor systems. Results show that the SoDaD
  method provides a simple approach for POD estimation
  with limited data, while the RP method offers
  enhanced modeling fidelity by utilizing repeated
  measurements. These methods are applicable when a scalar damage
  index or similar response is used to make a detection decision.
\end{abstract}


\newpage
\tableofcontents
\newpage

\section{Introduction}\label{intro}
\subsection{Background}
There is a high degree of interest in the development and
implementation of structural health monitoring (SHM) technology to
augment traditional nondestructive inspection (NDI) methods to
improve safety and reduce maintenance costs. SHM has the potential
to be used in many areas of application from critical components in
aircraft to pipelines. The effectiveness of SHM, however, needs to
be demonstrated by using properly conducted performance data and
successful field
operations~\citep{munns2003analysis,roach2018addressing}. Key to the
performance assessments of SHM are methods to estimate the
probability of detection (POD) levels for SHM sensors.

The purpose of this paper is to review the basic statistical
concepts of POD and to show how these concepts can be used to
develop appropriate statistical methods for a class of SHM-POD
studies in which sensors are used to detect cracks in local or small
zonal areas where sensors have been installed. We illustrate the
methods using data from three different recent SHM-POD
studies. Potentially, there are many different kinds of SHM-POD
studies that could be conducted depending on the particular
application. In our concluding remarks, we describe the need to
develop alternative methods for use in more global kinds of SHM. Our
work and the corresponding studies were motivated by aerospace
applications where permanently attached sensors will be used to
detect cracks after initiation and some growth.

\subsection{Applicability of MIL-HDBK-1823A to SHM}
Given the existence of damage (e.g., a crack) that should be
detected, POD is the probability that, for a given inspection
opportunity, the inspection will detect that damage. For traditional
NDI POD, the probability distribution is estimated from experimental
data, as described in~\citet[][Appendix~G]{milhdbk1823a2009}.

A frequently asked question is whether and how it might be possible
to adapt the POD methods and procedures given in MIL-HDBK-1823A to
SHM technologies. The nature of SHM differs from that seen in
traditional NDI in that the position of SHM sensors is fixed, and
because they are permanently installed, SHM data can be sampled
frequently over time as flaws grow and/or evolve. Thus,
generalizations of the MIL-HDBK-1823A methods will be required to
quantify POD for SHM.

\subsection{Related work}
\citet{shook2008simulation} point out that repeated-measures
observations from SHM-POD studies will not be independent, and
develop a statistical model to take dependency into account when
estimating POD. \citet{forsyth2016structural}, in the context
of POD for SHM, reviews the important fact that in POD studies
involving cracks, the dominant source of variability is
crack-to-crack variability due to crack morphology.
\citet{schubertkabban2015probability} describe a statistical
random-coefficient model for SHM POD that is similar to the model
presented in Section~\ref{section:methodology.shm.pod}
of this paper, but which uses only a
random slope. \citet{seaver2013workshop} describe the
need for a ``statistically valid'' assessment of an SHM system's
ability to detect flaws before SHM systems will be accepted for
organizations like the Department of Defense.
\citet{falcetelli2022probability} reviewed the evolution of
reliability metrics for SHM systems, emphasizing POD along with
emerging localization and sizing metrics to address the unique
uncertainties and variability inherent in SHM
applications. \citet{rentala2024pod} evaluated
SHM-specific POD methodologies for civil engineering applications,
showing that models such as size-of-damage-at-detection
(SoDaD) and random intercept Linear
Mixed-Effects Model provide more realistic reliability
estimates than traditional NDI POD
methods. \citet{gobat2021shm} developed a
reduced-order modeling strategy for SHM by integrating Proper
Orthogonal Decomposition and Domain Decomposition, enabling accurate
and real-time structural damage assessment with substantial
computational speedup.
\citet{SAE2025ARP6821} is an industry standard developed by an SAE
committee that included authors of this paper. The work leading to
this standard provided an early investigation of methods for
assessing the damage-detection performance of SHM systems and helped
motivate the development of the SHM-POD statistical methods
described in this paper.

\subsection{Overview}
To address the challenges associated with repeated-measures data in
SHM-POD studies, this paper considers two complementary statistical
approaches. The first approach is the SoDaD method,
which simplifies the analysis by using only the
crack length at the first detection event for each crack-sensor
combination. This method avoids the complexities associated with
dependent repeated measurements and provides a computationally
simple framework for POD estimation.  The second approach is a
random-parameter (RP) regression model that extends the traditional
$\hat{a}$ versus $a$ method to account for repeated measurements.
This method is a generalization of the widely used $\hat{a}$ versus
$a$ method described in
\citet[][Appendix~G]{milhdbk1823a2009}.  The RP method quantifies
variability by estimating separate intercepts and slopes for each
crack-sensor combination along with the joint distribution of the
slopes and intercepts in the population of all similar crack-sensor
combinations. These two approaches have
different trade-offs between simplicity and modeling fidelity and
form the basis for the subsequent development and comparison of
SHM- POD estimation methods in this paper.

The paper compares the SoDaD and RP methods
by applying them in three
case studies involving Piezoelectric Transducer (PZT), Carbon
Nanotube (CNT), and Comparative Vacuum Monitoring (CVM) sensor
systems. The results demonstrate that the SoDaD method provides a
simple and robust framework for POD estimation, particularly when
the number of crack-sensor combinations is limited. In contrast, the
RP method more fully uses the available
repeated-measures data, providing a more complete characterization of
variability in crack-sensor combinations.

Overall, this paper makes three main contributions. First, it
clarifies why conventional NDI POD methods cannot be directly
applied to SHM studies with fixed sensors and repeated
measurements. Second, it develops and compares two statistically
valid approaches for SHM-POD estimation: the
SoDaD method and the
RP regression model method. Third, it illustrates these methods with
three representative SHM sensor systems using PZT, CNT, and CVM
sensors. By linking POD concepts from MIL-HDBK-1823A to the
repeated-measures structure of SHM data, this paper provides a
practical statistical framework for evaluating SHM detection
capability and supporting future implementation of SHM and
corresponding model-assisted probability of detection (MAPOD) methods in
safety-critical applications.

The remainder of this paper is organized as follows.
Section~\ref{section:technical.background}
reviews the key concepts and assumptions underlying POD
studies, including the conventional statistical model used for POD
estimation in traditional NDI applications, detection thresholds,
false alarm probability, variability, uncertainty, and the
characteristics that distinguish SHM-POD studies from conventional
NDI POD studies. Section~\ref{section:methodology.shm.pod} presents the
SoDaD method and the RP
method for estimating POD from repeated-measures SHM data.
Section~\ref{section:application.case.studies} applies the
proposed methods to the three SHM-POD data sets. Section~\ref{section:comparison.sodad.rp} 
compares the SoDaD and RP methods in terms
of their assumptions, advantages, limitations, and practical
implementation. Section~\ref{section:mapod} discusses how the proposed
statistical framework can support MAPOD. Finally, Section~\ref{section:concluding.remarks}
provides concluding remarks and identifies areas for future
research.

\section{Technical Background for POD and SHM POD}
\label{section:technical.background}
This section provides the statistical background for probability of
detection (POD) analysis in traditional NDI and its extension to SHM
applications. It reviews the conventional $\hat{a}$ versus $a$
methodology, introduces key metrics such as $a_{90}$ and
$a_{90/95}$, and discusses the roles of variability, uncertainty,
detection thresholds, and probability of false alarm. The section
concludes by highlighting the unique characteristics of SHM-POD
studies, particularly the challenges associated with
repeated-measures data.

\subsection{Definition of POD}
Given the existence of damage that should be detected, POD is the
probability that a given inspection opportunity will detect that
damage. POD is usually expressed as a function of damage
characteristics (e.g., crack length). When there are other important
factors affecting POD, the implicit assumption (not always met) is
that these factors have been suitably considered and incorporated,
according to an appropriate joint probability distribution
describing the variation of those factors in actual inspection
applications. Moreover, these factors need to cover the range that
could be experienced during the time SHM measurements would be
taken. For example, if it is planned to take SHM measurements at an
ambient temperature that can range from $-18^\circ\mathrm{C}$ to
$49^\circ\mathrm{C}$ ($0^\circ\mathrm{F}$ to $120^\circ\mathrm{F}$),
then it is necessary to account for that variability in the POD
study and the computation of POD.

In general, the POD is computed as the probability that the
inspection method output (NDI or SHM) falls into the detection
region (the detection part of the line in the case of a scalar
signal), as a function of chosen fixed characteristics (such as
crack length). The probability distribution should reflect the
variability in all of the other important signal-affecting factors
in the inspection (described in
Section~\ref{section:sources.of.variability}). In general,
finding this probability distribution presents one of the most
challenging parts of a POD study. Conducting experiments that
correctly capture all the important sources of variability for SHM
technologies, however, will be challenging and costly.

\subsection{\texorpdfstring{The Simple   $\hat{a}$ versus $a$ Method
    in POD}{The Simple ahat versus a Method in POD}}
\label{section:simple.ahat.vs.a}
\begin{figure}[tbp]
\begin{tabular}{cc}
(a) & (b) \\[-1.8ex]
  \includegraphics[width=0.5\columnwidth,
    trim=0 22.9pt 0 22.9pt,clip]{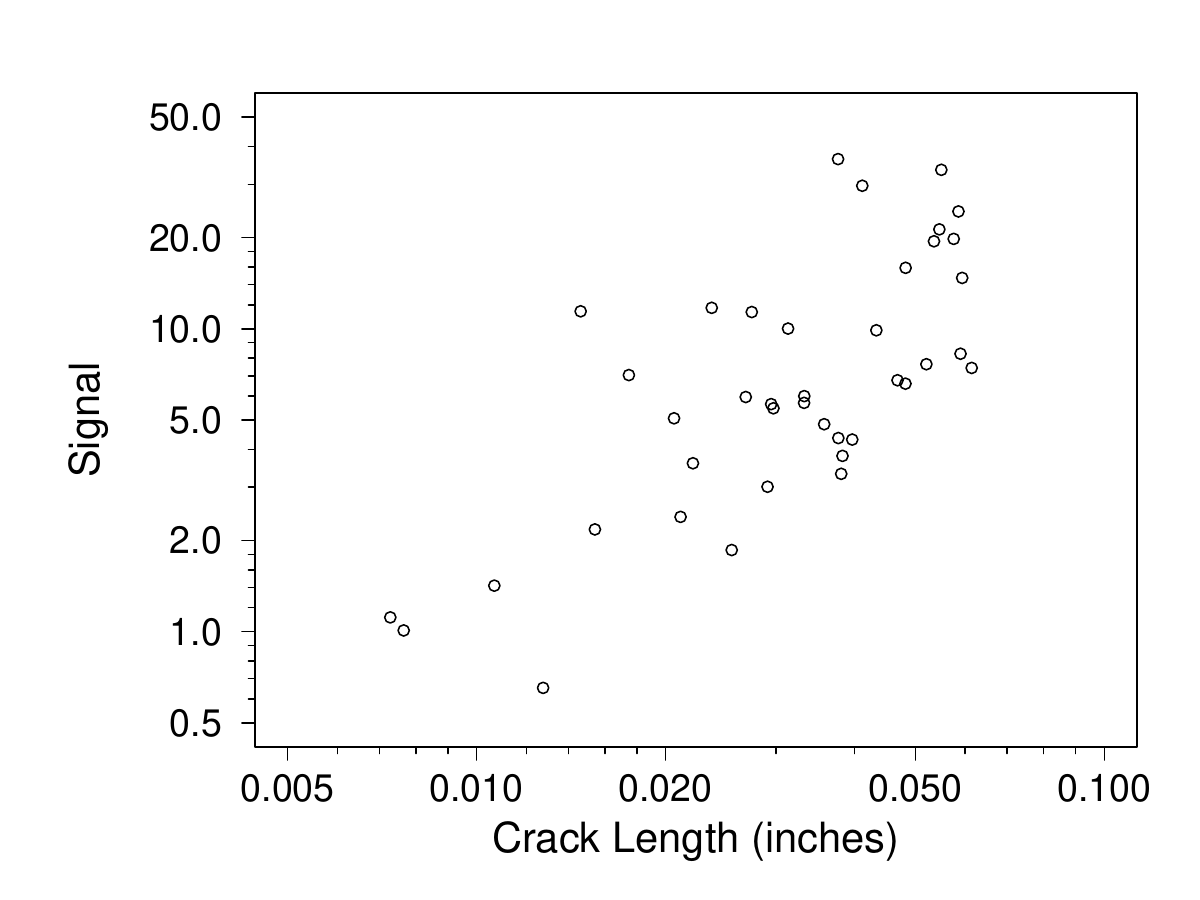} &
  \includegraphics[width=0.5\columnwidth,
    trim=0 20.0pt 0 20.0pt,clip]{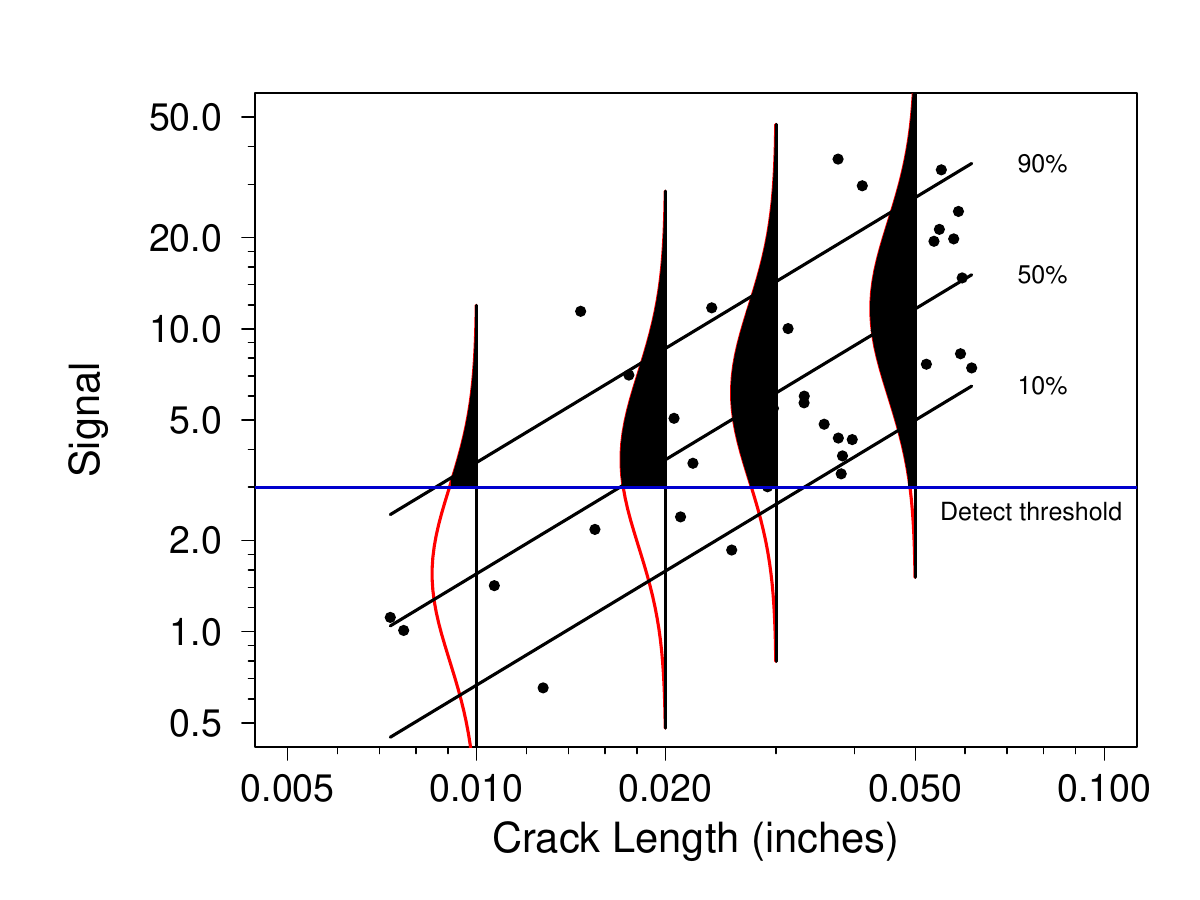}
\end{tabular}
\caption{$\hat{a}$ versus $a$ data from an eddy current inspection
  of boltholes~(a)
  and the fitted $\hat{a}$  versus  $a$ model showing how POD is computed~(b).}
\label{figure:ahat.versus.a.data.model}
\end{figure}

The $\hat{a}$ versus $a$ method is described, for example, in
\citet[][Appendix~G]{milhdbk1823a2009}. This method is
based on the underlying statistical model that there is a population
of cracks and that the POD study is based on data collected from a
sample of these cracks under conditions that are similar to actual
inspections. Figure~\ref{figure:ahat.versus.a.data.model}a
is a plot of typical $\hat{a}$
(signal response in arbitrary units) versus (crack length) data on a
log-log scale. The data are from an eddy current inspection of bolt
holes. Note the large amount of variability in the signal response
$\hat{a}$ for cracks of similar crack length $a$.
The   $\hat{a}$ versus $a$ statistical model is a simple regression model
\begin{equation}
\begin{aligned}
\response_i
&= \beta_0 + \beta_1 \length_i + \varepsilon_i,
\qquad i = 1,\ldots,n,\\
\varepsilon_i &\sim \Norm(0,\sigma).
\end{aligned}
\end{equation}
where $\response_i$ is the random signal response, $\length_i$ is the
length of crack $i$ (or some other relevant
quantification of target size), $\beta_0$ is the intercept, $\beta_1$ is the
slope, and $\epsilon_i$ is a statistical error term. The error term describes
the deviation between the random response value and the linear model
for crack $i$, and $\sigma$ is the error standard deviation
describing the spread of the random response values around the true
model line $\beta_0 + \beta_1 \times \length_i$. The deviations can
arise from various sources of variability, such as flaw morphology
and noise in the measurement process. Then the available data (a
single response from each of the cracks and the crack length) are
used to estimate the model parameters $\beta_0$, $\beta_1$, and $\sigma$
as illustrated in Figure~\ref{figure:ahat.versus.a.data.model}b.

\begin{figure}[tbp]
\centering
\includegraphics[width=0.7\columnwidth,
  trim=0 22.3pt 0 22.3pt,clip]{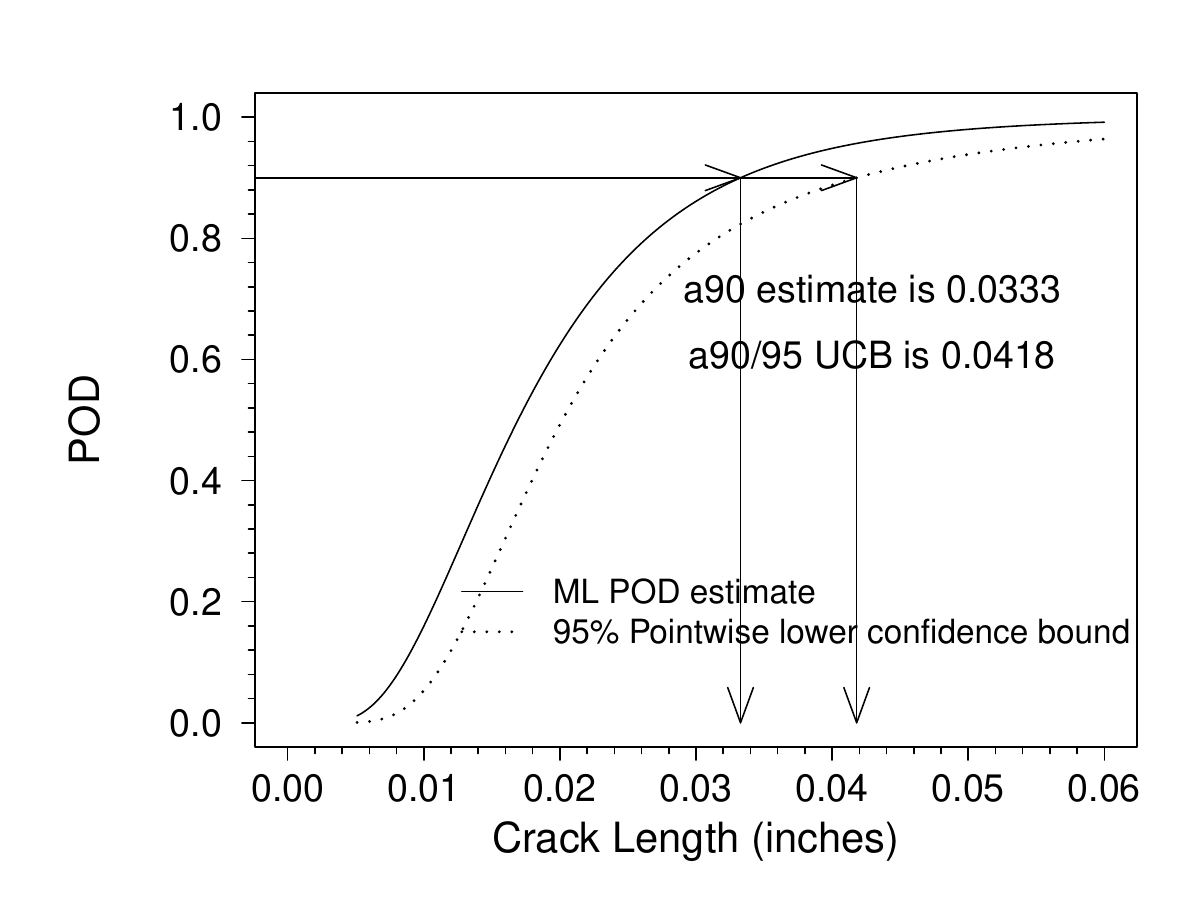}
\caption{The estimate of POD with a set of one-sided pointwise 95\% lower confidence bounds. The $\hat{a}_{90}$  and $a_{90/95}$  values are also shown.}
\label{figure:image47}
\end{figure}

The probability of detection (POD) is the probability of
having a response that is greater than the detection threshold
$\athresh$ and is computed as
\begin{equation}
\text{POD}(length)
=
\Pr({\response}>\athresh) = 1-\Phi_{\norm}(z)
\end{equation}
where
\begin{equation}
z = \frac{a_{th} - (\beta_0 + \beta_1\times \length)}{\sigma}
\end{equation}
and $\Phi_{\norm}(z)$ is the standard normal (Gaussian) cumulative
distribution function. To estimate POD, $\beta_0$, $\beta_1$, and
$\sigma$ in~(2) are replaced by their maximum likelihood (ML)
estimates. Figure~\ref{figure:image47} is a plot of the estimate of POD
as a function of crack length, along with a set of 95\% lower
confidence bounds for POD. The value $a_{90}$ is the crack length
that will be detected with probability 0.90 and also the 0.90
quantile of the POD distribution. Then $\hat{a}_{90}=0.033$ is
an estimate of the crack length where the POD is 0.90. The
$a_{90/95}=0.042$ value is a 95\% upper confidence bound for $a_{90}$,
representing the largest crack that might be missed in an
inspection. The upper confidence bound $a_{90/95}$ is a common
scalar metric that is used to describe inspection
capability. Details for how to do these statistical
computations (including with censored observations that sometimes
arise) are given in \citet[][Appendix~G]{milhdbk1823a2009}.

\subsection{Variability and Uncertainty}
In discussing POD and other statistical applications, it is
important to distinguish between variability and
statistical uncertainty (i.e., uncertainty due to limited
data). Variability generally arises due to the stochastic
nature of physical phenomena, including damage characteristics and
inspection variables. The spread in the data points in
Figure~\ref{figure:ahat.versus.a.data.model}a for a given crack length illustrates this
variability.

Statistical uncertainty arises when there is limited information (i.e., limited
data) about a quantity of interest (e.g., POD of a crack of a
particular size for a given ultrasonic testing (UT)
inspection). This uncertainty is quantified by the lower confidence
bounds for POD in Figure~\ref{figure:image47}. In general, more data
will reduce statistical uncertainty, but not variability. If instead of 40
cracks there had been 160 cracks in the study, the estimate would be
almost the same, but the distance between the estimate and the lower
confidence band would be cut, approximately, in half. For further
discussion, see
\citet{li2012distinguishing,li2015quantile}. Statistical
confidence intervals and one-sided confidence bounds are used to
describe statistical uncertainty. Statistical tolerance intervals and prediction intervals are
used to describe a combination of variability in a distribution and
the uncertainty in describing the distribution due to limited
data.
\citet[][Chapter~2]{meeker2017statistical} provides detailed explanations
about such statistical intervals.

\subsection{Detection Threshold and Probability of a False Alarm}
In any inspection process, there will be a need to set a detection
threshold. In a simple situation where the SHM signal is a scalar,
then, as in \citet[][Appendix~G]{milhdbk1823a2009},
the detection threshold is also a
scalar. For example, as illustrated in Figure~\ref{figure:image47},
there will be a detection call if the signal response is greater
than $\athresh=3$. Smaller values of $\athresh$ will result in larger
values of POD, but will increase the probability of a false alarm
(PFA).

In NDI applications the threshold is generally set high enough that
the probability of a PFA is acceptably low. The acceptable level of
PFA will depend on the particular application, but
in most applications, there
will be a need to have a negligible PFA. Then POD can be
evaluated to determine if it is acceptable or not. The detection
capability (i.e., POD and $a_{90/95}$) for different inspection methods can be
compared \textit{only} if the corresponding PFA values are the same. PFA
would be computed in exactly the same way as POD, except using the
corresponding probability distribution of signals in the absence of
damage (i.e., the distribution of noise). Statistical methods for
doing this are described in \citet[][Appendix~G]{milhdbk1823a2009}.

\subsection{Characteristics of a Traditional POD Study}
The purpose of a POD study is to collect the necessary information
about model parameters so that an estimate of POD and a
quantification of statistical uncertainty can be computed. For
traditional NDI, a POD study consists of a set of targets (e.g.,
flat-bottom holes in a block, cracks in flat plates, or synthetic
hard alpha inclusions in a titanium block).
\citet{milhdbk1823a2009}
recommends that at least 40 targets be used in an $\hat{a}$ versus
$a$ POD study. Each target is inspected once, and the resulting data
can be used to estimate POD, as illustrated in Section~\ref{section:simple.ahat.vs.a}. It
is possible to conduct a POD study
with fewer than 40 targets, but then there would be less precision
for estimation of POD and less information for identifying the
appropriate statistical model (whether and how to transform the data
and which distribution to use to describe the statistical error term).

To properly quantify POD from a POD study, it is essential that all
of the important sources of variability that could affect detection
are explicitly captured. Omitting influential sources of variability
in a POD study could result in overestimating the probability of
detecting smaller cracks or underestimating the potential for false
alarm indications.

\subsection{Considerations for an SHM-POD study}
\label{considerations.shm.pod.study}
An SHM-POD experiment for some aerospace applications will generally
consist of fixed sensors being placed on specimens (e.g., flat
plates) with starter-cracks (e.g., electrical discharge machining
(EDM) notches). Then the cracks could be grown in fatigue over time
by applying cyclic mechanical loads to the plate. SHM signal data
would be taken periodically over time and related to the length of
the crack at that time. Similarly, for pipeline applications, fixed
sensors could be used to monitor a corrosion process where SHM
signals would be related to the amount of metal loss. SHM examples
are presented in Section~\ref{section:application.case.studies}.

The SHM-POD experiment should accurately simulate the actual SHM
process. Again, it is important that the experiment capture relevant
sources of variability. For example, the variability in cracks grown
in the experiment should accurately represent the variability seen
in actual cracks. The observational units in SHM-POD studies will be
crack/sensor combinations (where there may be an array of sensors in
some applications). To simplify the discussion in the rest of this
paper, however, we will generally refer to a ``crack'' as the
observational unit.

In any SHM application, there will be an important consideration of
how to map the SHM signal(s) into a detect/no-detect decision at
each inspection opportunity (however an inspection opportunity is
defined), typically referred to as a damage index. The POD will then
depend on the (joint) probability distribution of the inputs to that
decision-making mechanism. Data from one or more SHM sensors may be
mapped into one or more scalar damage indices that can be used for
decision making; however, each damage index would produce separate
individualized POD results. In this paper, we will assume that the
decision-making response is a scalar and that the crack length,
which can be measured with reasonable accuracy,
adequately describes crack properties.

Because SHM-POD experiments generally consist of fixed sensors with
repeated measures on cracks that are growing over time, the
traditional \citet{milhdbk1823a2009} $\hat{a}$ versus
$a$ model cannot be used without generalizing the statistical
methods. This paper describes two alternative and appropriate
statistical methods and these are described and illustrated in the
next two sections of this paper.

\subsection{Factors Affecting Detection Sensitivity and Sources of
  Variability in SHM}
\label{section:sources.of.variability}
Factors relating to damage and system properties that could affect
SHM signals include:
\begin{itemize}
    \item Damage size, shape, and orientation (including changes in
      these characteristics over time). We note that this is
      typically the dominant source of variability in traditional
      NDI for crack detection
      and it is expected that this will be true also for SHM
      applications.
    
    \item Damage location relative to sensor location (e.g., the
      distance between the sensor and the damage).
    
    \item Environmental variables such as temperature and humidity.
    
    \item Mechanical variables such as strain conditions (due to
      varying amounts of fuel on board, etc.).
    
    \item Variability in sensor signal responses due to
      sensor-to-sensor manufacturing variability.
    
    \item Change in the structural configuration where the sensors
      are located as a function of time and that could have an
      effect on the SHM signal.
    
    \item Changes in sensor performance over time due to maintenance
      repairs, re-painting, etc.
    
    \item Sensor aging and degradation.
    
    \item Sensor adhesive and other characteristics relating to
      installation-to-installation variability.
\end{itemize}
Then, for those factors that are not assumed to be held constant
across inspections or compensated for by a calibration operation, it
is essential that there be an accurate characterization of the joint
probability distribution. For example, in traditional NDI,
crack-to-crack variability arising from differences in crack
morphology (different cracks with nominally the same length can have
signal responses that vary enormously) tends to be the dominant
source of variability, and this is also expected to be the case in
SHM applications.

Many factors are involved in obtaining viable SHM data for POD
calculations. Some of these factors depend on the SHM system itself
and some depend on the type of testing, the complexity of the test
article used in the assessment, and the type, location, and
orientation of damage being detected. Factors to be considered
include, but are not limited to, determining the boundaries for the
SHM system applications, producing validation tests that are
representative of the actual structure, establishing proper damage
detection thresholds, utilizing data with appropriate signal content
compared to system noise, and data analysis methods. These factors
should be considered when designing a POD study.

Perhaps the biggest difference between NDI and SHM is that one can
store NDI equipment in a controlled environment, but SHM sensors
(and maybe hardware) will be exposed to a range of environmental and
mechanical static and dynamic extremes. It is necessary to make sure
that POD is just as valid after 1/5/10 years of flight as it is
when installed.

\section{Methodology for SHM-POD}
\label{section:methodology.shm.pod}
This section presents two statistical methods for estimating the
POD in SHM
applications, with particular attention to the challenges associated
with repeated-measures data. The first approach, the
SoDaD method, uses only the \textit{size of the damage}
at detection for each observational unit (e.g.,
crack-sensor combination in crack detection). (Note that size is a
generic term and could refer to the length of a crack or an area in an
ultrasonic C-scan or X-ray image. Our examples involve crack detection, so we
will use the term crack length.)

By
reducing the data to a single observation per specimen, the SoDaD method
avoids the need to model the dependence among repeated measurements
and offers a simple framework for POD estimation.  The second
approach is the RP regression model, which makes
use of the relevant set of signal-response measurements collected during
crack growth. This model extends the traditional $\hat{a}$ versus
$a$ framework by incorporating random intercepts and slopes to
represent crack-to-crack variability.

\subsection{The Size-of-Damage-at-Detection (SoDaD)
  Method to estimate POD}
\label{section:sodad.method.pod}
\subsubsection{Concept Overview}
Statistical performance assessments of damage detection sensors that
are permanently mounted in a fixed position must be handled
differently from similar studies using hand-held or other deployed
transducers that are moved along the structure being inspected. In
the case of in-situ SHM sensors, the damage of interest originates,
and may even propagate, into the region being monitored by the SHM
sensor. Performance analyses then consider the response of the
sensor or damage detection and its relationship with the size of the
damage when detected.

The SoDaD method for repeated inspections of cracks growing under or
near fixed sensors provides a simple, statistically valid method to
compute POD for SHM crack-detection applications. This method was originally
suggested by Floyd Spencer (personal communication, 2006) and first
applied to POD assessments of SHM systems in
\citet{roach2007use} and \citet{roach2009real}. This method
uses only the crack-length values when cracks are first
detected. For our crack-detection applications, we will refer to
these values as ``length at detection'' (LaD) values.
Similar to other POD applications, the underlying
statistical model is that there is a population of crack/sensor
combinations and that the POD study is based on a sample of these
crack/sensor combinations. Each crack has a length, random from
crack to crack, at which the crack will be detected. Because only
one observation is taken from each crack/sensor combination, the
issue of dealing with the dependency of repeated measures data does
not arise.

\subsubsection{Definition of the LaD values}
LaD can be defined in several different ways. With constant
monitoring, LaD could be defined as the crack length when the SHM
response reaches the detection threshold. When the SHM response is
measured periodically (as would be the case in most actual SHM
applications), the common practice is to define LaD as the crack
length at the time of the first inspection where the SHM response has
exceeded the detection threshold. Although it might be argued that
interpolation should be used to approximate more closely the actual
crossing length, defining LaD by the crack length at the first
inspection exceeding the detection threshold (for an appropriately
chosen inspection interval) would more closely mimic
crack detection is periodic
monitoring in actual SHM applications.

\subsubsection{Computation of POD and Lower Confidence Bound in LaD}
Under the assumption that the size-of-damage-at-detection values
(e.g., crack length LaD values) can
be described by a normal (Gaussian) probability distribution, POD
can be computed as follows. From the available sample of cracks, let
$x_1, x_2, \ldots, x_n$ denote the observed
crack lengths at detection for the $n$ cracks. Also, let $\xbar$
denote the sample mean and let $s$ denote the sample standard
deviation of the $x_1, x_2, \ldots, x_n$ values. Then the POD
estimate for a crack having a specified length denoted by
$length$ is
\begin{equation}
\begin{aligned}
\POD(\length)
&= \Pr(X \le \length) = \Phi_{\norm}
\left(
\frac{\length - \xbar}{s}
\right).
\end{aligned}
\end{equation}
where $\Phi_{\norm}(\cdot)$ is the standard normal
distribution cumulative distribution function (cdf). A corresponding
lower confidence bound on the probability of detecting a crack of a
specified length can be computed by using a standard statistical
method based on the noncentral $t$ distribution. The details of this
method are described in
\citet[][Section~4.5]{meeker2017statistical} R function \texttt{normTailCI}
in R package \texttt{StatInt} is available to do the needed
computation. Then this procedure is repeated for different values of
$\length$ over the desired range.
Under the assumption that the LaD values can
be described by a lognormal distribution, POD can be computed in a
similar manner as
\begin{equation}
\begin{aligned}
\POD(\length)
&= \Pr(X \le \length) = \Phi_{\norm}
\left(
\frac{\log(\length) - \xbar}{s}
\right),
\end{aligned}
\end{equation}
where
$\xbar$
is the sample mean and
$s$
is the sample standard deviation of the logarithms of the LaD values.

Other location-scale and log-location-scale distributions can also
be used to describe the LaD distribution. These pairs of
distributions include (but are not limited to) the logistic and
log-logistic, the smallest extreme value and Weibull, and the
largest extreme value and Fr\'{e}chet distributions. These distributions
are described in
\citet[][Appendix~C]{meeker2017statistical} and some of
them are employed in the numerical examples in
Section~\ref{section:application.case.studies}.
It is important to check the
adequacy of these different distributions (as we will do in
our examples) and the effect that they have on POD estimation
because there is usually not a sufficient amount of information in
the data to confirm that a particular distribution should be used.

Because of the close relationship between confidence intervals for
probability distribution quantiles and tail probabilities, the
computation of the lower confidence bounds for POD in the SoDaD
method can also be done by using statistical methods for computing a
one-sided tolerance bound, as described in
\citet{roach2009real}.
In previous applications of this tolerance bound calculation,
this approach was called
the ``One-Sided Tolerance Interval'' (OSTI) method.

More specifically, $a_{90/95}$ is a 95\% upper confidence bound on
$a_{90}$, the crack length at which POD is 0.90. This upper confidence
bound can be obtained as a one-sided upper confidence bound on the
0.90 quantile of the LaD distribution. Equivalently, it is a
one-sided upper tolerance bound having content 0.90 and confidence
level 0.95 for the LaD distribution. Methods for computing this
confidence (or tolerance) bound are given in
\citet[][Section~4.4]{meeker2017statistical}. The \Rsoftware{}
function \texttt{normQuantileCI} in the \Rsoftware{} package
\texttt{StatInt} is available to do the needed computation. Tables of
factors for computing such tolerance bounds are also available in
\citet{krishnamoorthy2009statistical},
\citet{meeker2017statistical}, and some engineering statistics
textbooks. Corresponding estimation, confidence-bound, and
confidence-interval methods for other location-scale and
log-location-scale distributions are described and illustrated in
\citet[][Chapter~14]{meeker2017statistical}.

\subsection{\texorpdfstring{A Random-Parameter (RP) Generalization
    of the   $\hat{a}$  versus $a$   Signal-Response Model}{A
    Random-Parameter (RP) Generalization of the ahat versus a
    Signal-Response Model}}
\label{section:rp.method.pod}\label{section:rp.model}
\subsubsection{Basic Idea and the RP Model}
\label{section:rp.model.basic.idea}

Although the $\hat{a}$ versus $a$ method described in
Section~\ref{section:technical.background} of this paper
and more fully in \citet{milhdbk1823a2009}
is not applicable for SHM applications with repeated measures
on cracks, a suitable statistical generalization is available. This
method is based on the same underlying statistical model that there
is a population of crack/sensor combinations, that the POD study
is based on data collected from a sample of these crack/sensor
combinations, and that a signal response (or some transformation of
the signal response) can be described as a linear function of crack
length (or some transformation of crack length). Then it is possible to
fit a linear random-parameter (RP) regression model that describes the
crack-to-crack variability in the intercepts and slopes. This model
assumes that \textit{each crack/sensor specimen} in the population has its
own intercept and slope and we estimate the joint distribution of
intercepts and slopes. That is,
\begin{equation}
\begin{aligned}
\response_{ij}
&= \beta_{0i}
+ \beta_{1i}
\left(
\length_{ij}
- \lengthbar
\right)
+ \varepsilon_{ij},\\
&\qquad i=1,\ldots,n,\quad j=1,\ldots,m_i .
\end{aligned}
\end{equation}
where $\lengthbar$ is the sample mean of all of the
measured length values in the data set, $\beta_{0i}$ is the intercept (mean
response at $\lengthbar$), and $\beta_{1i}$ is the
slope, respectively, for crack/sensor combination $i$.
Also, $\varepsilon_{ij}$, independent of the random parameters,
is the error term for reading $j$ from
crack/sensor combination $i$, representing the deviation between the
measured value and the true line for crack/sensor combination
$i$. The $\beta_{0i}$ and $\beta_{1i}$ values are assumed to have a
bivariate normal probability distribution and $\varepsilon_{ij} \sim
\Norm(0,\sigma_{\varepsilon})$ values are assumed to be
independent of the random $\beta_{0i}$ and $\beta_{1i}$ values. Then
the available repeated-measures data from the sample of
crack/specimen combinations are used to estimate the six parameters
$\mu_{\beta_0},\; \mu_{\beta_1},\; \sigma_{\beta_0},\;
\sigma_{\beta_1},\; \rho, \;\text{and}\; \sigma_{\varepsilon}$.
These parameters can be interpreted as follows.
\begin{enumerate}
\item
  $\mu_{\beta_0}$ is the population mean of the
  random intercepts and, because crack size is centered at
  $\lengthbar$ is the population mean response at crack size.
\item $\sigma_{\beta_0}$ is the standard deviation of intercepts
      at the crack length $\lengthbar$.
    \item $\mu_{\beta_1}$ is the population mean of the slopes.
    \item $\sigma_{\beta_1}$ is the standard deviation of the slopes.
    \item $\rho$ is the correlation between the slopes and intercepts.
    \item $\sigma_{\varepsilon}$ is the standard deviation of the
      spread of the $\response$ values around the lines
      for each crack/sensor combination.
\end{enumerate}

\subsubsection{Computation of POD for the RP method}
\label{section:rp.model.pod}
The POD for a randomly selected crack/sensor combination from the
population
under the RP model is, for a crack of length $\length$,
\begin{equation}
\begin{aligned}
\label{equation:rp.model.pod}
\POD(\length)
&= \Pr(\response > \athresh)
= 1 - \Phi_{\norm}(z)
\end{aligned}
\end{equation}
where $\delta= \length - \lengthbar$, and
\begin{equation}
\begin{aligned}
z
&=
\frac{
\athresh
- (\mu_{\beta_0}
+ \mu_{\beta_1}\delta)
}{
\sigma_{\response}(\length)
},\\
\sigma_{\response}(\length)
&=
\Big[
\sigma_{\beta_0}^2
+ \delta^2 \sigma_{\beta_1}^2 + 2\delta\rho_{\beta_0\beta_1}\sigma_{\beta_0}\sigma_{\beta_1}
+ \sigma_{\varepsilon}^2
\Big]^{1/2}.
\end{aligned}
\label{eq:rp2}
\end{equation}
Here $\mu_{\beta_0} + \mu_{\beta_1} \left(\length -
\lengthbar \right)$ is the mean of the
$\response$ values at crack length $\length$ and
$\sigma_{\response}(\length)$ is the corresponding standard deviation of
the $\response$ values at crack length $\length$.

\subsubsection{Using Bayesian Methods to Estimate POD and Compute Lower Confidence Bounds}
\label{section:rp.model.bayesian}
Modern methods of Bayesian inference provide a formal method to
combine information from data with prior information from the
physics of inspection or other established knowledge with data from
physical experiments or other sources. More specifically, Bayes'
theorem shows how to combine a likelihood (information from data)
with prior information to obtain a posterior distribution
representing updated knowledge. Bayesian methods are particularly
convenient with random-parameter statistical models
(also known as hierarchical models).

Bayesian estimation with weakly
informative (i.e., proper but diffuse) prior distributions for the parameters
provides an appealing alternative to maximum likelihood estimation
for RP models. With a flat joint prior distribution,
the posterior distribution is proportional to the likelihood
function and Bayesian inferences (if there is a reasonable amount of
information in the data) will be approximately equal to more
traditional likelihood-based methods of inference.

The computations in this paper were done with the
\Rsoftware{} package \texttt{rstan},
providing an interface between \Rsoftware{}~\citep{rcoreteam2019r} and
\texttt{Stan}~\citep{stan2018usersguide}. Specific codes have been written in
\Rsoftware{} and \texttt{Stan} to implement the methods
described in Sections~\ref{section:rp.model.basic.idea}
and \ref{section:rp.model.pod}. These have been
included in the \Rsoftware{} package \texttt{RPOD}. The
output from \texttt{Stan} provides a large number of draws (typically 10 to
20 thousand draws are sufficient to reduce Monte Carlo error enough
so that at least two significant digits of accuracy are available in
final answers) from the marginal posterior distributions of the
model parameters $\mu_{\beta_0},\; \mu_{\beta_1},\;
\sigma_{\beta_0},\; \sigma_{\beta_1},\; \rho, \;\text{and}\;
\sigma_{\varepsilon}$. These marginal posterior draws are then used
to compute draws from the marginal distributions of quantities of
interest such as POD and $a_{90/95}$, from which point estimates and
credible intervals or bounds are easy to obtain.

The priors for $\mu_{\beta_0}$ and $\mu_{\beta_1}$ are weakly
informative (approximately noninformative)
normal distributions with a very large
standard deviation, relative to the observed spread in the data.
For $\rho$ the prior is also a weakly informative
normal distribution with a large standard deviation but truncated
outside of $(-1,1)$. The prior distributions for $\sigma_{\beta_0}$,
$\sigma_{\beta_1}$, and $\sigma_{\varepsilon}$ are
weakly informative half-Cauchy
distributions with a scale parameter that is large relative to the
sample standard deviations computed from the data using a simple
two-stage estimation method.

\section{Application of the SHM-POD Methods in Empirical Case
  Studies}
\label{section:application.case.studies}
This section applies the SoDaD and RP methods to three SHM
case studies involving PZT, CNT, and CVM sensor systems. For each
data set, POD curves and their 95\% lower confidence bounds are
estimated using the SoDaD approach based on the distribution of
detection lengths, while the RP method utilizes the full set of
repeated signal-response measurements to characterize crack-to-crack
variability. The results from both methods are compared across the
three case studies.

\subsection{Application to the Piezoelectric Transducer (PZT) SHM Data}
\label{section:pzt.application}
\subsubsection{PZT background and data}
Piezoelectric Transducer (PZT) sensors can be attached to existing
structures without changing the local and global structural dynamics
\citep[as described in][]{kumar2006insitu}. PZT sensors
can act as both transmitters and receivers. As transmitters,
piezoelectric sensors use electrical excitation to generate elastic
waves in the surrounding material. As receivers, they receive
elastic waves and transform them into electrical signals. It is
possible to install arrays of active-sensors, in which each element
takes, in turn, the role of transmitter and receiver, and thus scan
large structural areas using ultrasonic waves. Complete reflection,
partial reflection, scattering, or other detectable effects on the
ultrasonic waves can be used as the basis for damage detection.

A network of PZT sensors was installed on a generalized beam
structure which is representative of structures found in rotorcraft
cabin frames and bulkheads. Starter notches were used to control the
damage such that only one crack was being monitored at a time. Then
tension-tension fatigue tests were used to grow cracks. Data from
the network were processed to provide a damage index (DI). Details
are omitted here but are available in \citet{roach2016wide}.

\begin{figure}[tbp]
\begin{tabular}{cc}
(a) & (b) \\[-1.8ex]
  \includegraphics[width=0.5\columnwidth,
    trim=0 22.9pt 0 0,clip]{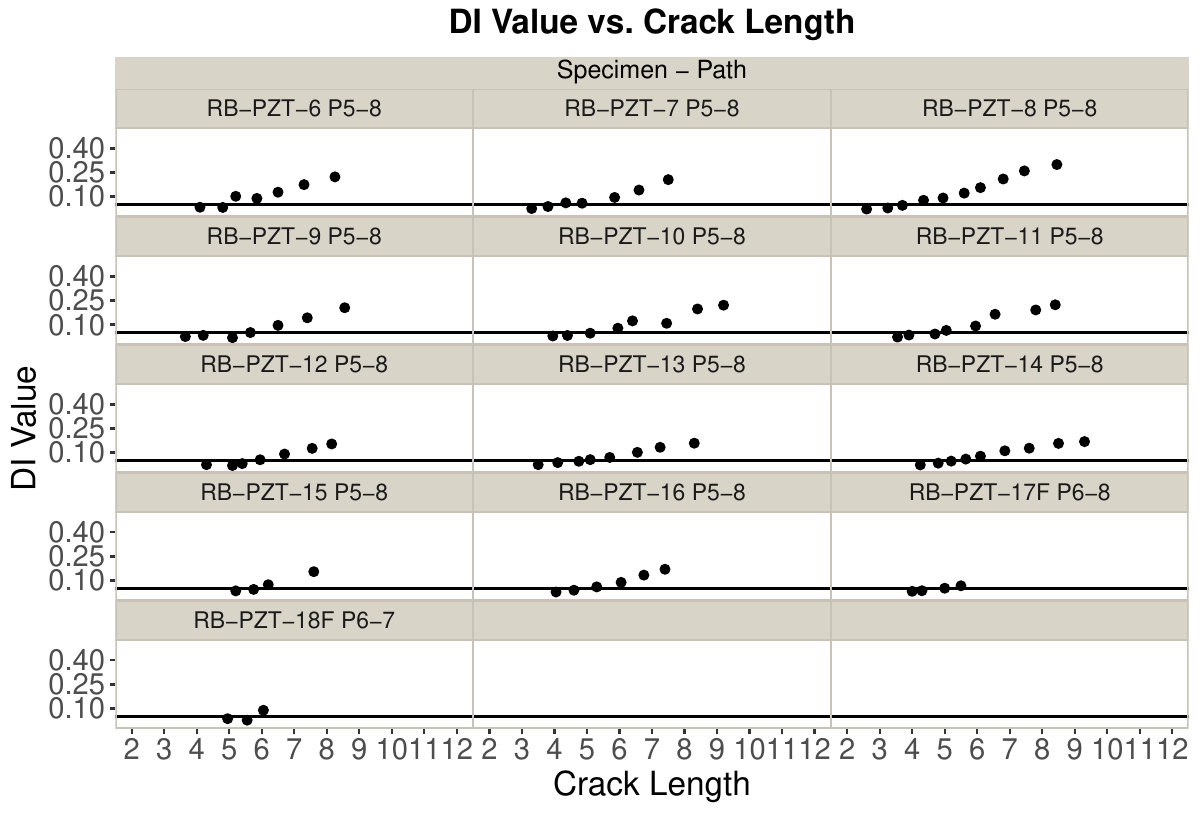} &
  \includegraphics[width=0.5\columnwidth,
    trim=0 20.0pt 0 20.0pt,clip]{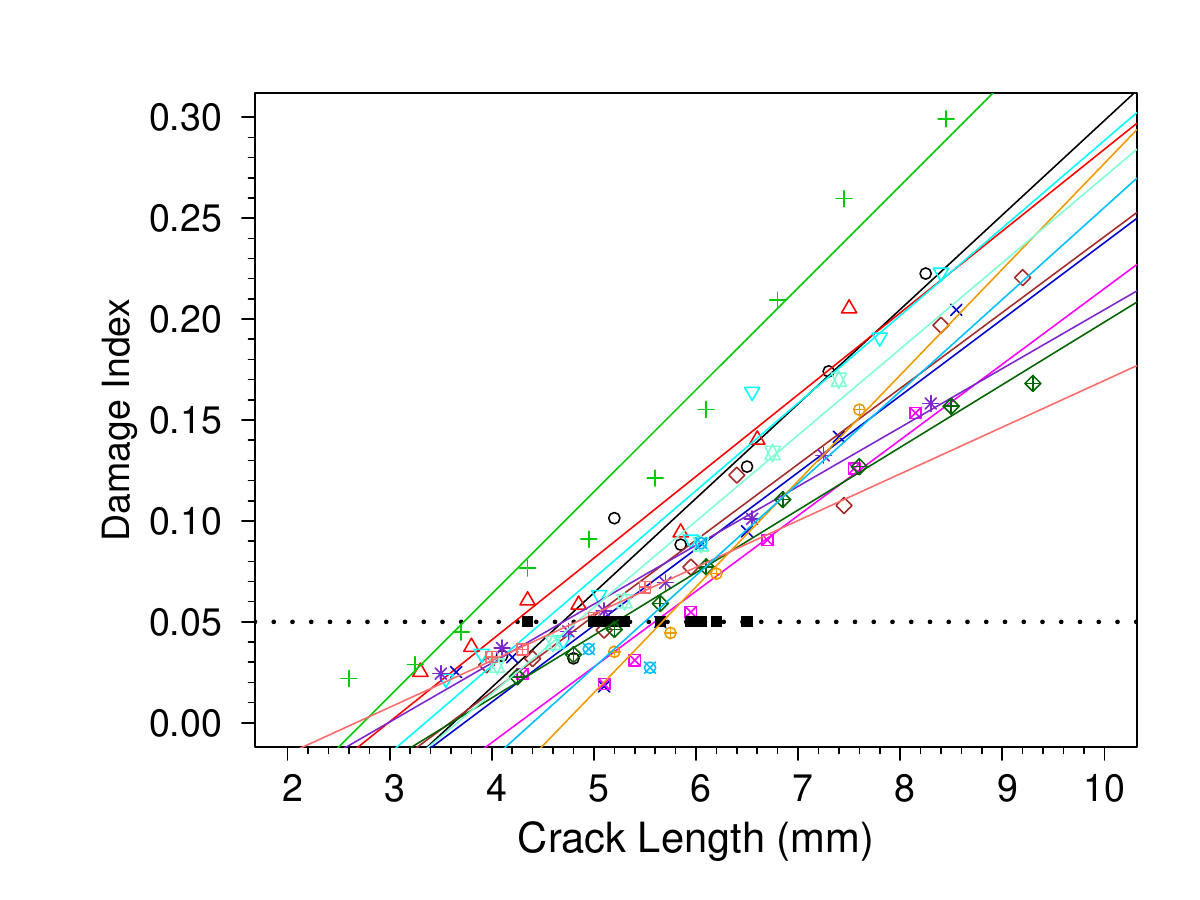}
\end{tabular}
\caption{Plot of the PZT \DI{} versus crack length. Each plot
  is labeled with the specimen ID and the horizontal lines indicate
  the detection threshold of 0.05~(a) and a
  plot of the PZT POD data with a linear regression line for
  each crack/sensor combination. The LaD values
  corresponding to the detection threshold of 0.05 are also shown~(b).}
\label{figure:pzt.data.crossing}
\end{figure}

Figure~\ref{figure:pzt.data.crossing}(a) is a plot of the PZT \DI{} data
captured around the predetermined detection threshold $a_{th}=0.05$
for crack detection from the sensor network.
This threshold was
chosen conservatively so that the probability of a false alarm would
be, effectively, zero. For the range of crack lengths in the data in
Figure~\ref{figure:pzt.data.crossing}(a),
the \DI{} is approximately linear in crack length
for crack lengths $> 6~\mathrm{mm}$ ($0.24~\mathrm{in.}$), which is
an important consideration for the RP method that was
introduced in Section~\ref{section:rp.method.pod}.
These data will be used to compute
POD using the SoDaD method described in
Sections~\ref{section:sodad.method.pod} and the RP method
described in Section~\ref{section:rp.method.pod}.

\subsubsection{Estimates of POD from the SoDaD (LaD) Method for the
  PZT data}
\label{section:pzt.sodad.pod.estimates}
Figure~\ref{figure:pzt.data.crossing}(b) is a plot of the PZT data with
individual fitted regression lines.
The points along the horizontal line at $\DI{}=0.05$
indicate the LaD for each crack/sensor combination, defined as the crack
length after the first time that the \DI{} crosses the
detection threshold, simulating periodic interrogation of the sensor
network that is commonly used in actual applications.
Table~\ref{table:pzt_lad_slope} provides a summary of the data.

\begin{table}[tbp]
\centering
\caption{Crack LaD and estimated slope for each
  crack/sensor combination in the PZT data}
\label{table:pzt_lad_slope}
\renewcommand{\arraystretch}{1.08}
\footnotesize
\begin{tabular}{lll}
\hline
Specimen   & LaD (mm) & Slope \\
\hline
RB-PZT-7   &  4.35 & 0.041 \\
RB-PZT-8   &  4.35 & 0.051 \\
RB-PZT-17F &  5.00 & 0.023 \\
RB-PZT-11  &  5.05 & 0.043 \\
RB-PZT-13  &  5.10 & 0.029 \\
RB-PZT-6   &  5.20 & 0.047 \\
RB-PZT-16  &  5.30 & 0.043 \\
RB-PZT-14  &  5.65 & 0.031 \\
RB-PZT-10  &  5.95 & 0.038 \\
RB-PZT-12  &  5.95 & 0.037 \\
RB-PZT-18F &  6.05 & 0.046 \\
RB-PZT-15  &  6.20 & 0.052 \\
RB-PZT-9   &  6.50 & 0.038 \\
\hline
\end{tabular}
\end{table}

Probability plots are used to assess and compare the adequacy of
distributional models for data. If the plotted points for a plot
corresponding to a particular distribution are approximately linear,
one can say that the data are consistent with the distribution.
\begin{figure*}[!tbp]
\centering
\begin{minipage}[t]{0.4\textwidth}
    \includegraphics[width=\textwidth,trim=0 20.4pt 0 20.4pt,clip]{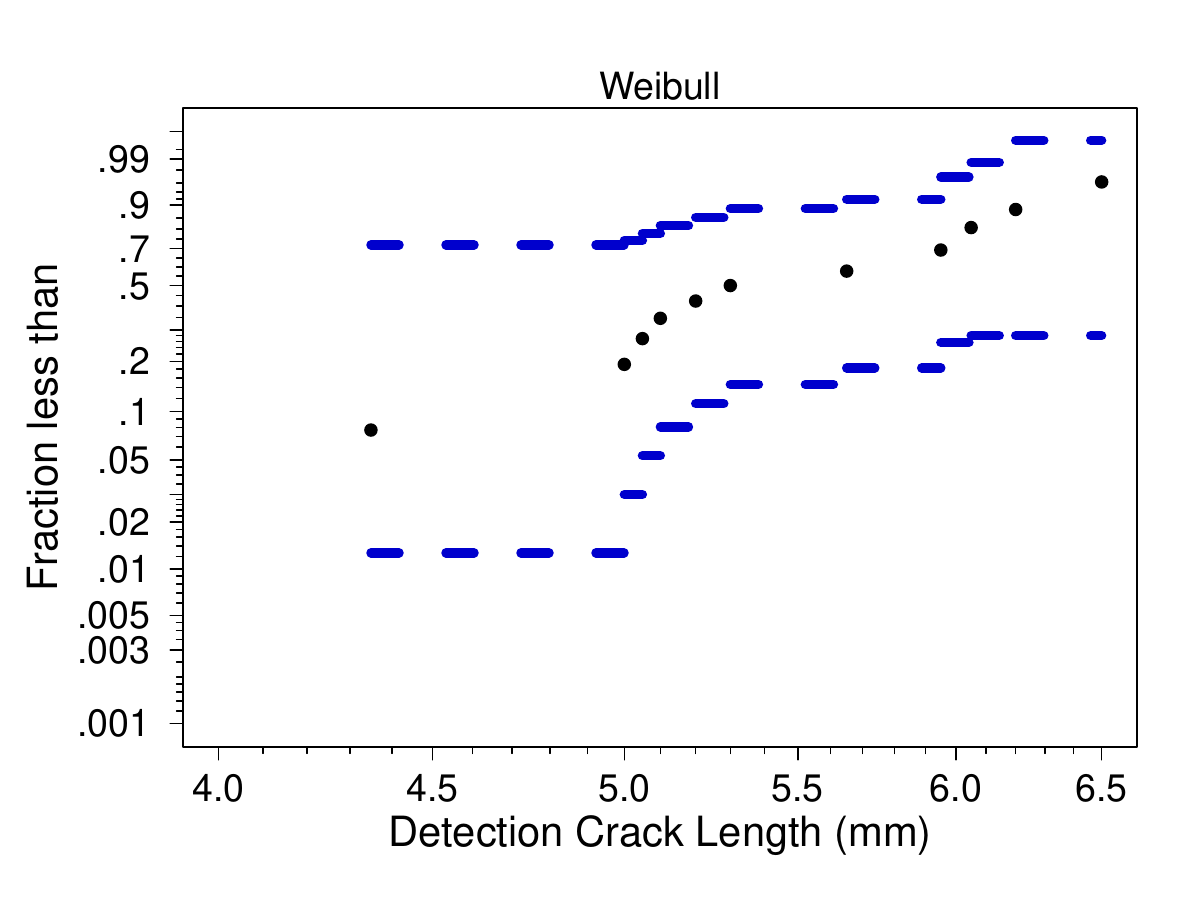}
\end{minipage}
\hspace{0.01\textwidth}
\begin{minipage}[t]{0.4\textwidth}
    \includegraphics[width=\textwidth,trim=0 13.7pt 0 13.7pt,clip]{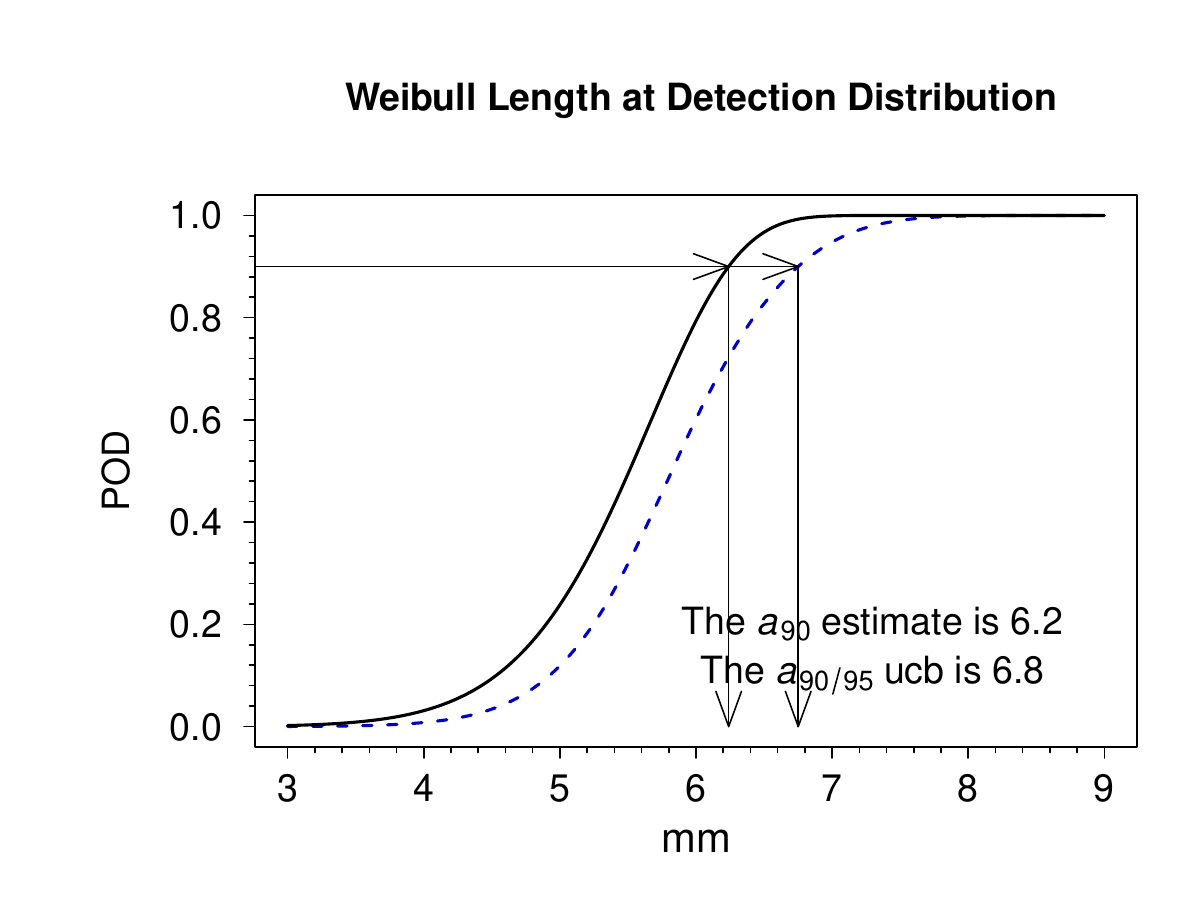}
\end{minipage}

\begin{minipage}[t]{0.4\textwidth}
    \includegraphics[width=\textwidth,trim=0 21.7pt 0 21.7pt,clip]{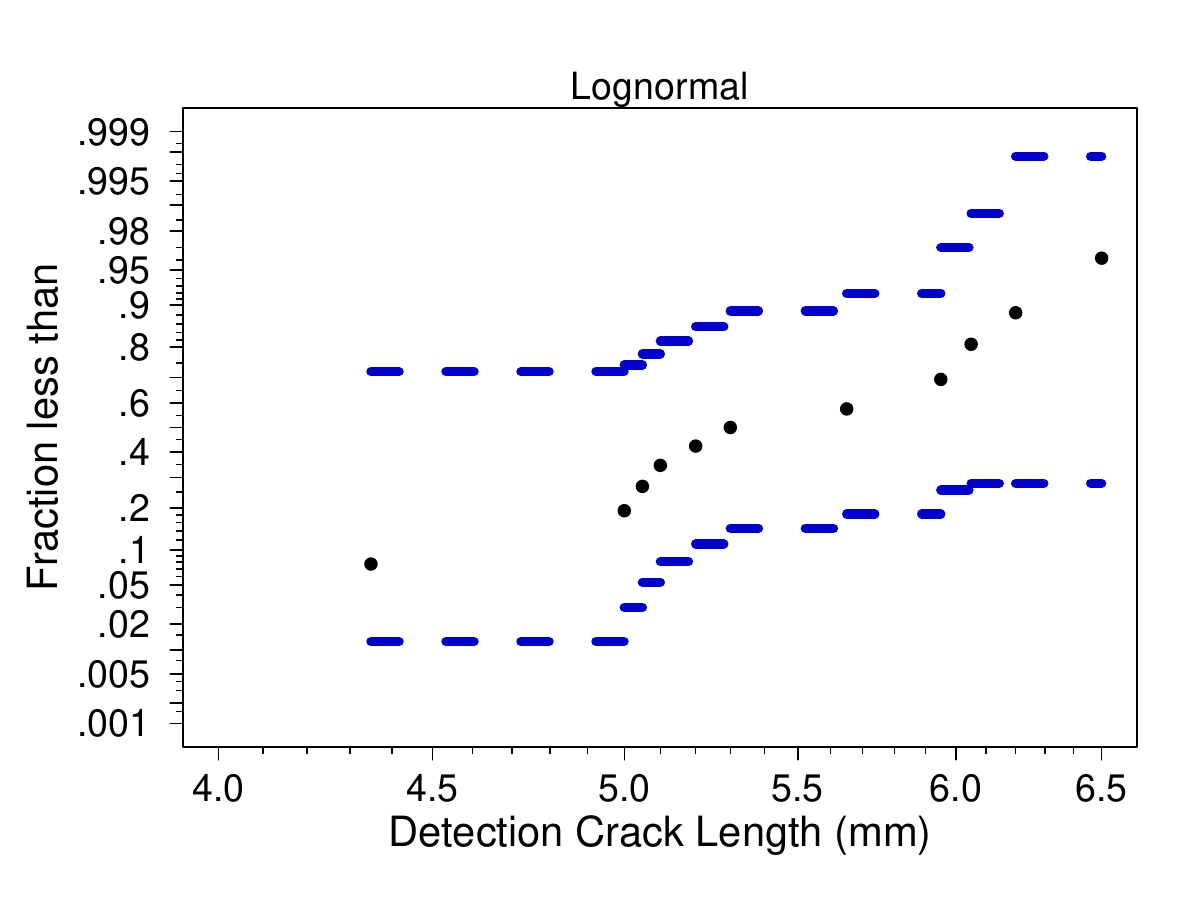}
\end{minipage}
\hspace{0.01\textwidth}
\begin{minipage}[t]{0.4\textwidth}
    \includegraphics[width=\textwidth,trim=0 14.8pt 0 14.8pt,clip]{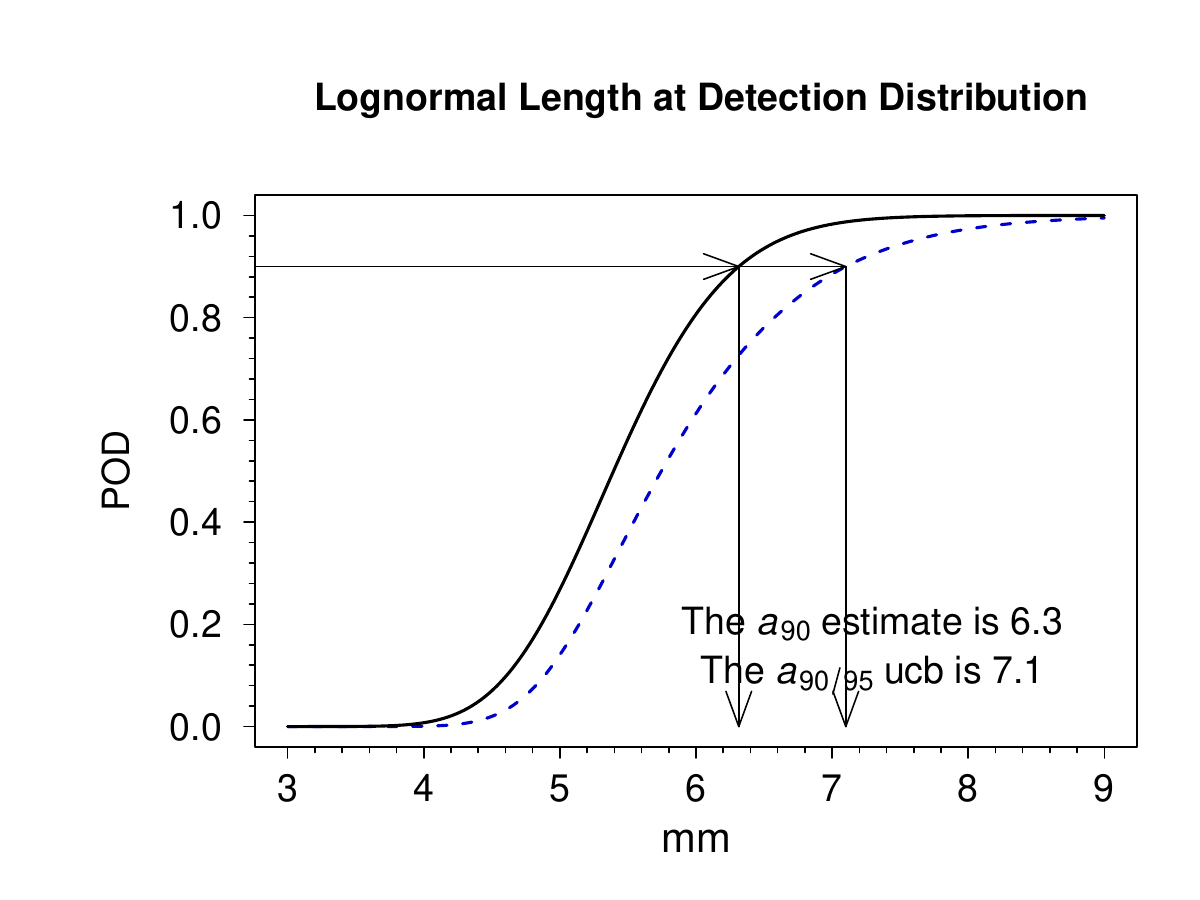}
\end{minipage}

\begin{minipage}[t]{0.4\textwidth}
    \includegraphics[width=\textwidth,trim=0 16.3pt 0 16.3pt,clip]{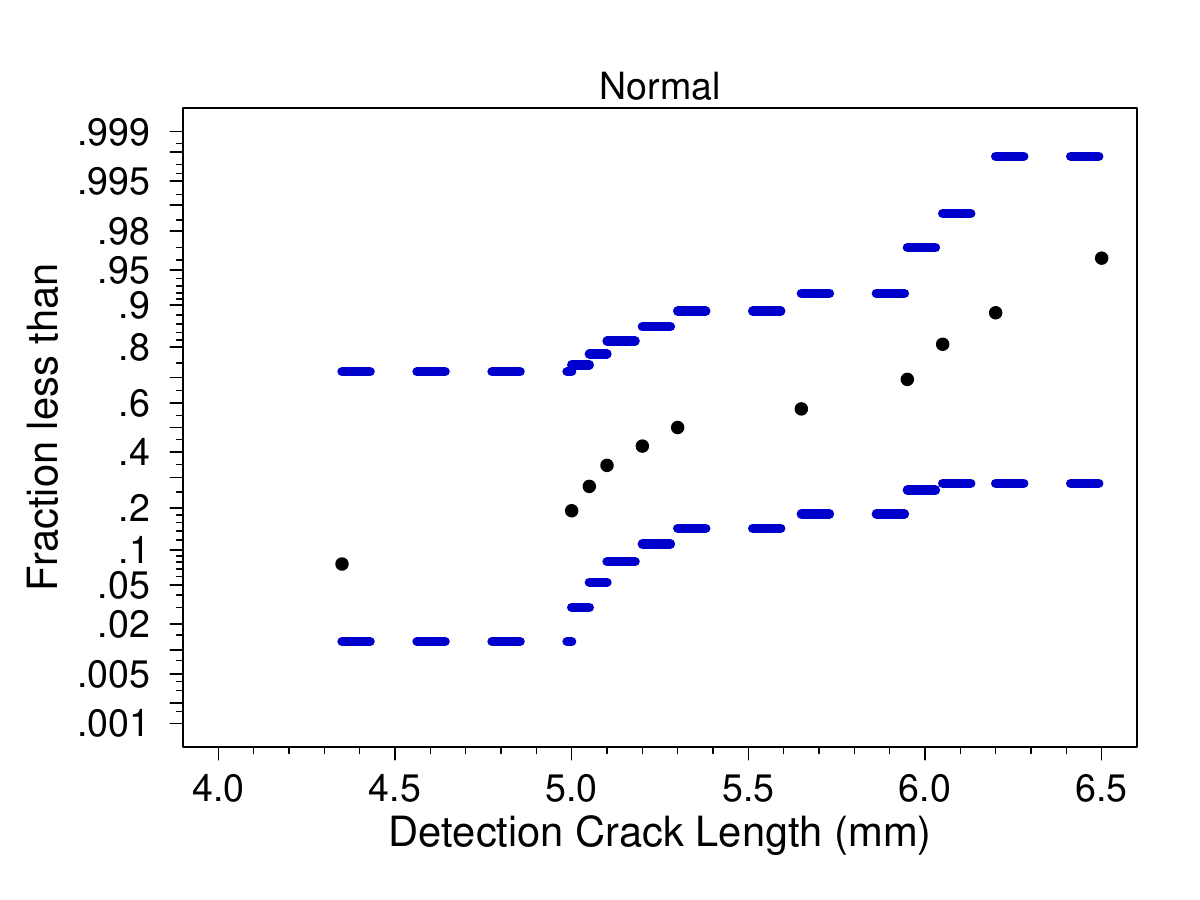}
\end{minipage}
\hspace{0.01\textwidth}
\begin{minipage}[t]{0.4\textwidth}
    \includegraphics[width=\textwidth,trim=0 15.9pt 0 15.9pt,clip]{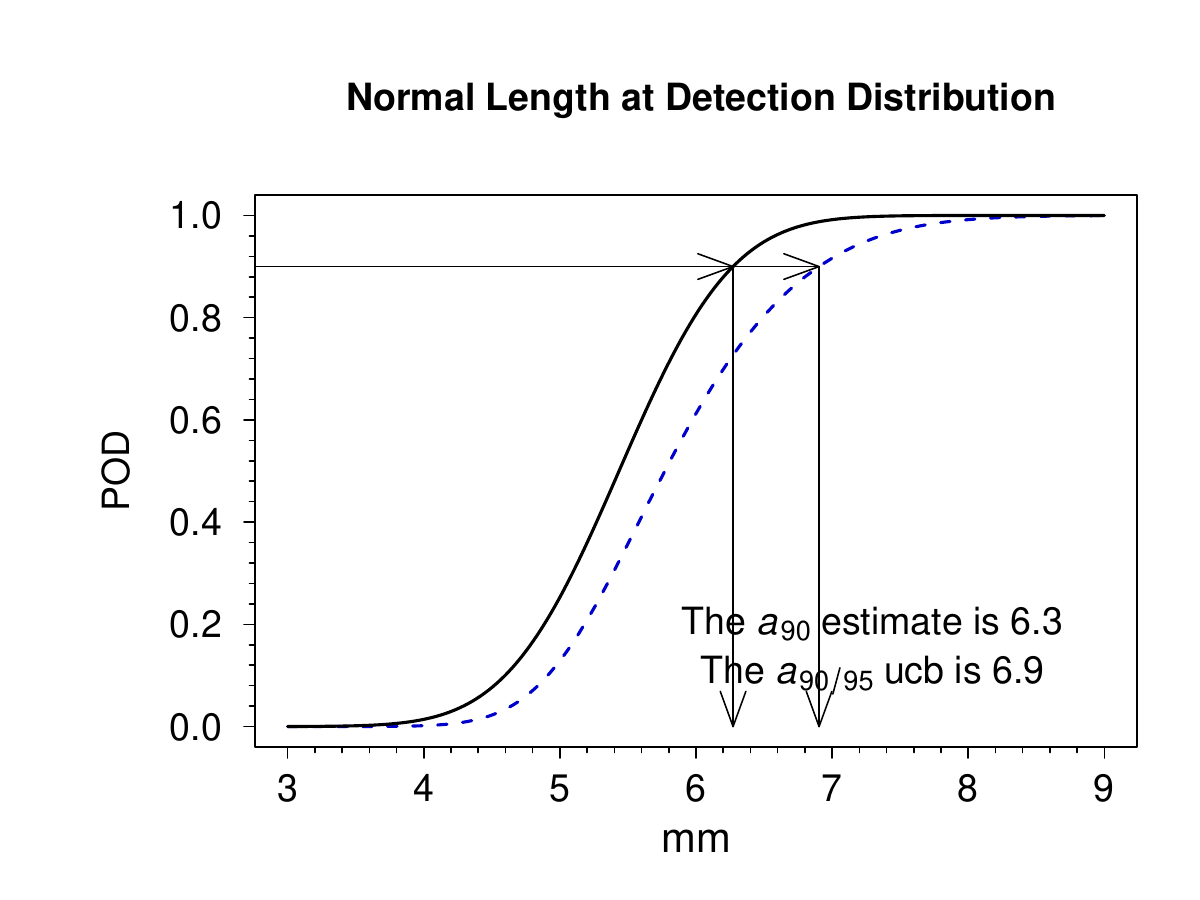}
\end{minipage}

\begin{minipage}[t]{0.4\textwidth}
    \includegraphics[width=\textwidth,trim=0 22.7pt 0 22.7pt,clip]{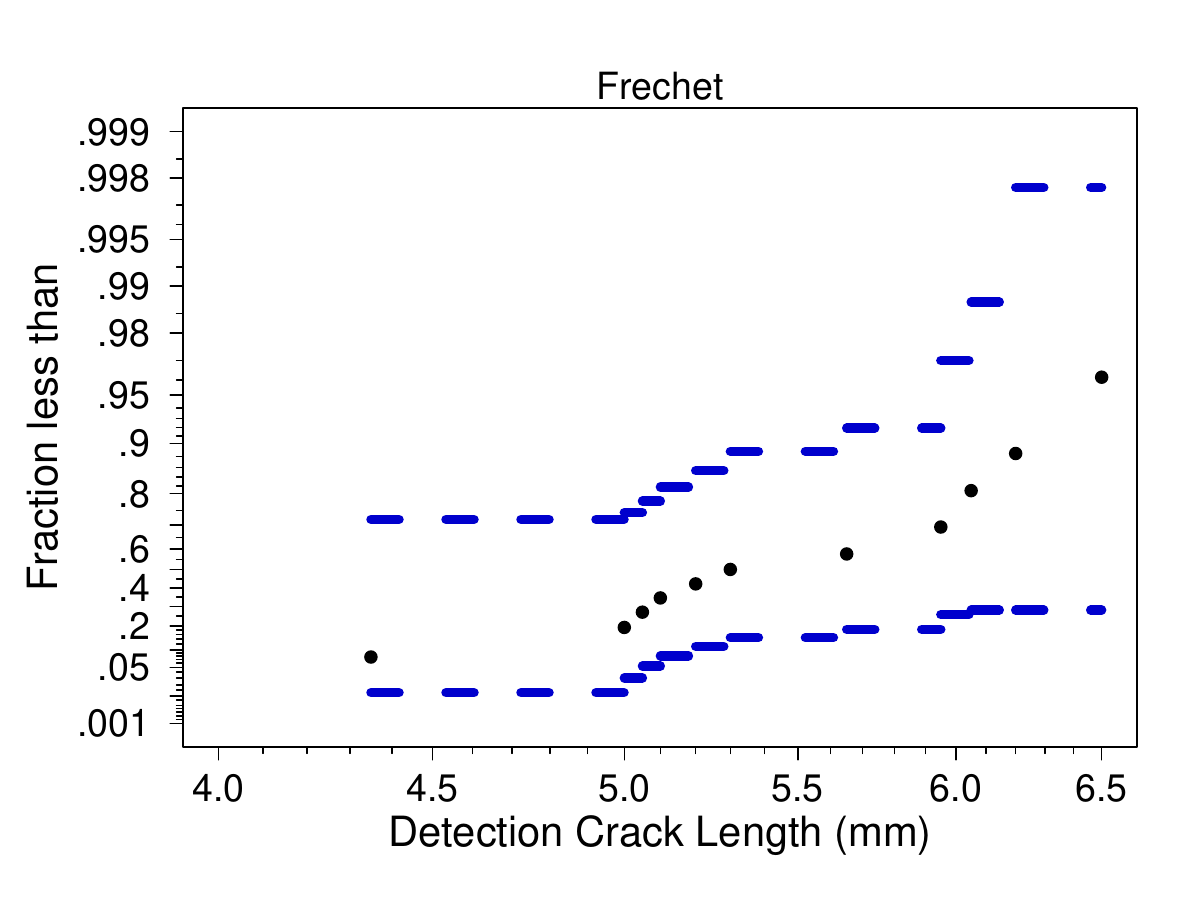}
\end{minipage}
\hspace{0.01\textwidth}
\begin{minipage}[t]{0.4\textwidth}
    \includegraphics[width=\textwidth,trim=0 14.7pt 0 14.7pt,clip]{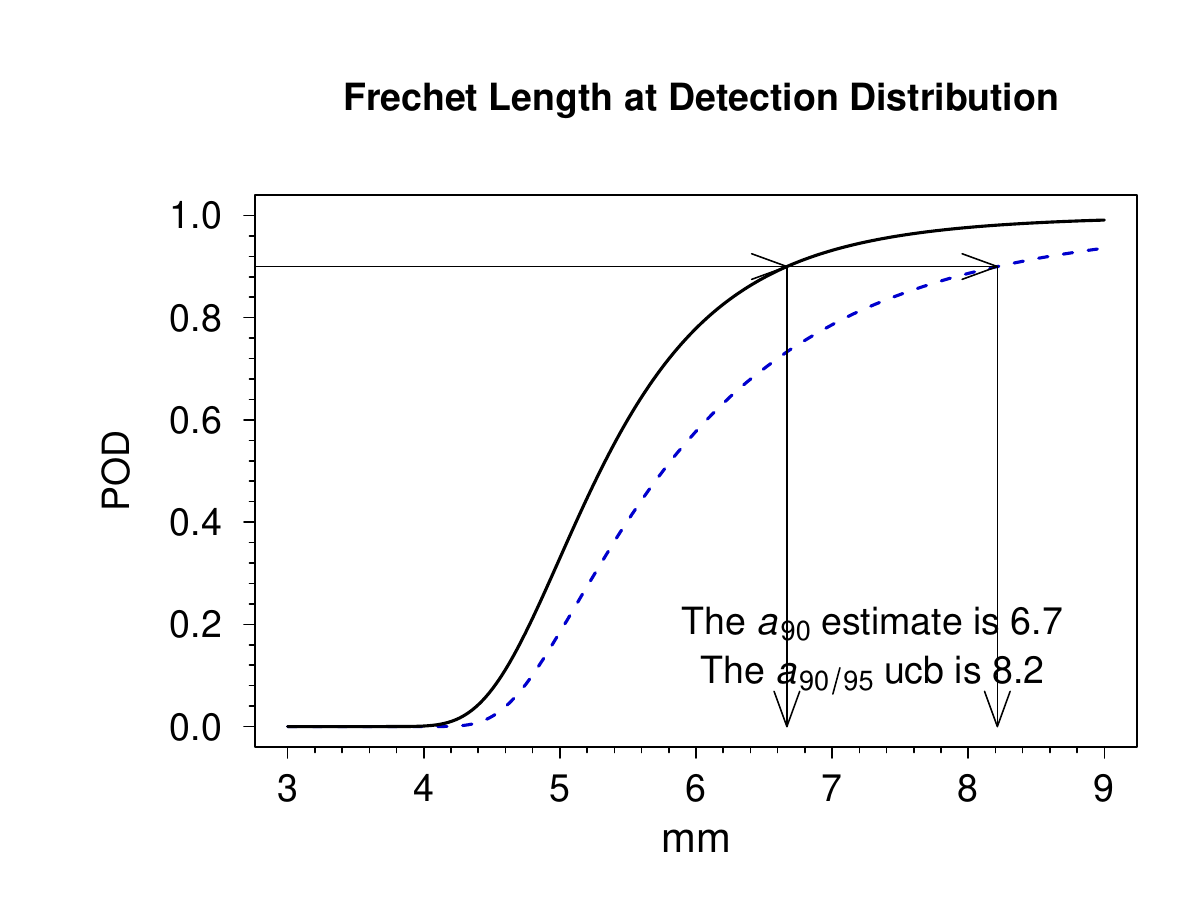}
\end{minipage}

\caption{In the left column are probability plots with 95\%
  simultaneous confidence bands for the LaD
  values from the PZT data for the (from top to bottom) Weibull,
  lognormal, normal, and largest extreme value distributions. Plots
  on the right provide estimates and 95\% lower confidence bounds
  for POD for the same four distributions.}
\label{figure:pzt_prob_pod_distributions}
\end{figure*}
More specifically, if one can draw a straight line through 95\%
simultaneous confidence bands plotted on a probability plot for a
particular distribution, then the data are consistent, at the 95\%
level of confidence, with that distribution. See
\citet[][Chapter~6]{meeker2022statistical} for more
information about probability plots and simultaneous confidence
bands.

The plots on the left in Figure~\ref{figure:pzt_prob_pod_distributions}
are probability plots with 95\% simultaneous confidence bands for
the LaD values from the PZT data for the
(from top to bottom) Weibull, lognormal, normal, and largest extreme
value distributions. Note that what is being plotted in these four
probability plots is exactly the same. What is changing are the
scales of the plots. These kind of plots were made for six different
probability distributions (see Table~\ref{table:pod_estimates_lad_rp})
and, to save space, the four best fitting
or most relevant were chosen, for each example, for the figures.

The plots show that none of the
distributions can be ruled out. In such cases, unless knowledge of
the physics of inspection would suggest which
distribution is more appropriate, a
reasonable choice would be the distribution that gives the largest
value of $a_{90/95}$. The need to choose a distribution for LaD
values is
similar to the need to choose a distribution for the $\hat{a}$
versus $a$ method described in \citet[][Appendix~G]{milhdbk1823a2009}.

Plots on the right-hand side of
Figure~\ref{figure:pzt_prob_pod_distributions} provide estimates and
95\% lower confidence bounds for POD for the same four
distributions. The UCB (95\% upper confidence bound on the crack
length that will be detected with probability 0.90) is also indicated
on these plots. In this application, there is little difference
between the normal and lognormal distributions, partially because
the dynamic range of the data (ratio of the maximum to the minimum)
is small, so that the log transformation has little effect on the
statistical estimates from the model. The larger $a_{90/95}$ for the largest
extreme value distribution is due to the heavier upper tail of that
distribution.

\subsubsection{Estimates of POD from the Random-Parameters Model for the
  PZT data}
\label{section:pzt.rp.pod.estimates}
\begin{figure}[tbp]
\begin{tabular}{cc}
(a) & (b) \\[-0.3ex]
  \includegraphics[width=0.5\columnwidth]{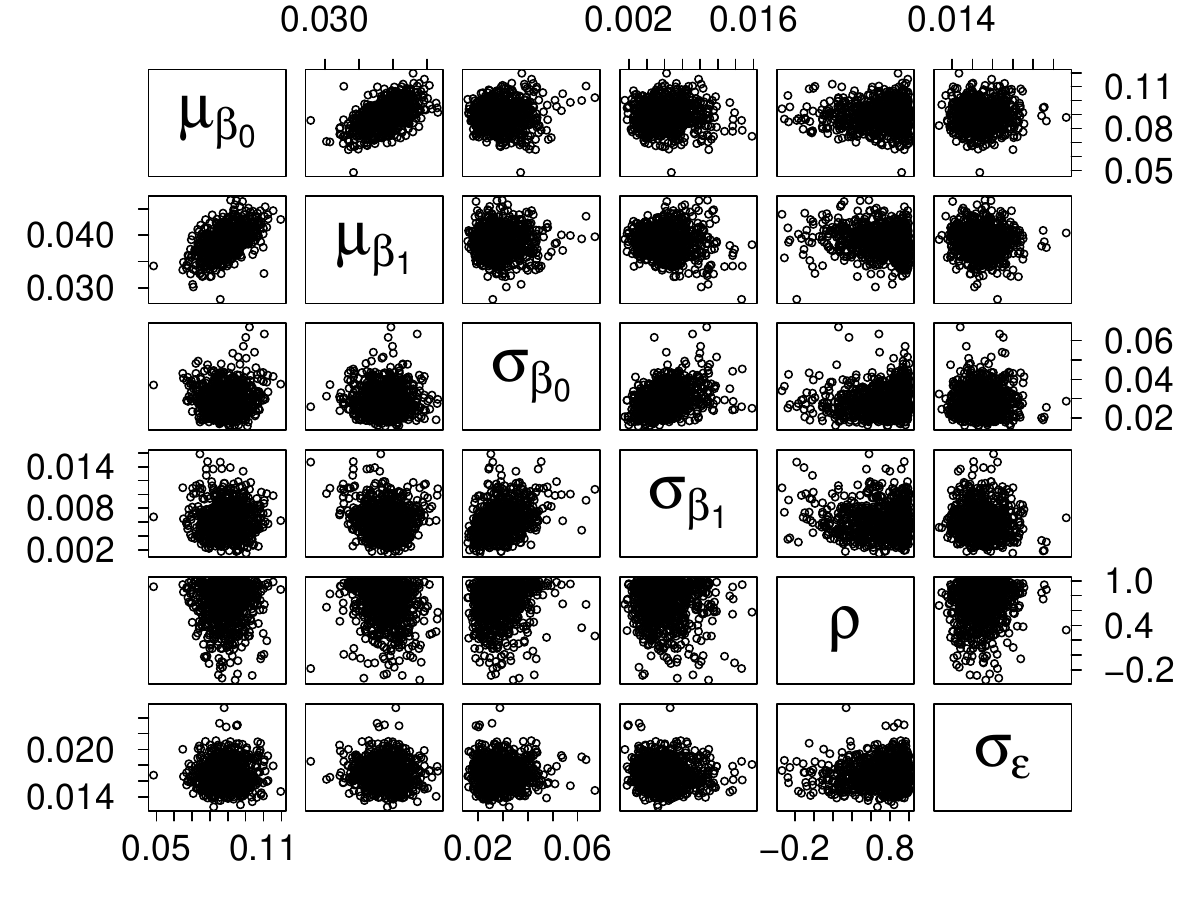} &
  \includegraphics[width=0.5\columnwidth]{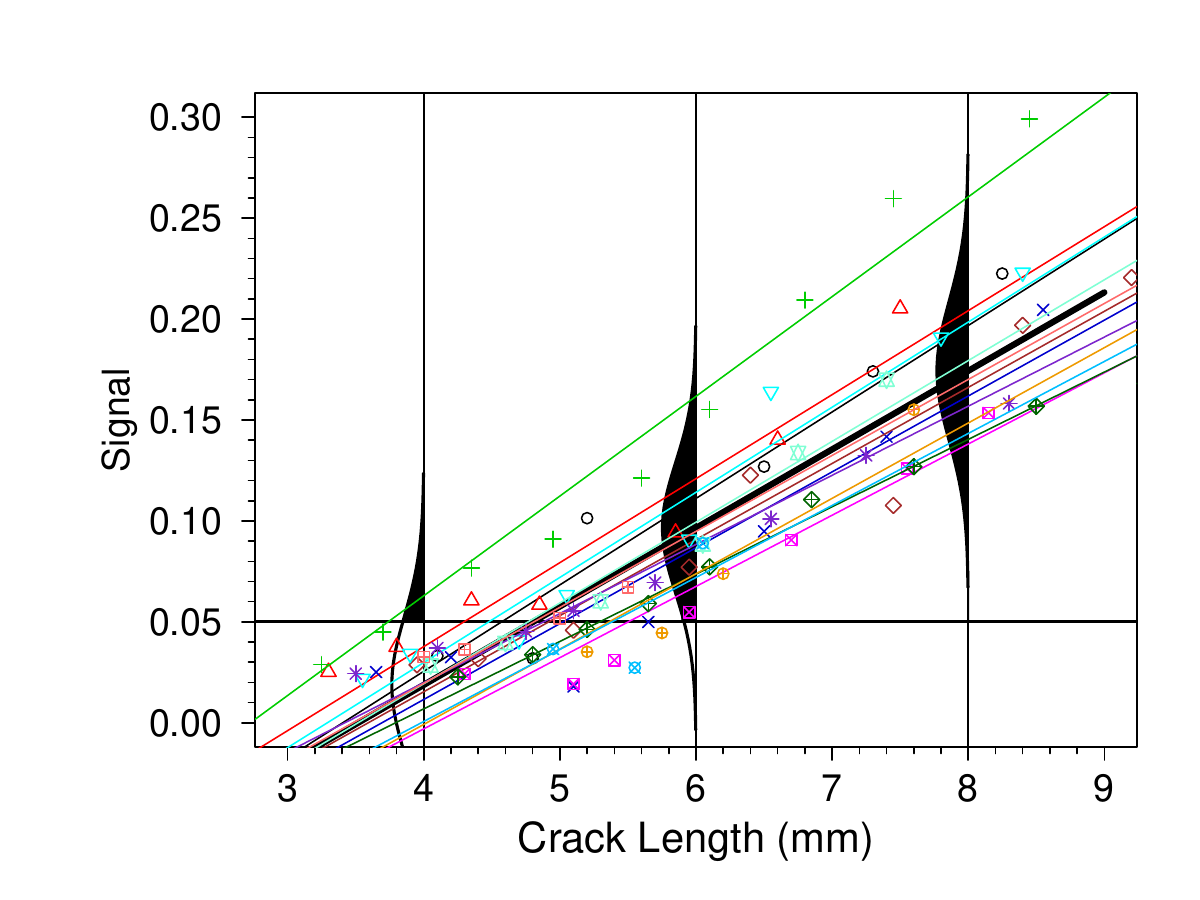}
\end{tabular}
\caption{A scatterplot matrix of the draws from the joint posterior
  distribution of the model parameters for the PZT data~(a) and a
  plot of the corresponding fitted RP model with POD indicated by
  the shaded area under the normal densities
  (bottom)~(b).}
\label{figure:random_effects_pair_response_PZT}
\end{figure}

The scatterplot matrix in
Figure~\ref{figure:random_effects_pair_response_PZT}a provides a graphical
summary of draws from the joint posterior distribution
for the RP model described in Section~\ref{section:rp.method.pod}.
The distributions
are well behaved and the commonly used MCMC diagnostics
(details not provided here) suggest that the
draws will accurately describe the joint posterior
distribution. Along the diagonal are histograms of the marginal
posterior distributions for each of the six parameters. The
off-diagonal plots are scatter plots of the bivariate marginal
posterior distributions for each pair of parameters. We can see some
positive correlation between the estimates of $\mu_{\beta_0}$ and
$\mu_{\beta_1}$. On the other hand, we see no correlation between
$\sigma_{\varepsilon}$ and either $\mu_{\beta_0}$ or
$\mu_{\beta_1}$, as predicted by statistical theory.

Draws from the marginal distributions of $\beta_{0i}$ and
$\beta_{1i}$, $i = 1, \ldots, n$ are also provided. The medians of
these distributions are estimates of the intercepts and slopes,
respectively for the individual crack/sensor combinations and these
were used to plot the regression lines in
Figure~\ref{figure:random_effects_pair_response_PZT}b. The probability
densities and the thicker line in the center give the response
distribution and estimated mean response as a function of crack length
and were computed using the marginal posterior draws of the model
parameters $\mu_{\beta_0}$, $\mu_{\beta_1}$, $\sigma_{\beta_0}$,
$\sigma_{\beta_1}$, $\rho$, and $\sigma_{\varepsilon}$.

These posterior draws are also substituted
into~(\ref{equation:rp.model.pod}) and
(\ref{eq:rp2}) for a given value of length to provide draws from the
posterior marginal distribution of the POD function at that value of
length. Then an estimate of POD is obtained from the median of these
draws. The lower 95\% credible (confidence) bound is obtained from
the 0.05 quantile of these marginal posterior draws. The
computations are repeated for different values of length.
The results are plotted in
Figure~\ref{figure:random_effects_pair_response_PZT}b. The
$\hat{a}_{90}$ and $a_{90/95}$ values are a little smaller than
those from the SoDaD method because of the implicit interpolation to
estimate the crossing length, relative to the SoDaD method which
uses the crossing length at the first inspection after the damage
index crosses the detection threshold.

\subsection{Application to the Carbon Nanotube (CNT) SHM Data}
\label{section:cnt.application}
\subsubsection{CNT background and data}
This section describes the analysis of the carbon nanotube (CNT)
sensor POD data using our SHM-POD methods, first introduced in
Section~\ref{section:technical.background}.
CNT networks involve literally trillions of CNTs forming
the electrical path, each contributing to an overall network
resistance as a member in parallel and in series with adjacent
members.
As a surface-breaking crack grows anywhere under the CNT network,
the measured resistance will change as a function of the crack
length.  For the experiments described here, $50 \times 20$ mm CNT
sensors were bonded to $300 \times 25 \times 3$ mm 6061-T6 Aluminum
bars with a strain gauge epoxy. A $0.25 \times 1.5 $ mm EDM
crack-starter notch was cut
into one edge of each specimen.
The specimens were loaded in 4-point-bending with 25 mm
between the inner rollers and 200 mm between the outer rollers, and
cycles at 80\% yield strain (3300 microstrain). Resistance was
measured every 1000 cycles in the unloaded condition, and crack
length was estimated by simply solving the prior equation for crack
length using a known relationship between crack length and
resistance. More details about CNT sensors and the experiment are
given in \citet{kessler2015carbon} and \citet{kessler2019detection}.
\begin{figure}[tbp]
\begin{tabular}{cc}
(a) & (b) \\[-0.80ex]
  \includegraphics[width=0.5\columnwidth,
    trim=0 22.9pt 0 22.9pt,clip]{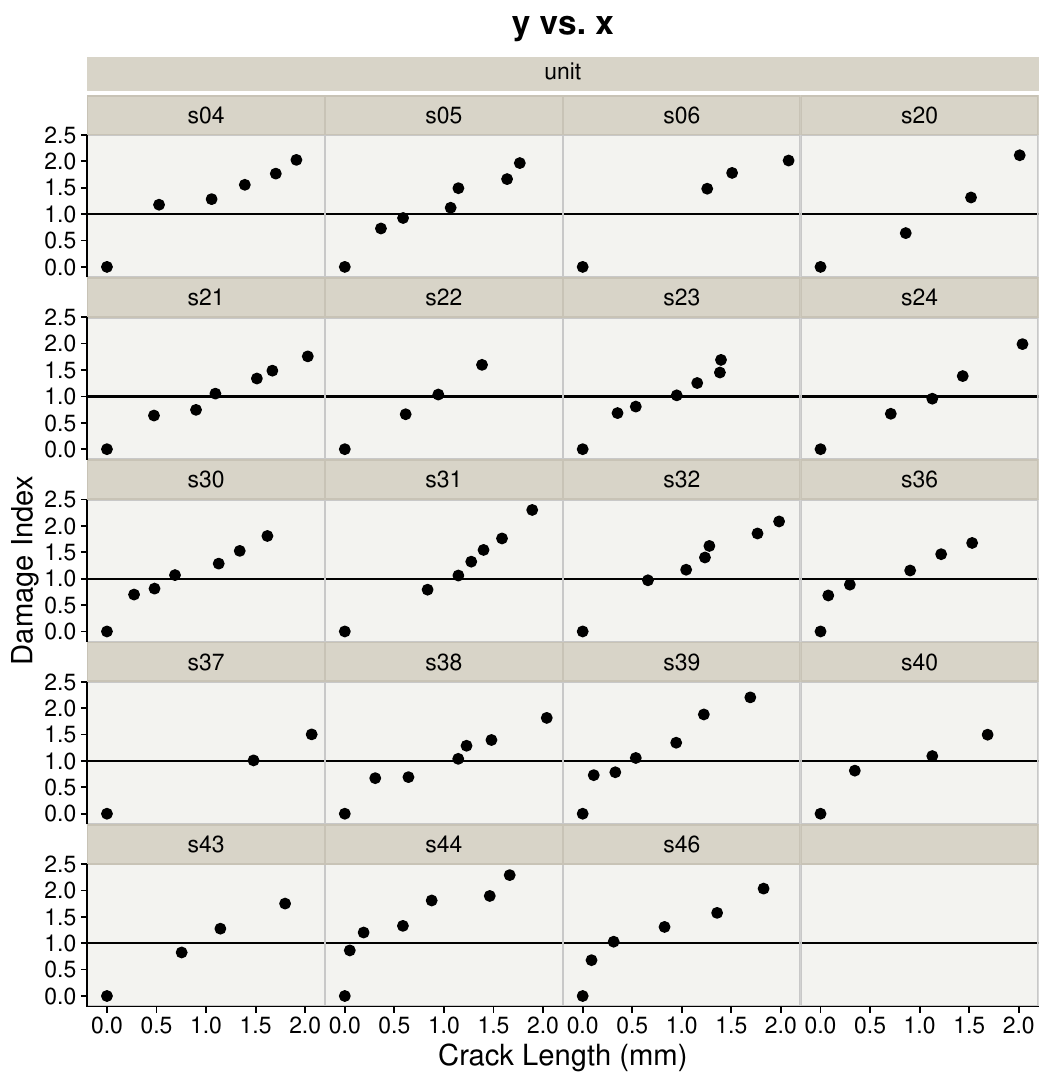} &
  \includegraphics[width=0.5\columnwidth,
    trim=0 20.0pt 0 20.0pt,clip]{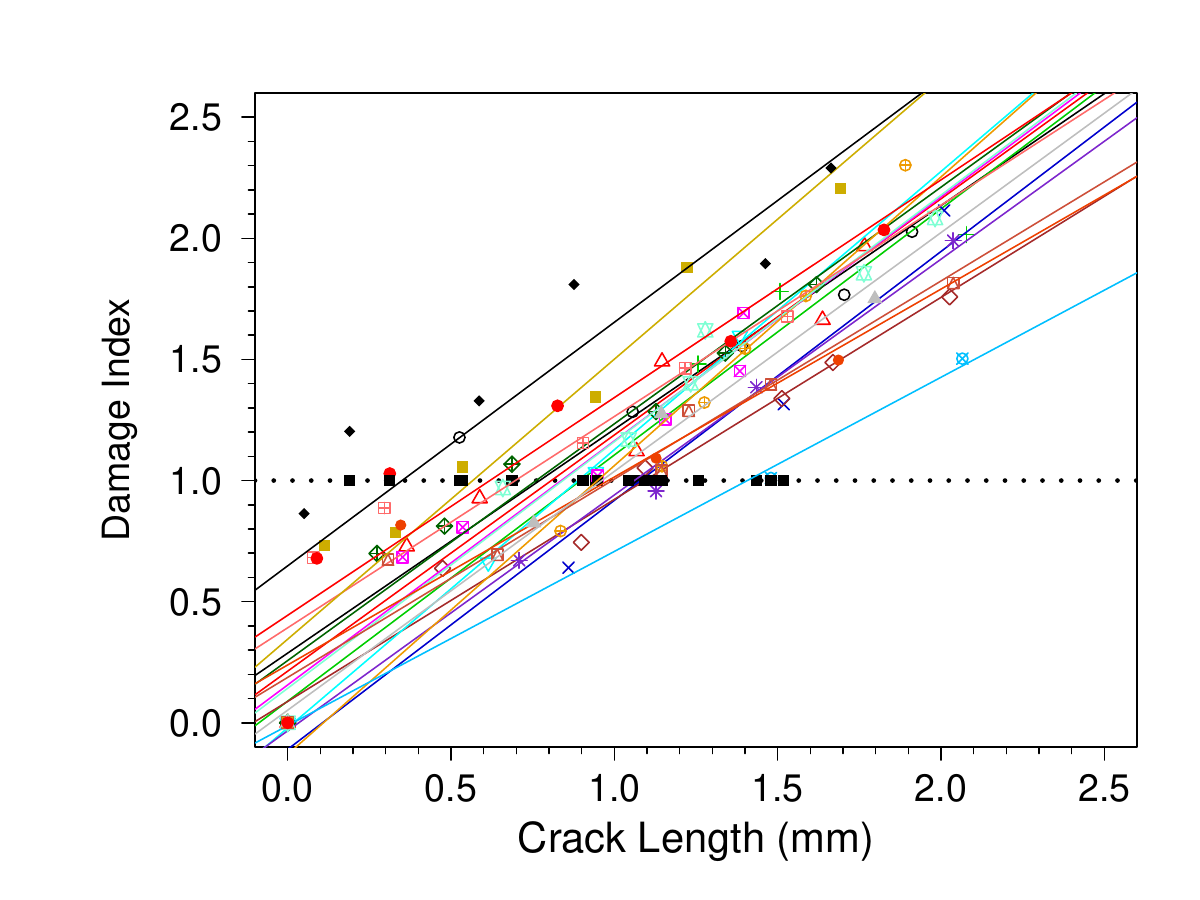}
\end{tabular}
\caption{Plot of the CNT \DI{} versus crack length. Each plot
  is labeled with the specimen ID and the horizontal lines indicate
  the detection threshold of 1.00~(a) and a
  plot of the CNT POD data with a linear regression line for
  each crack/sensor combination. The LaD values
  corresponding to the detection threshold of 1.00 are also shown~(b).}
\label{figure:cnt.data.crossing}
\end{figure}
The statistical methods used in this section are
the same as those used in Sections~\ref{section:pzt.sodad.pod.estimates}
and~\ref{section:pzt.rp.pod.estimates} for the
PZT data.

\subsubsection{Estimates of POD from the SoDaD (LaD) Method for the
  CNT data}
\label{section:cnt.sodad.pod.estimates}
Figure~\ref{figure:cnt.data.crossing}a is a plot of the CNT sensor POD
data for cracks less than $2.1~\mathrm{mm}$ ($0.083~\mathrm{in.}$)
in length for each crack/sensor combination. The horizontal line at
$\DI{}=1$ is the detection threshold.
Figure~\ref{figure:cnt.data.crossing}b is a plot of the CNT data with
individual fitted regression lines for each crack/sensor
combination. The points along the horizontal line at $\DI{}=1$
indicate the lengths at detection (LaD) for each crack/sensor
combination (again defined as the length of the crack at the first
inspection after the \DI{} exceeds the
threshold). Table~\ref{table:cnt_lad_slope}  contains the numerical
values of the LaD and slopes.

\begin{table}[tbp]
\centering
\caption{LaD and estimated slope for each
  crack/sensor combination in the CNT data}
\label{table:cnt_lad_slope}
\renewcommand{\arraystretch}{1.08}
\begin{tabular}{lll}
\hline
Specimen & LaD (mm) & Slope \\
\hline
s44 & 0.050 & 1.006 \\
s36 & 0.296 & 0.872 \\
s46 & 0.313 & 0.899 \\
s39 & 0.329 & 1.156 \\
s40 & 0.346 & 0.776 \\
s30 & 0.480 & 0.977 \\
s04 & 0.526 & 0.924 \\
s23 & 0.536 & 1.007 \\
s05 & 0.588 & 0.975 \\
s32 & 0.659 & 1.018 \\
s43 & 0.754 & 0.986 \\
s31 & 0.835 & 1.192 \\
s22 & 0.943 & 1.149 \\
s21 & 1.094 & 0.834 \\
s24 & 1.128 & 0.973 \\
s38 & 1.144 & 0.818 \\
s06 & 1.257 & 1.016 \\
s37 & 1.479 & 0.719 \\
s20 & 1.518 & 1.028 \\
\hline
\end{tabular}
\end{table}

\begin{figure*}[!tbp]
\centering
\begin{minipage}[t]{0.4\textwidth}
    \includegraphics[width=\textwidth,trim=0 19.0pt 0 19.0pt,clip]{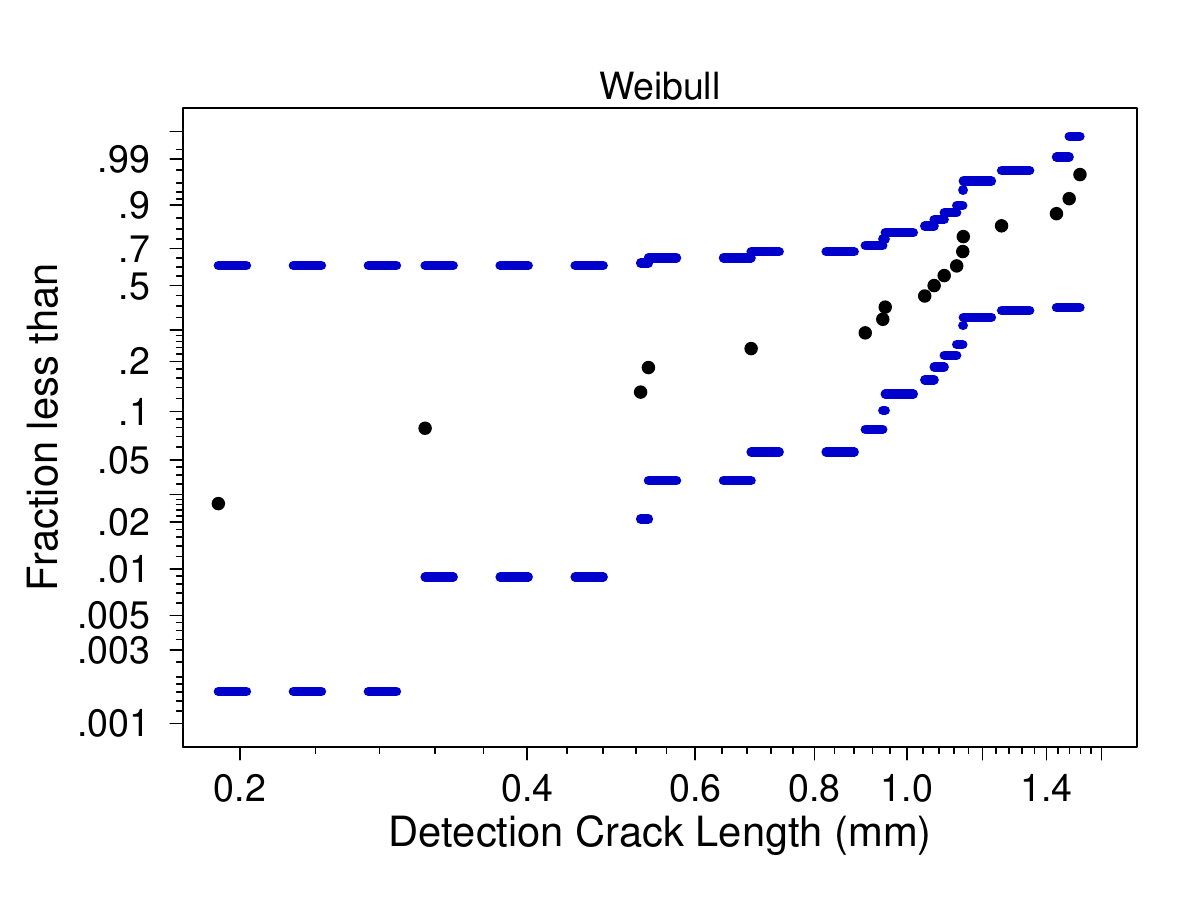}
\end{minipage}
\hspace{0.01\textwidth}
\begin{minipage}[t]{0.4\textwidth}
    \includegraphics[width=\textwidth,trim=0 14.0pt 0 14.0pt,clip]{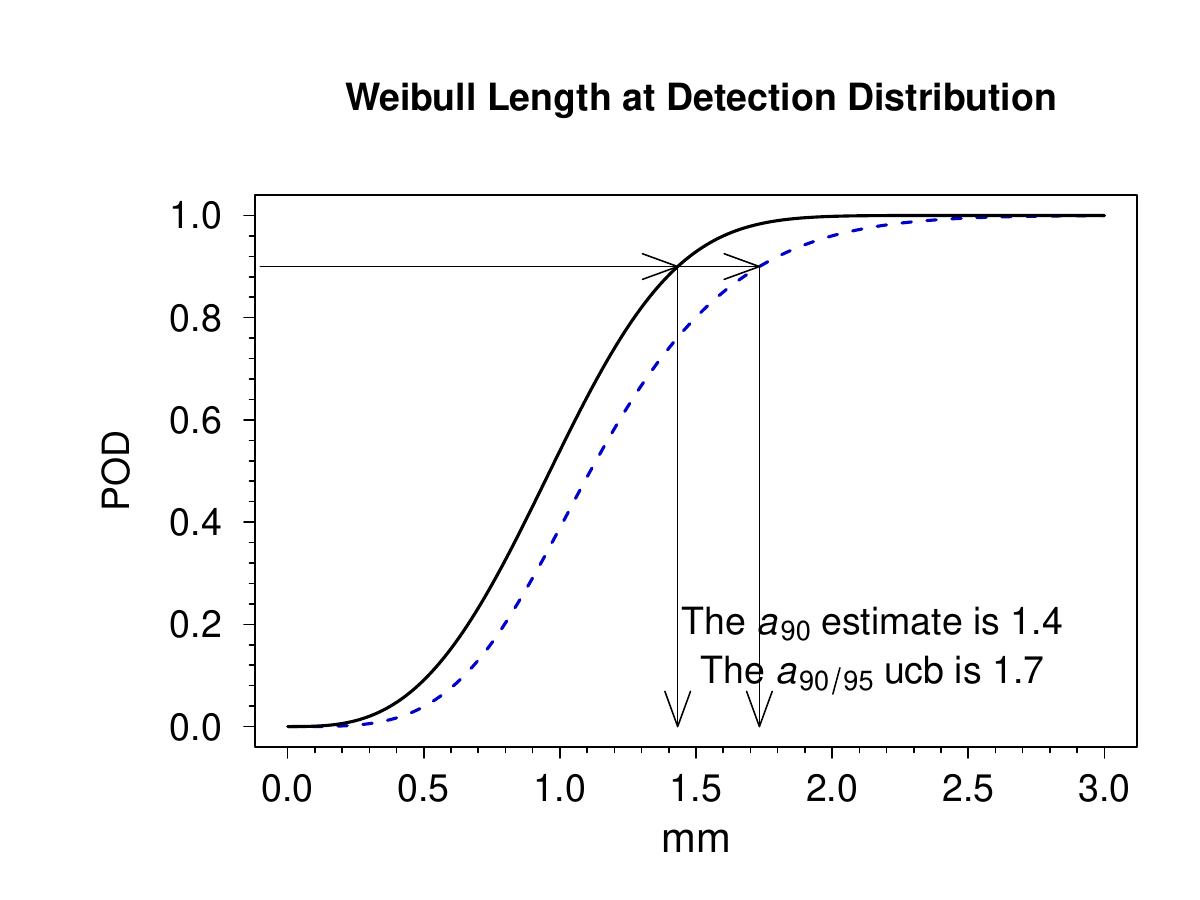}
\end{minipage}

\begin{minipage}[t]{0.4\textwidth}
    \includegraphics[width=\textwidth,trim=0 21.4pt 0 21.4pt,clip]{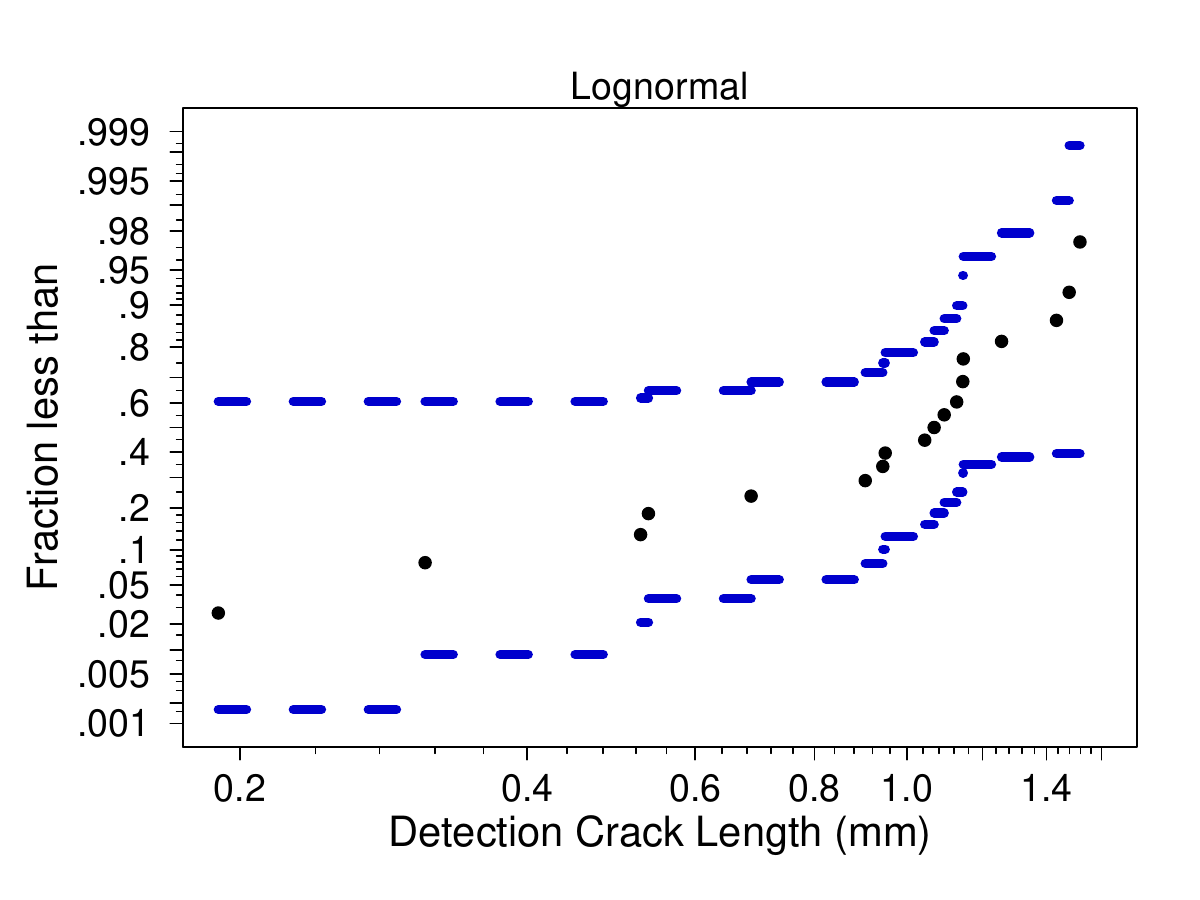}
\end{minipage}
\hspace{0.01\textwidth}
\begin{minipage}[t]{0.4\textwidth}
    \includegraphics[width=\textwidth,trim=0 18.9pt 0 18.9pt,clip]{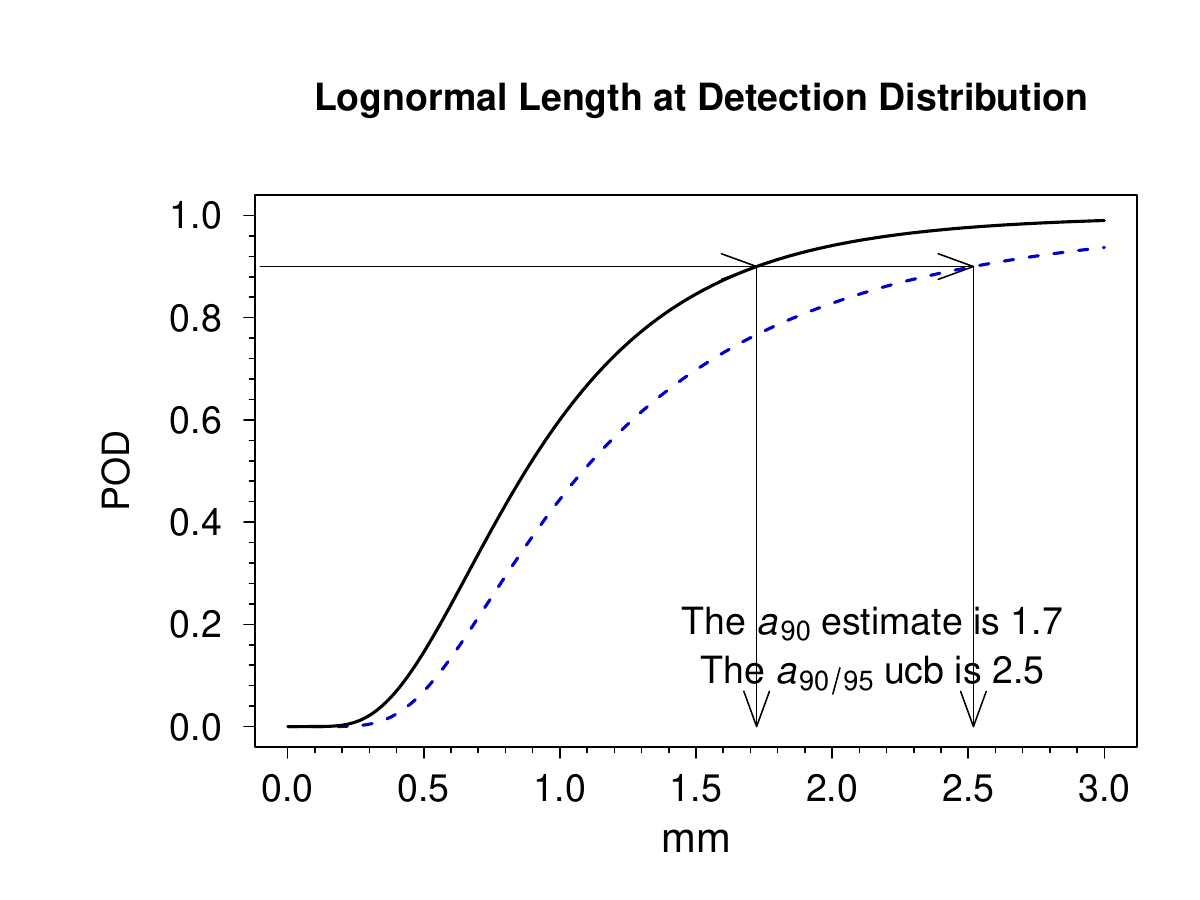}
\end{minipage}

\begin{minipage}[t]{0.4\textwidth}
    \includegraphics[width=\textwidth,trim=0 17.7pt 0 17.7pt,clip]{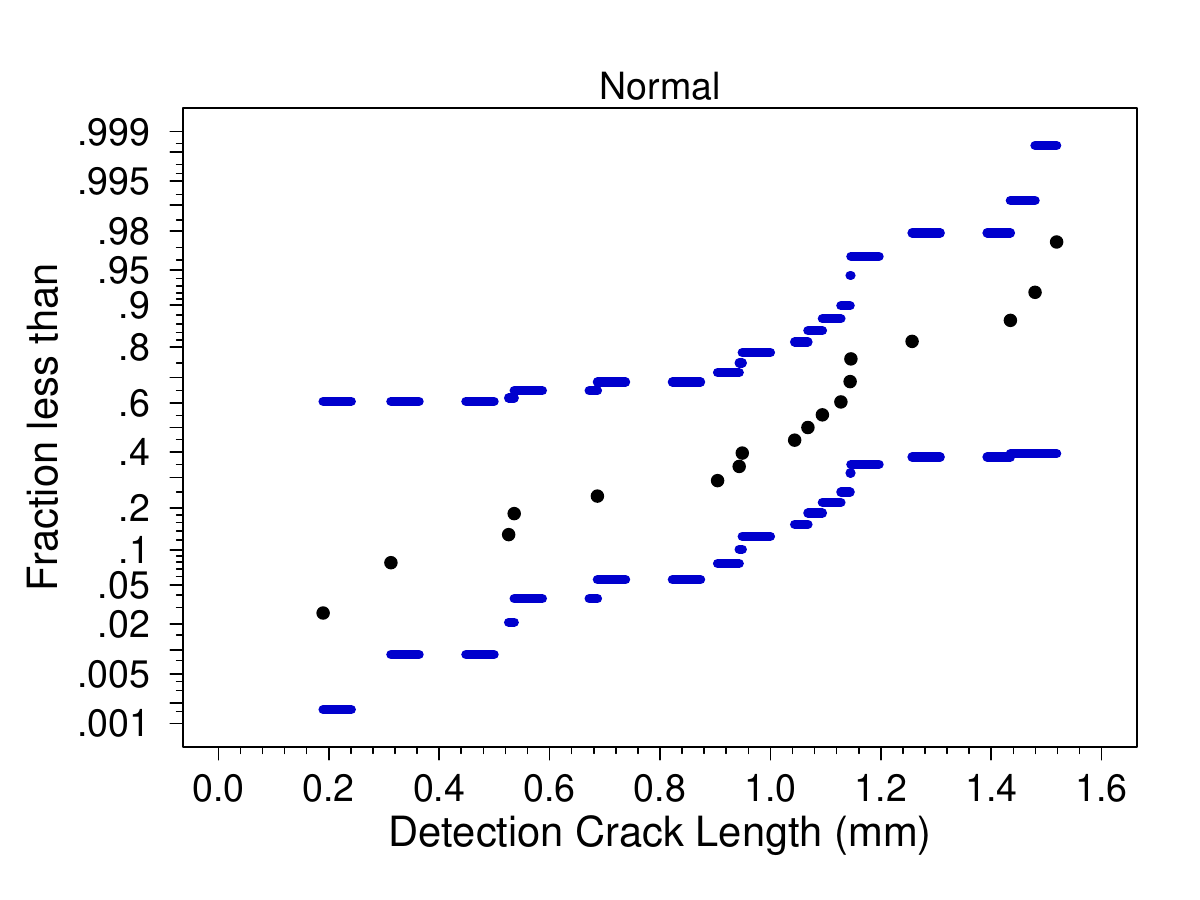}
\end{minipage}
\hspace{0.01\textwidth}
\begin{minipage}[t]{0.4\textwidth}
    \includegraphics[width=\textwidth,trim=0 20.1pt 0 20.1pt,clip]{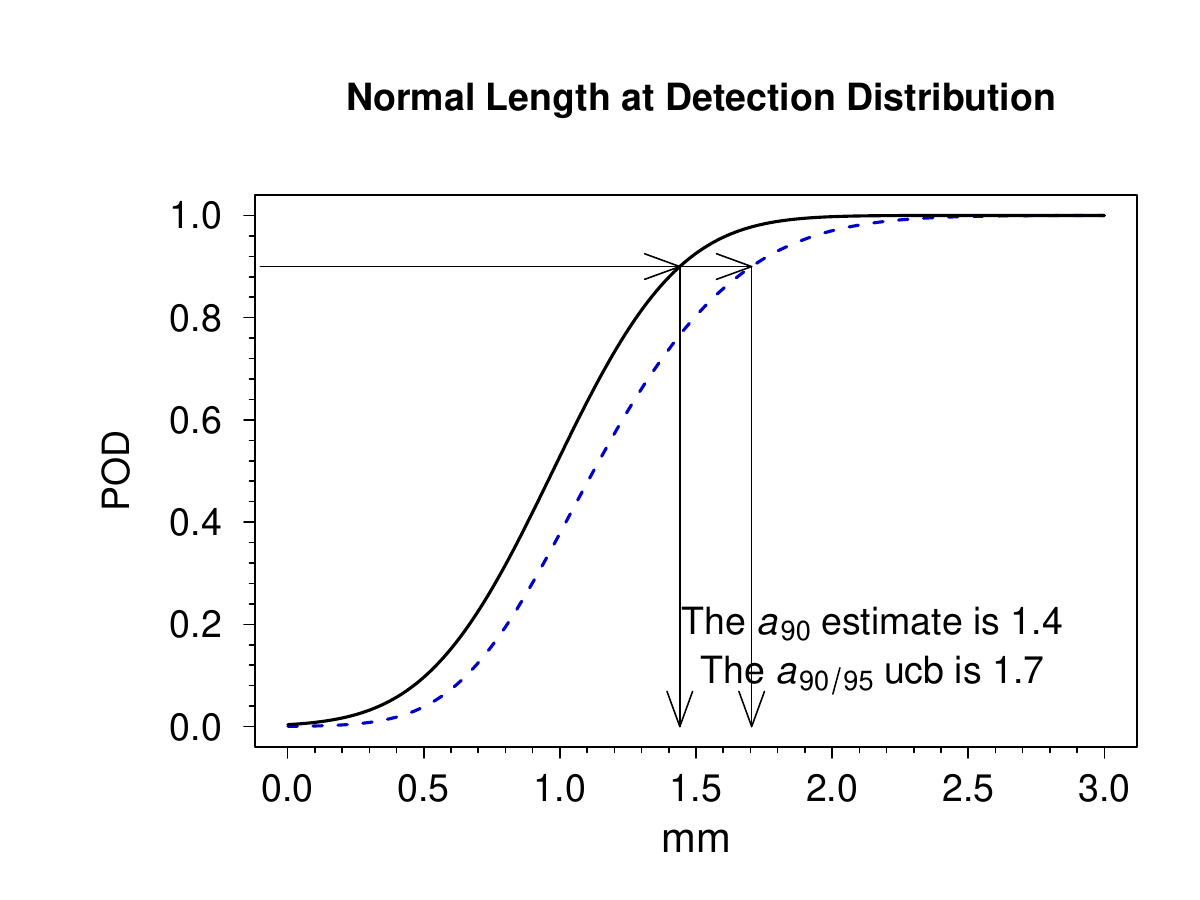}
\end{minipage}

\begin{minipage}[t]{0.4\textwidth}
    \includegraphics[width=\textwidth,trim=0 21.1pt 0 21.1pt,clip]{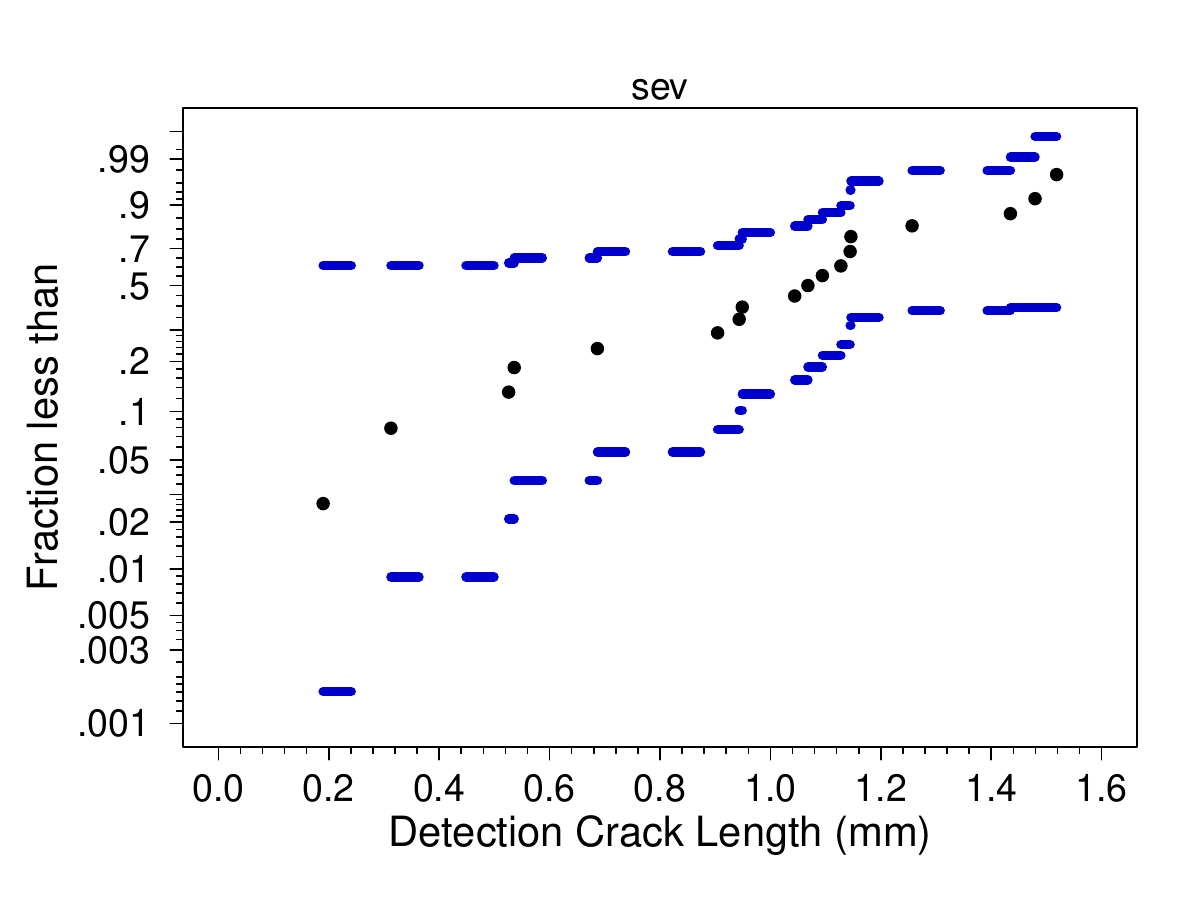}
\end{minipage}
\hspace{0.01\textwidth}
\begin{minipage}[t]{0.4\textwidth}
    \includegraphics[width=\textwidth,trim=0 17.1pt 0 17.1pt,clip]{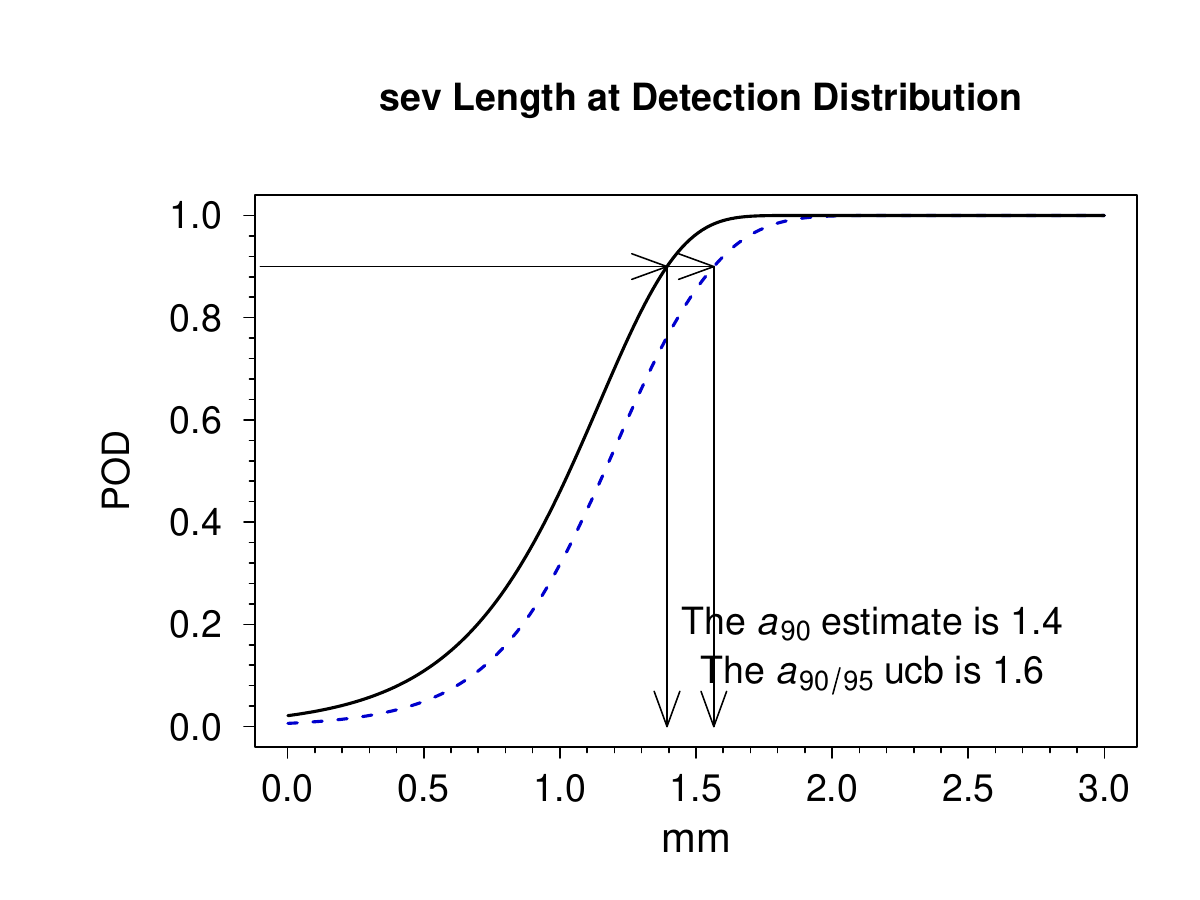}
\end{minipage}

\caption{In the left column are probability plots with 95\%
  simultaneous confidence bands for the LaD
  values from the CNT data for the (from top to bottom) Weibull,
  lognormal, normal, and smallest extreme value distributions. Plots
  on the right provide estimates and 95\% lower confidence bounds
  for POD for the same four distributions}
\label{figure:cnt_prob_pod_distributions}
\end{figure*}

The left-hand side of Figure~\ref{figure:cnt_prob_pod_distributions}
provides probability plots with 95\% simultaneous confidence bands
for the LaD values from the CNT data for the
(from top to bottom) Weibull, lognormal, normal, and smallest extreme
value distributions. Again, plots on the right provide estimates and 95\%
lower confidence bounds for POD for the same four
distributions. As in the PZT experiment,
the data are consistent with all four of these
distributions.
In contrast to the PZT data, there is a larger difference between
the fitted normal and lognormal distributions in
Figure~\ref{figure:cnt_prob_pod_distributions}, resulting is
substantially different values for the upper confidence bounds. The
effect of the log transformation in these data is stronger than it
was for the PZT data, due to the relatively large dynamic range of
the LaD values. The
difference between fitting these two distributions can be seen more
clearly in the probability plots in
Figure~\ref{figure:cnt_prob_pod_distributions}, showing ML estimates of
the respective distributions and 90\% two-sided intervals for POD
(so that the lower bounds correspond to one-sided 95\% bounds on
POD). Although the lognormal distribution provides a better fit to
the data, neither of these two distributions can be ruled out
statistically, as both distributions are consistent with the data.

\subsubsection{Estimates of POD from the Random-Parameters (RP) Model for the
  CNT data}
\begin{figure}[tbp]
\begin{tabular}{cc}
(a) & (b) \\[-0.30ex]
  \includegraphics[width=0.5\columnwidth]{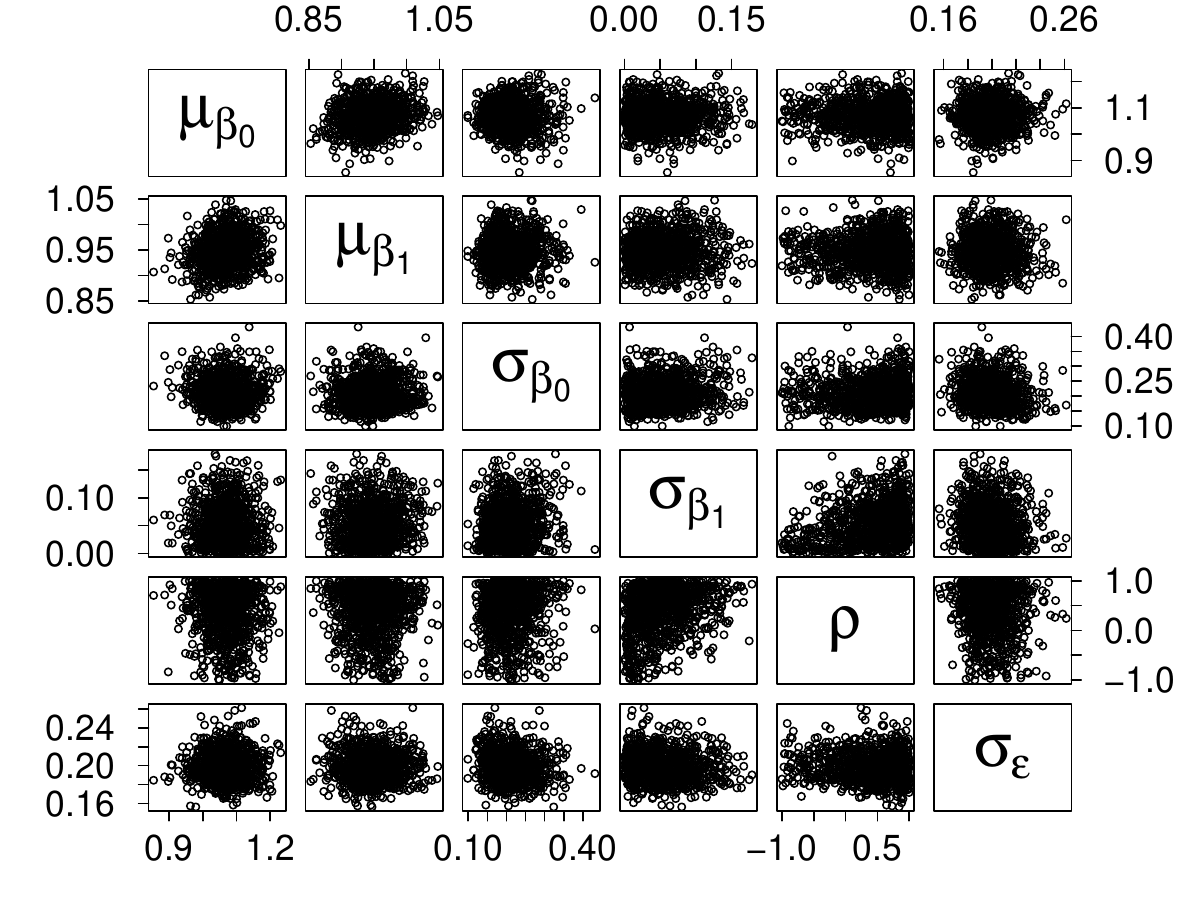} &
  \includegraphics[width=0.5\columnwidth,
    trim=0 20.0pt 0 20.0pt,clip]{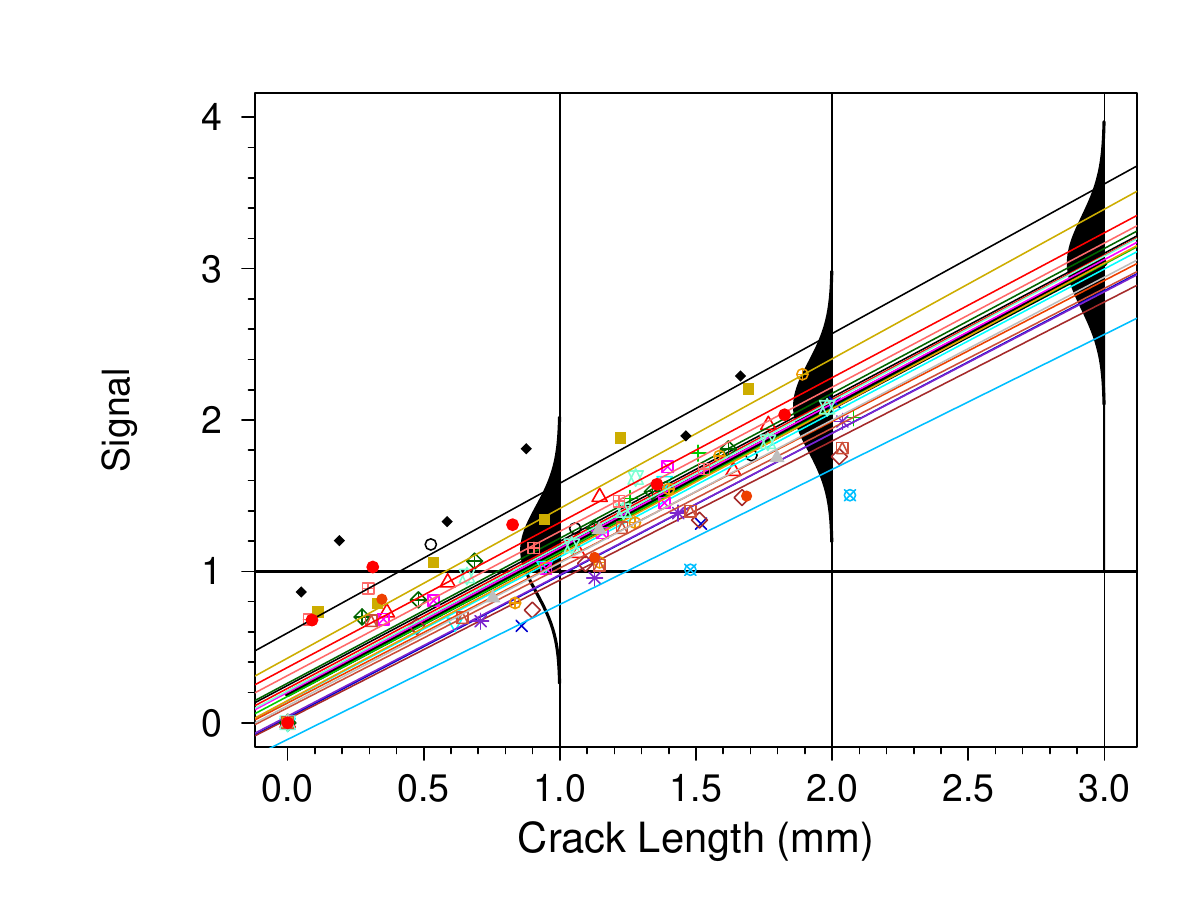}
\end{tabular}
\caption{A scatterplot matrix of the draws from the joint posterior
  distribution of the model parameters for the CNT data~(a) and a
  plot of the corresponding fitted RP model with POD indicated by
  the shaded area under the normal densities
  (bottom)~(b).}
\label{figure:random_effects_pair_response_CNT}
\end{figure}

Figure~\ref{figure:random_effects_pair_response_CNT}a is a scatterplot matrix
of the posterior draws which are well behaved and again, the commonly-used Markov chain Monte
Carlo (MCMC) diagnostics (details not given here) suggested that the
draws will accurately describe the posterior distribution.
Figure~\ref{figure:random_effects_pair_response_CNT}b
shows the estimated regression lines for each
sensor/crack combination and provides a visualization of how the
estimate of POD is computed. Note the considerable variability in
the signal response for any given crack length.

\begin{figure}[tbp]
\centering
\includegraphics[width=0.7\columnwidth,trim=0 1.1pt 0 1.1pt,clip]{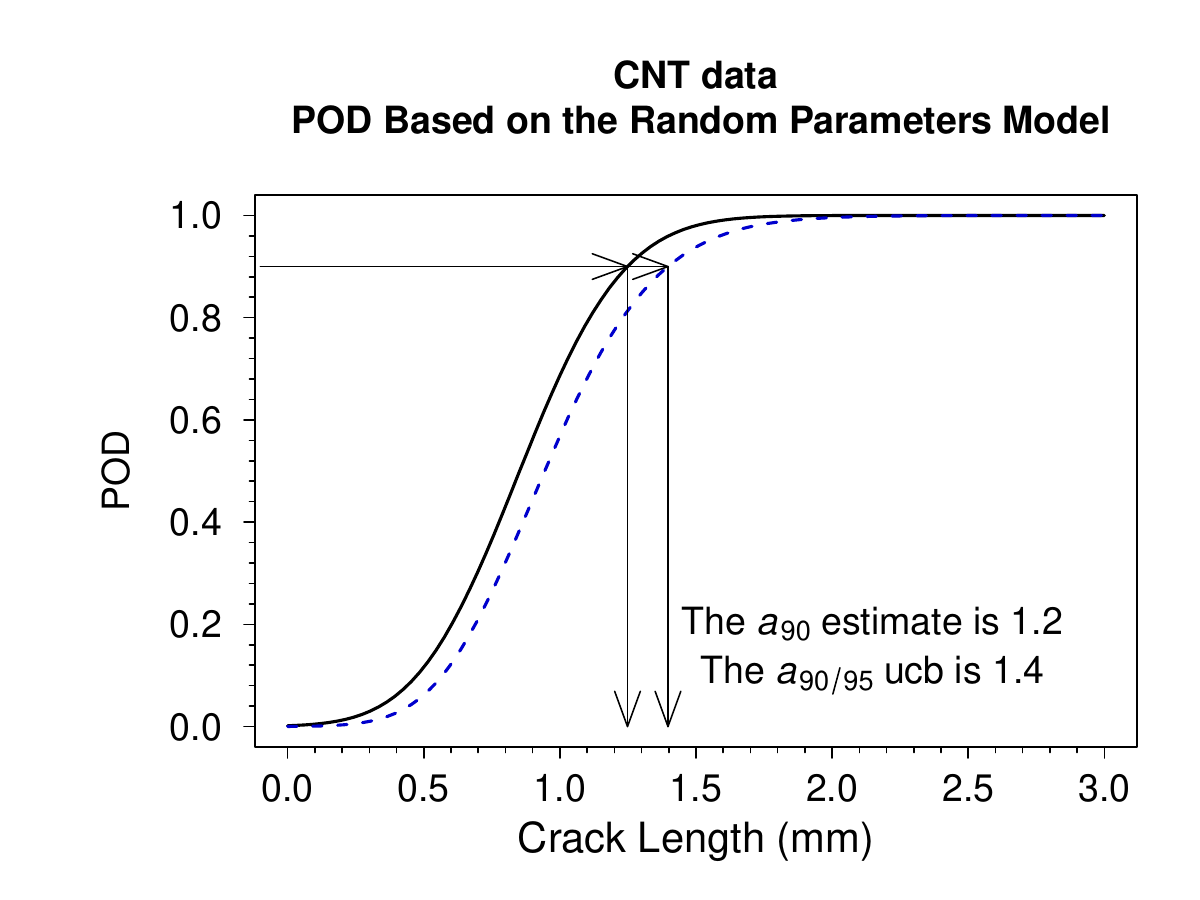}
\caption{The POD curve and lower confidence bound obtained from the
  draws from the joint posterior distribution of the
  RP model for the CNT data. }
\label{figure:image142}
\end{figure}

Figure~\ref{figure:image142} shows the corresponding plot of the
estimate of POD and the lower 95\% confidence bound on POD.
As with
the PZT example, the $a_{90}$ and $a_{90/95}$
values are somewhat smaller than the values
given by the SoDaD method in
Figure~\ref{figure:cnt_prob_pod_distributions}, in part, because the
RP model implicitly uses the point of crossing the
detection threshold as the definition the detection whereas the
SoDaD methods used the crack length at the observation after
crossing the detection limit.

\subsection{Application to the Comparative Vacuum Monitoring (CVM)
  SHM Data}
\label{section:cvm.application}
\subsubsection{CVM background and data}
\label{section:cvm.background.data}
 Comparative Vacuum Monitoring (CVM) sensors provide another method
 to detect cracks in structures. CVM is a pneumatic, elastomeric
 sensor with fine channels etched on its adhesive face. When the
 sensors are adhered to the structure under test, the fine channels
 and the structure itself form a manifold of galleries alternately
 at low vacuum and atmospheric pressure. When a crack develops, it
 forms a leakage path between the atmospheric and vacuum galleries,
 producing a measurable change in the vacuum level which is detected
 by the CVM monitoring device.
 
In the performance tests discussed here, CVM sensors were mounted
adjacent to a $5~\mathrm{mm}$ ($0.20~\mathrm{in.}$) edge notch on a
series of $600~\mathrm{mm} \times 40~\mathrm{mm} \times
2~\mathrm{mm}$ ($23.6~\mathrm{in.} \times 1.57~\mathrm{in.} \times
0.079~\mathrm{in.}$) aluminum-lithium (Al-Li) coupons. The CVM
sensor used a $20~\mathrm{mm}$ ($0.79~\mathrm{in.}$) long crack
intercept region with two $0.32~\mathrm{mm}$ ($0.013~\mathrm{in.}$)
wide sensing galleries to produce the crack detection response. Each
test specimen was subjected to tension-tension cyclic loading to
initiate and grow natural fatigue cracks. Vacuum levels were measured by
dCVM (a measure that is proportional to the ability to pull a vacuum
on the sensor galleries in a CVM system) every 1,000
cycles and a calibration exercise was used to determine the dCVM
value corresponding to sensor crack detection. Additional
information regarding CVM sensors and their use for crack detection
is provided in \citet{roach2009real,roach2018comparative}.

\begin{figure}[tbp]
\begin{tabular}{cc}
(a) & (b) \\[-0.60ex]
  \includegraphics[width=0.5\columnwidth,
    trim=0 22.9pt 0 22.9pt,clip]{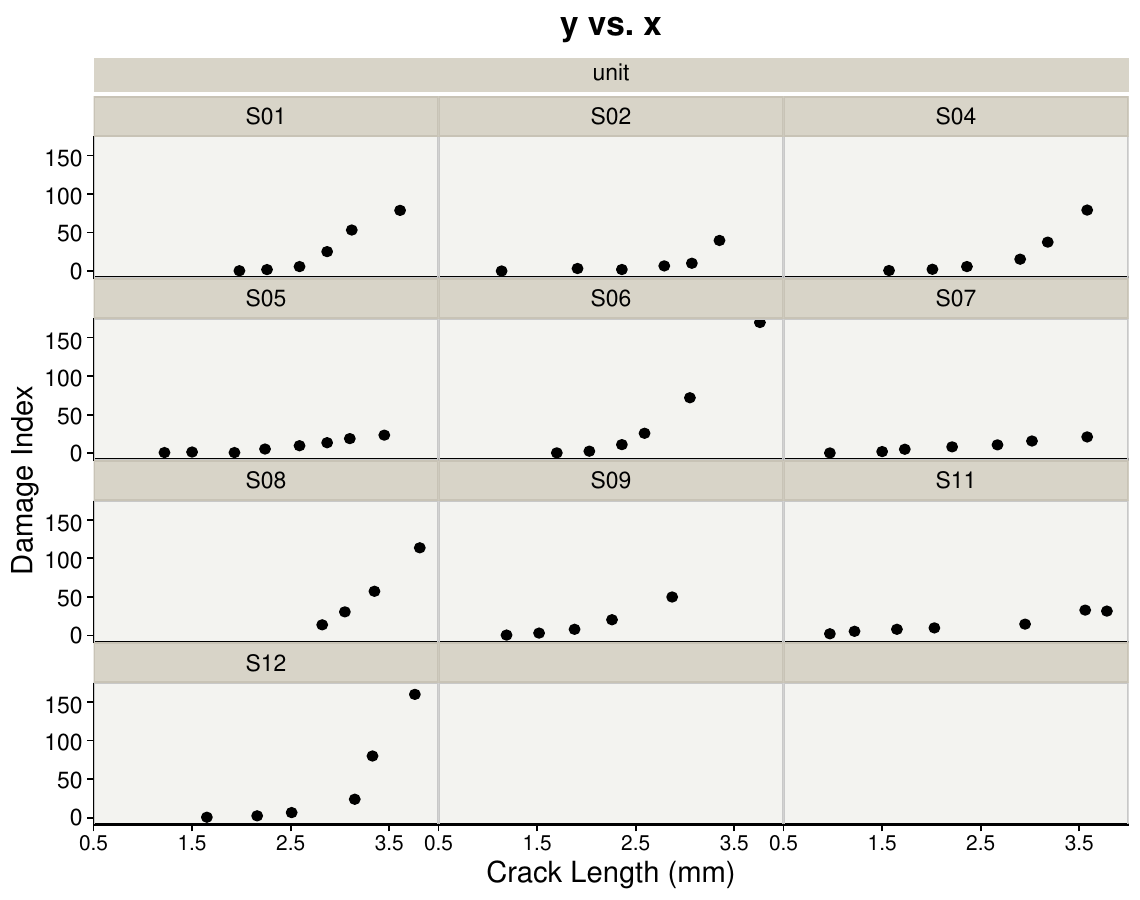} &
  \includegraphics[width=0.5\columnwidth,
    trim=0 20.0pt 0 20.0pt,clip]{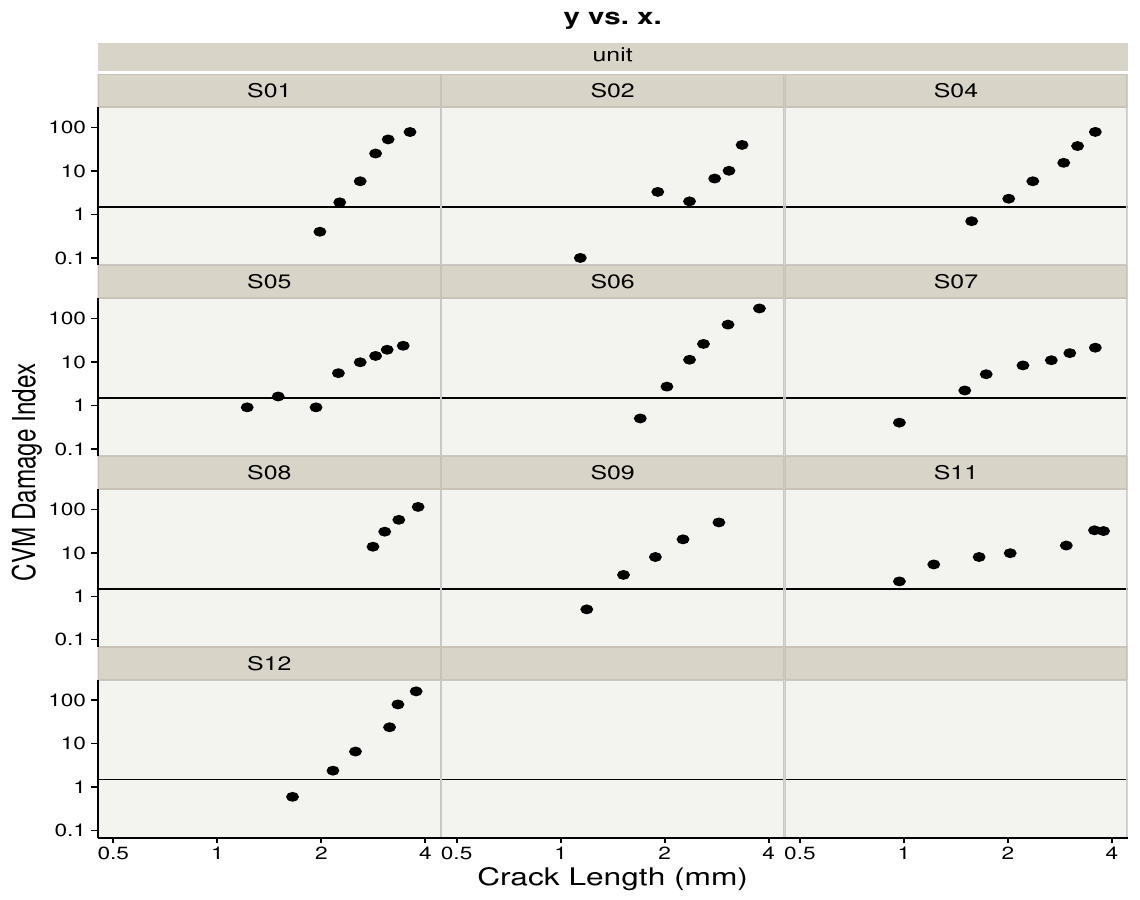}
\end{tabular}
\caption{Plot of the CVM \DI{} versus crack length on
  linear-linear scales~(a) and log-log scales~(b).
  Each plot is labeled with the specimen ID
  and the horizontal lines indicate the detection threshold of 1.5.}
\label{figure:cvm.scatter}
\end{figure}

There were only 10 specimens in the CVM experiment.
Figure~\ref{figure:cvm.scatter}(a) is a plot of the CVM data showing that the
\DI{} increases exponentially in crack
length. Figure~\ref{figure:cvm.scatter}{b} is a plot of the same data on log-log
scales. The log-log transformation makes the CVM \DI{}
approximately linear in crack length. The transformed data will be
analyzed in the following sections.
The analyses in this section are similar to those in
Section~\ref{section:pzt.application} (for the PZT data)
and Section~\ref{section:cnt.application} (for the CNT
data).

\subsubsection{Estimates of POD from the SoDaD (LaD) Method for the
  CVM data}
\label{section:cvm.sodad.pod.estimates}
\begin{figure}[tbp]
\centering
\includegraphics[width=0.7\columnwidth]{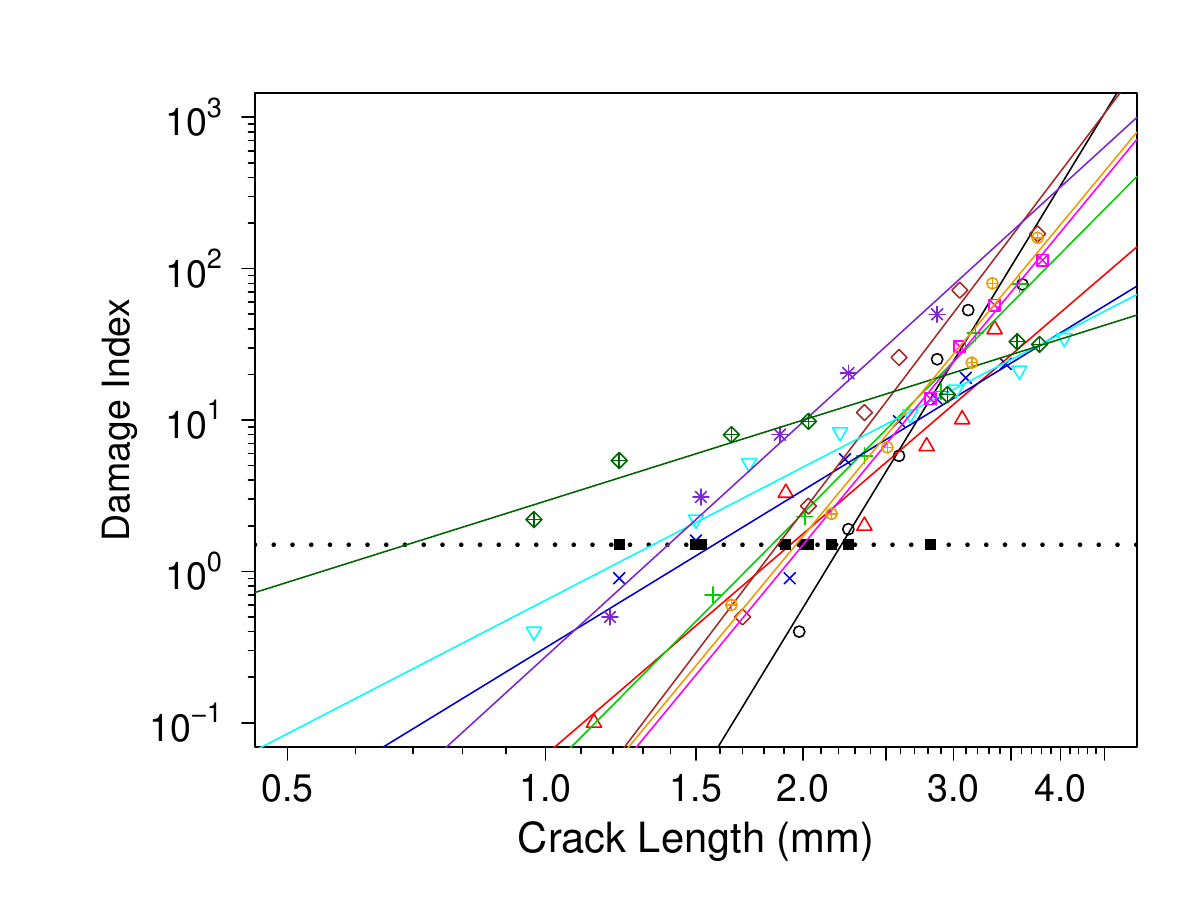}
\caption{Plot of the CVM POD data on log-log scales with a linear
  regression line for each crack/sensor combination. The LaD
  values corresponding to the detection threshold of 1.5
  are also shown.}
\label{figure:image143}
\end{figure}

Figure~\ref{figure:image143} shows the fitted regression line
for each specimen and the LaD values along the horizontal line at
the $\DI{}=0.405$ detection threshold.
Table~\ref{table:cvm_lad_slope} contains the numerical
values of the LaD and slopes.
\begin{table}[tbp]
\centering
\caption{LaD and estimated slope for each
  crack/sensor combination in the CVM data}
\label{table:cvm_lad_slope}
\renewcommand{\arraystretch}{1.08}
\begin{tabular}{lll}
\hline
Specimen & LaD (mm) & Slope \\
\hline
S11 & 0.97 & 10.520 \\
S09 & 1.19 & 29.508 \\
S05 & 1.22 & 10.562 \\
S07 & 1.50 & 10.305 \\
S04 & 1.57 & 35.192 \\
S12 & 1.65 & 67.398 \\
S06 & 1.70 & 83.751 \\
S02 & 1.91 & 12.934 \\
S01 & 2.26 & 51.553 \\
S08 & 2.82 & 101.651 \\
\hline
\end{tabular}
\end{table}
One interesting difference
is that, when compared to the PZT and the CNT data, there is more
variability in the slopes in the CVM data.

\begin{figure*}[!tbp]
\centering
\begin{minipage}[t]{0.4\textwidth}
    \includegraphics[width=\textwidth,trim=0 11.5pt 0 11.5pt,clip]{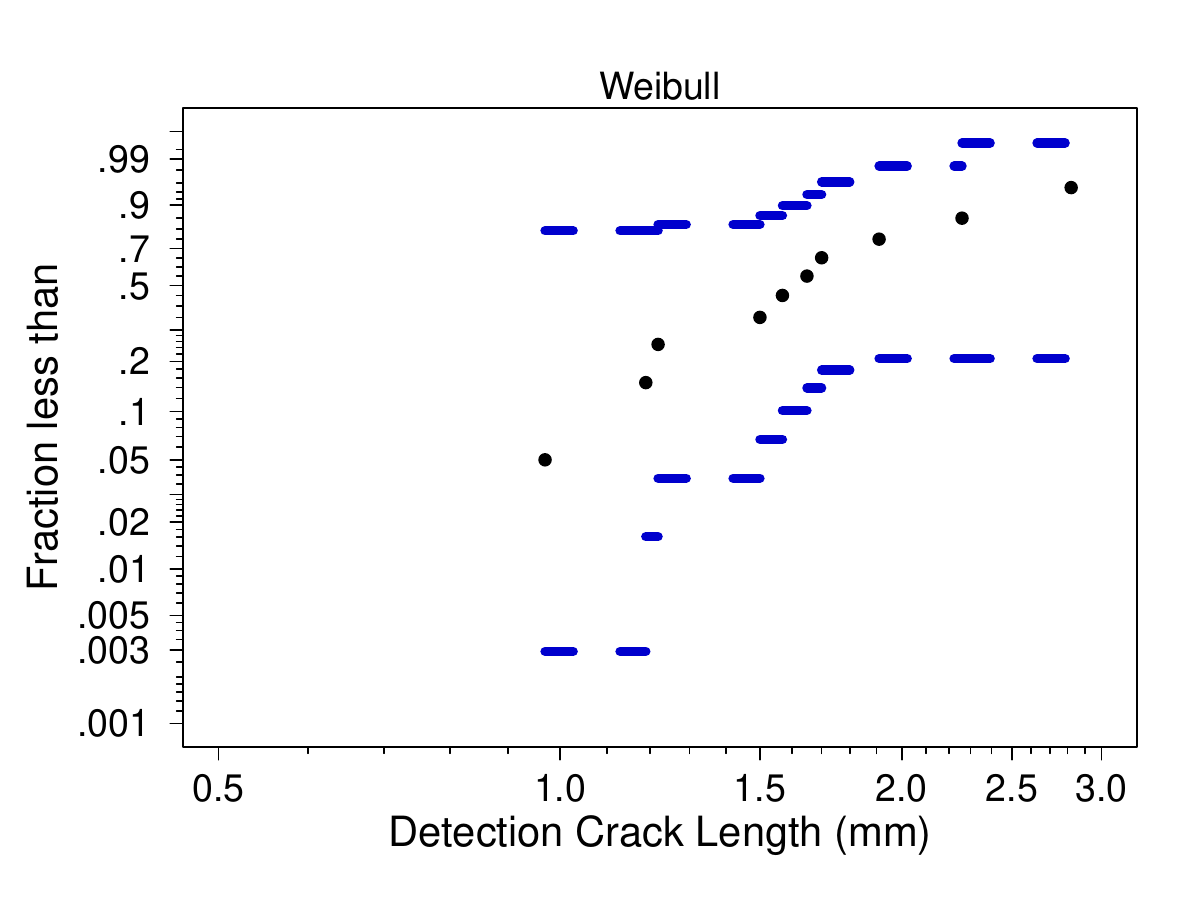}
\end{minipage}
\hspace{0.01\textwidth}
\begin{minipage}[t]{0.4\textwidth}
    \includegraphics[width=\textwidth,trim=0 7.9pt 0 7.9pt,clip]{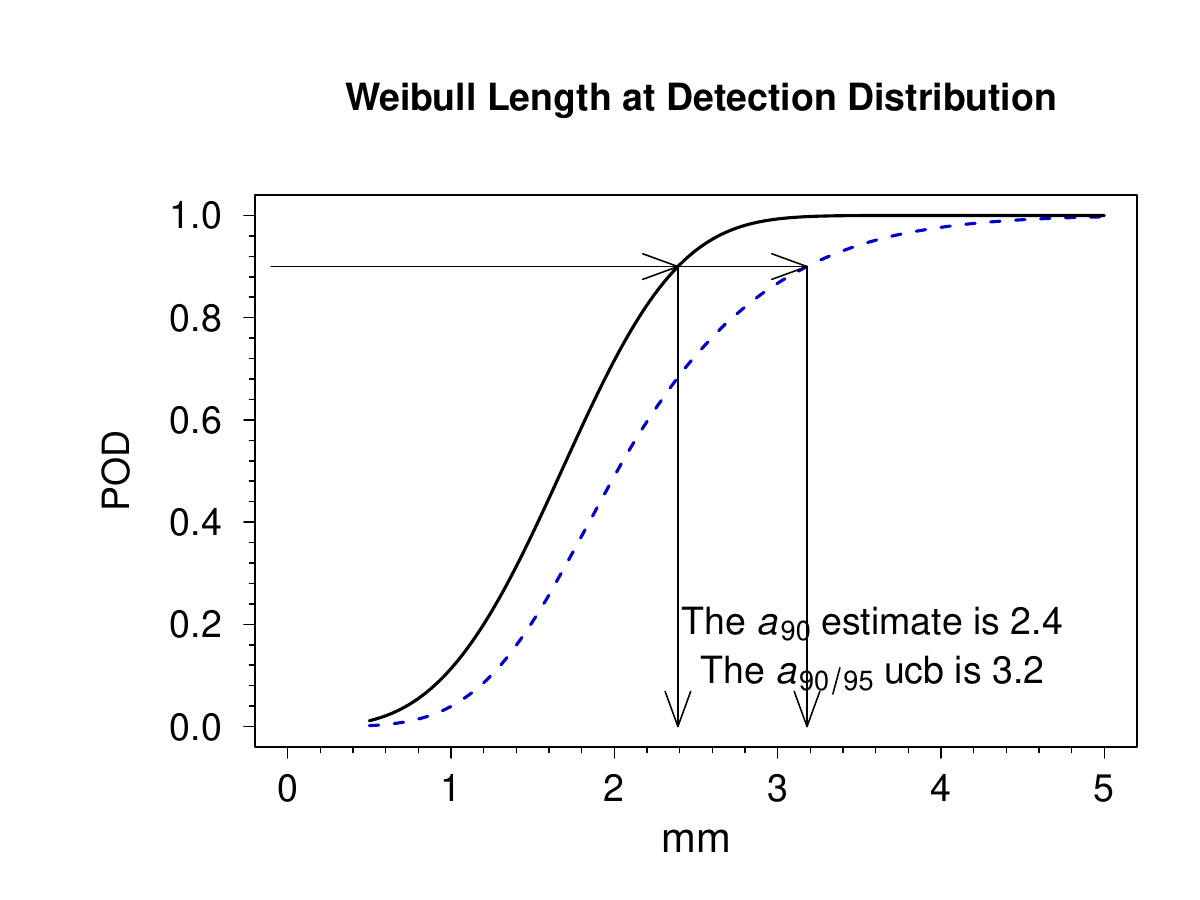}
\end{minipage}

\begin{minipage}[t]{0.4\textwidth}
    \includegraphics[width=\textwidth,trim=0 14.8pt 0 14.8pt,clip]{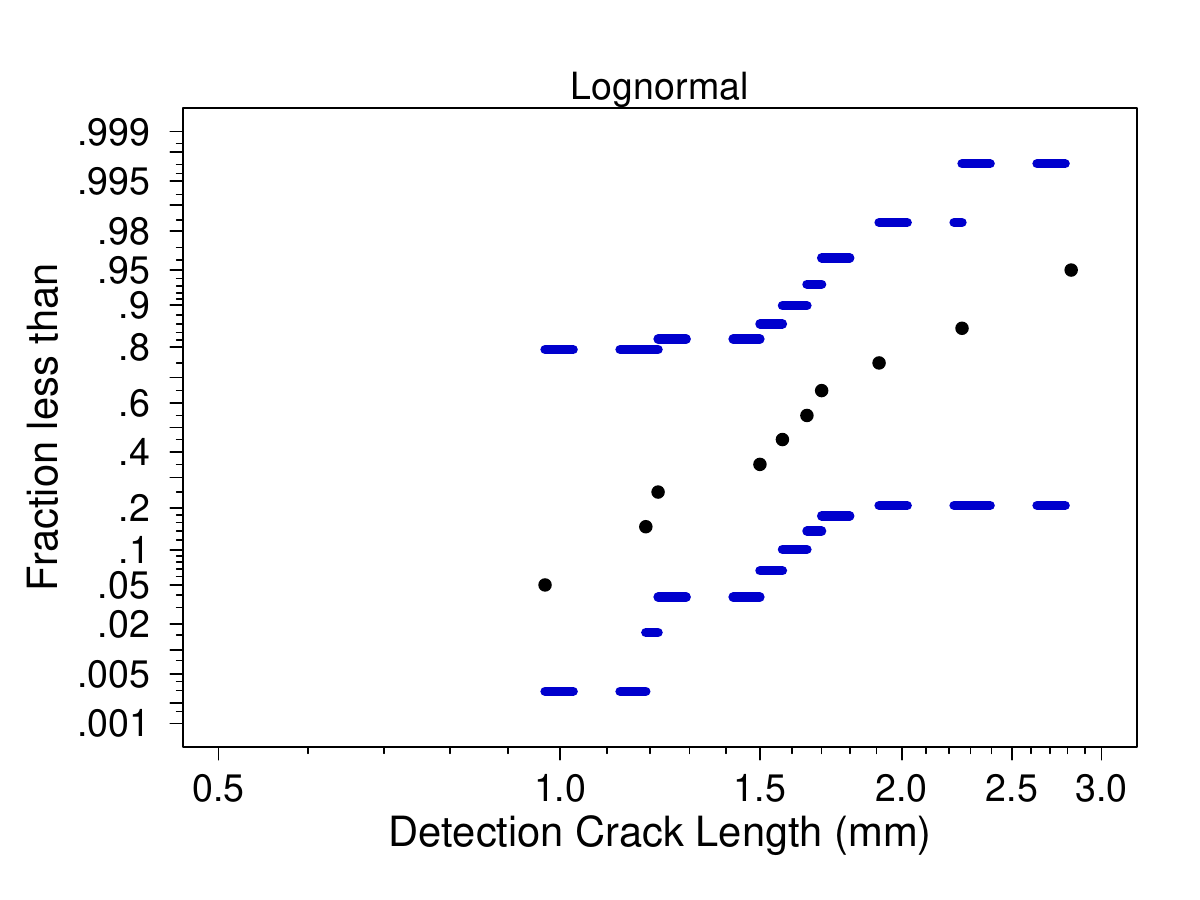}
\end{minipage}
\hspace{0.01\textwidth}
\begin{minipage}[t]{0.4\textwidth}
    \includegraphics[width=\textwidth,trim=0 6.5pt 0 6.5pt,clip]{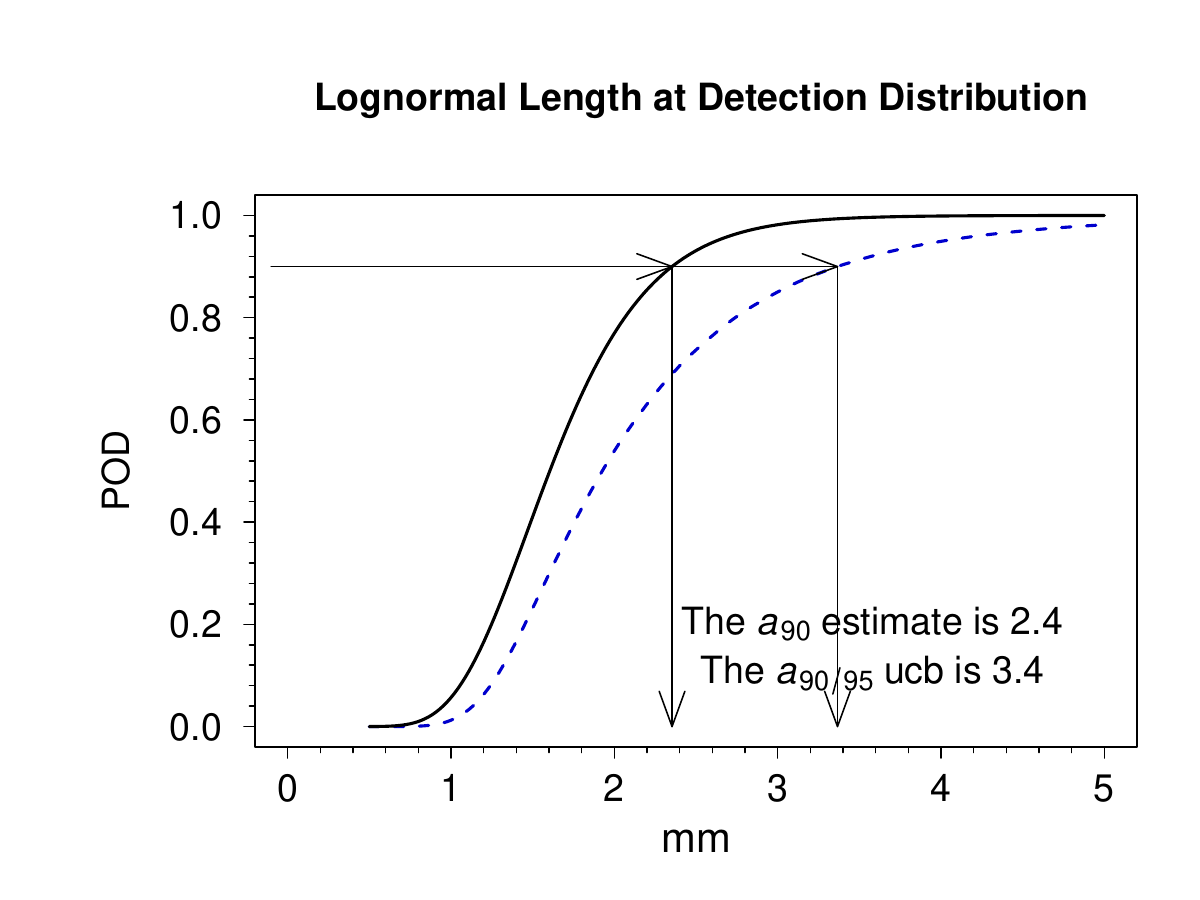}
\end{minipage}

\begin{minipage}[t]{0.4\textwidth}
    \includegraphics[width=\textwidth,trim=0 17.0pt 0 17.0pt,clip]{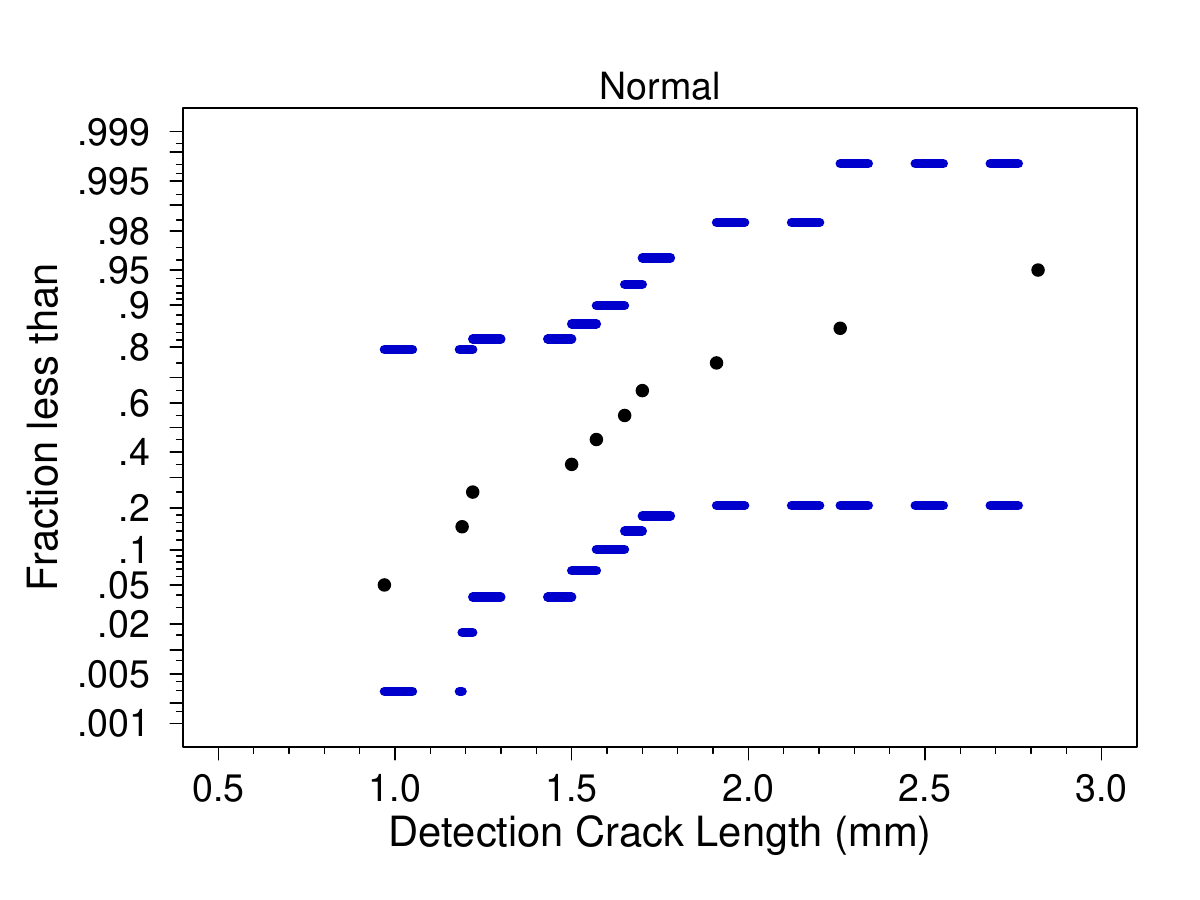}
\end{minipage}
\hspace{0.01\textwidth}
\begin{minipage}[t]{0.4\textwidth}
    \includegraphics[width=\textwidth,trim=0 7.0pt 0 7.0pt,clip]{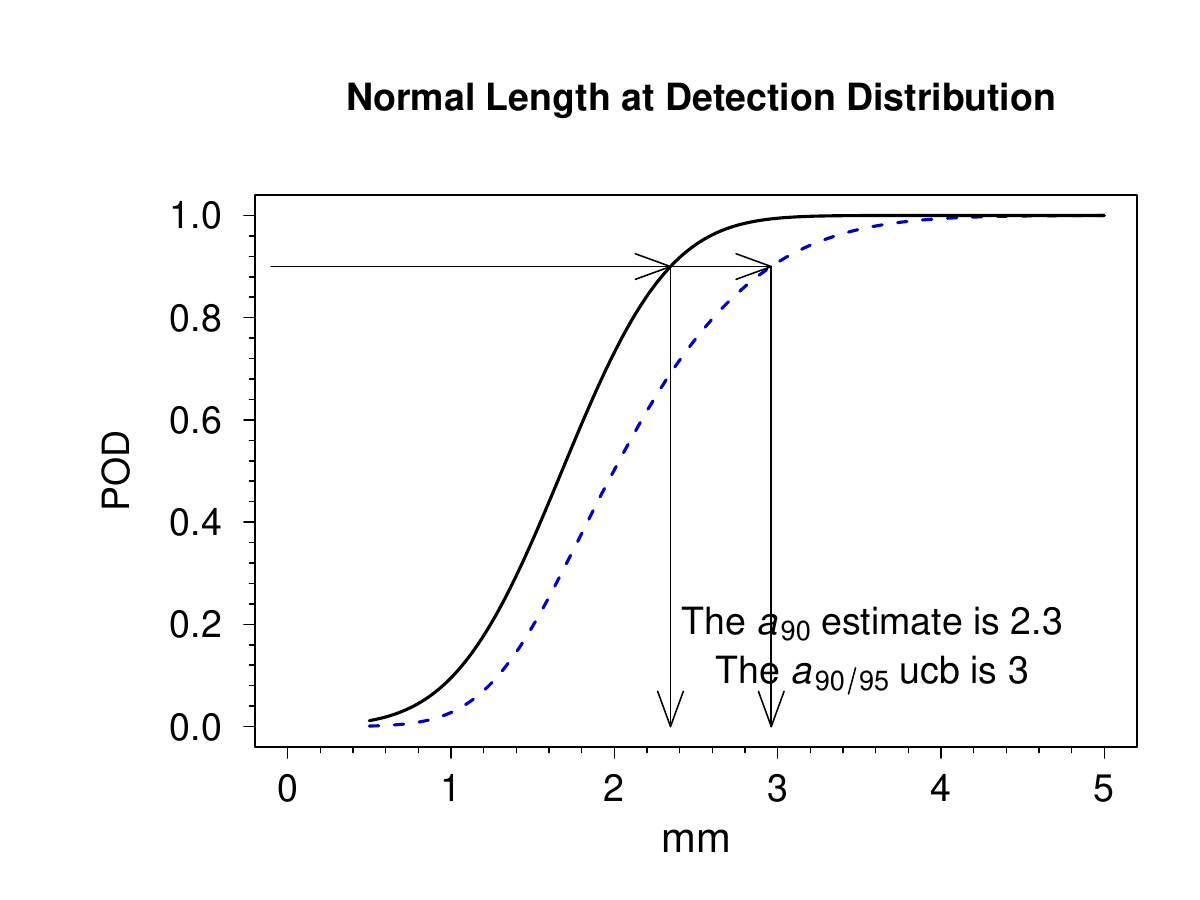}
\end{minipage}

\begin{minipage}[t]{0.4\textwidth}
    \includegraphics[width=\textwidth,trim=0 13.3pt 0 13.3pt,clip]{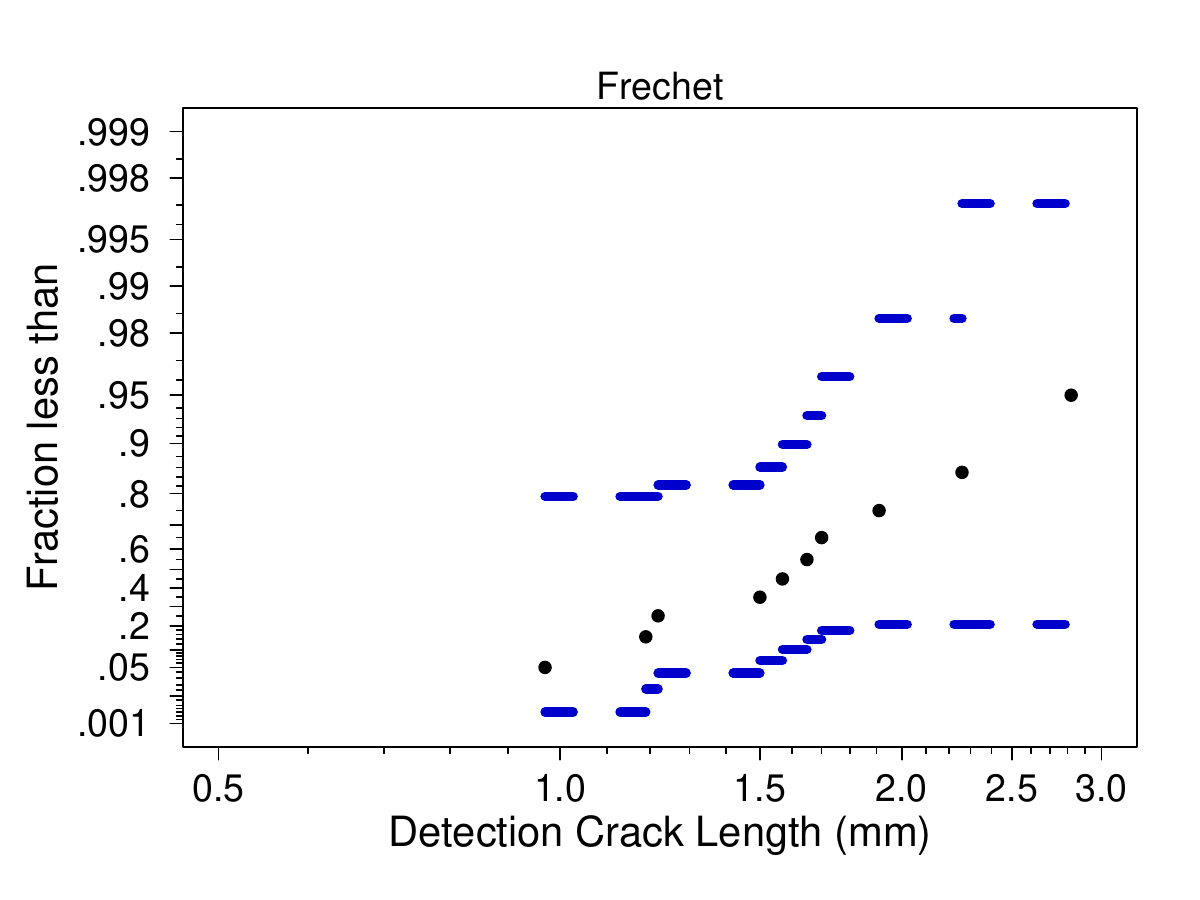}
\end{minipage}
\hspace{0.01\textwidth}
\begin{minipage}[t]{0.4\textwidth}
    \includegraphics[width=\textwidth,trim=0 6.0pt 0 6.0pt,clip]{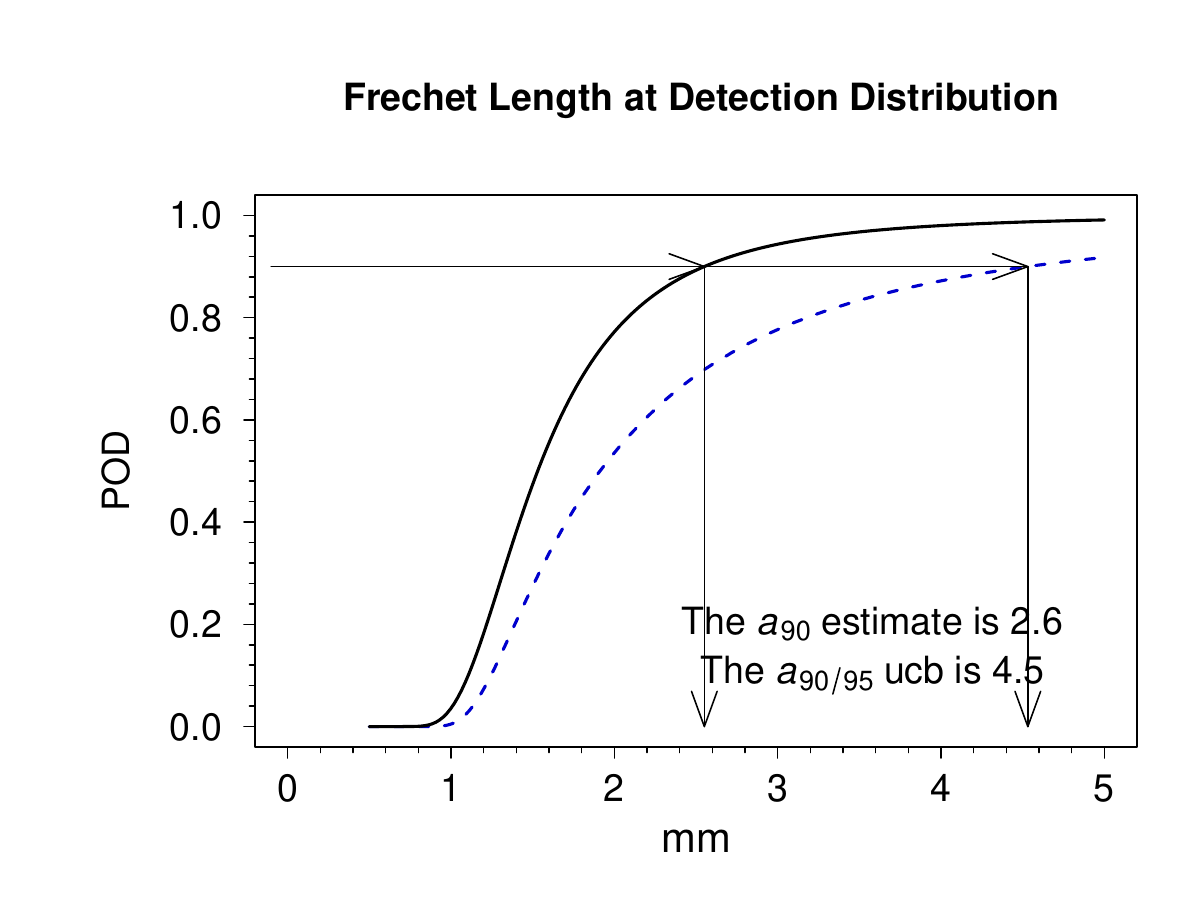}
\end{minipage}

\caption{ In the left column are probability plots with 95\%
  simultaneous confidence bands for the LaD
  values from the CVM data for the (from top to bottom) Weibull,
  lognormal, normal, and Fr\'{e}chet distributions. Plots
  on the right provide estimates and 95\% lower confidence bounds
  for POD for the same four distributions.}
\label{figure:cvm_prob_pod_distributions}
\end{figure*}

This section summarizes the analysis of the CVM LaD values for
four probability distributions. The
probability plots on the left-hand side of
Figure~\ref{figure:cvm_prob_pod_distributions} suggest that all four of
the distributions are consistent with the data. This is, in part,
because there were only 10 specimens and thus much
statistical uncertainty, leading to wide simultaneous confidence
bands in the probability plots. The POD plots on
the right-hand side show that there is little difference in the
estimates of $a_{90}$. Because it has a longer upper tail, the
Fr\'{e}chet distribution results in the largest $a_{90/95}$ value.

\subsubsection{Estimates of POD from the Random-Parameters Model for the
  CVM data}
\label{section:cvm.rp.pod.estimates}
\begin{figure}[tbp]
\begin{tabular}{cc}
(a) & (b) \\[-0.30ex]
  \includegraphics[width=0.5\columnwidth]{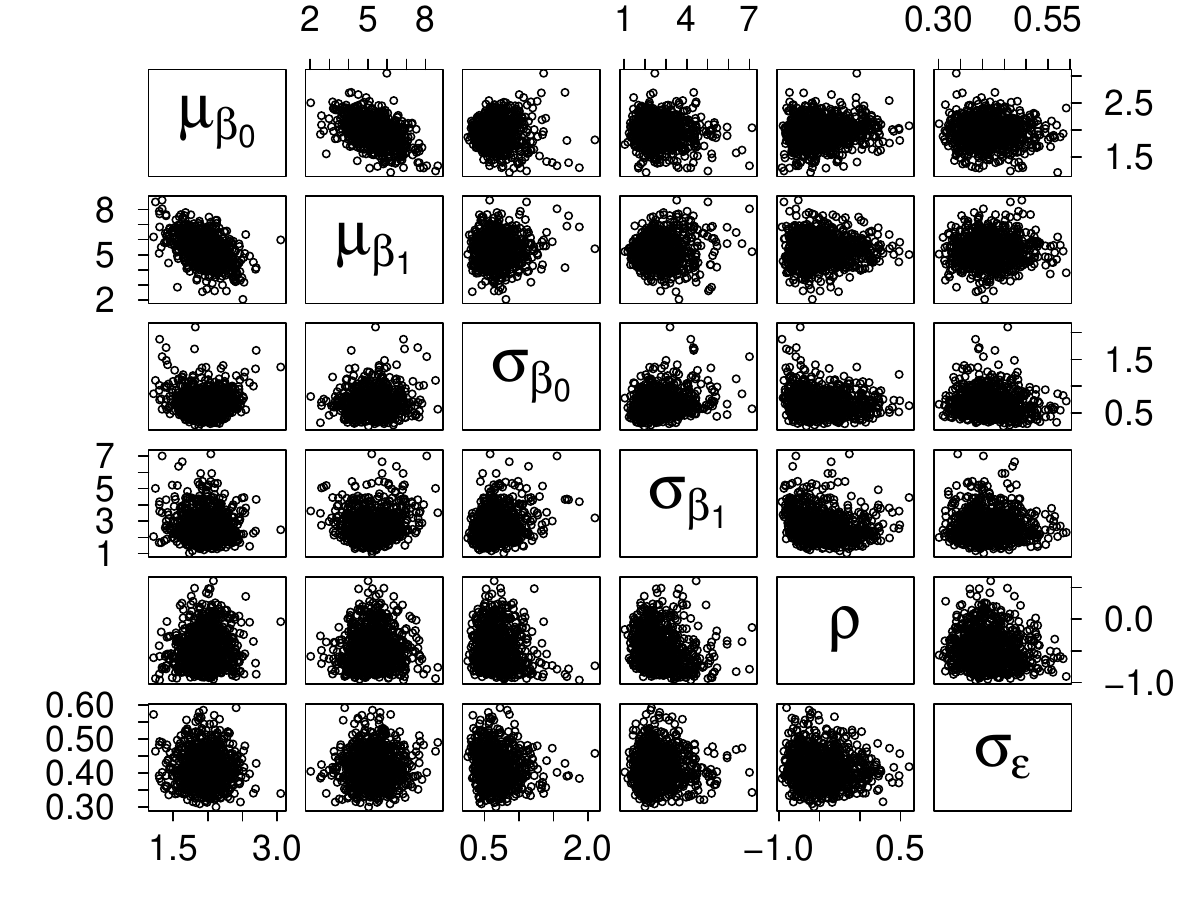} &
  \includegraphics[width=0.5\columnwidth,
    trim=0 20.0pt 0 20.0pt,clip]{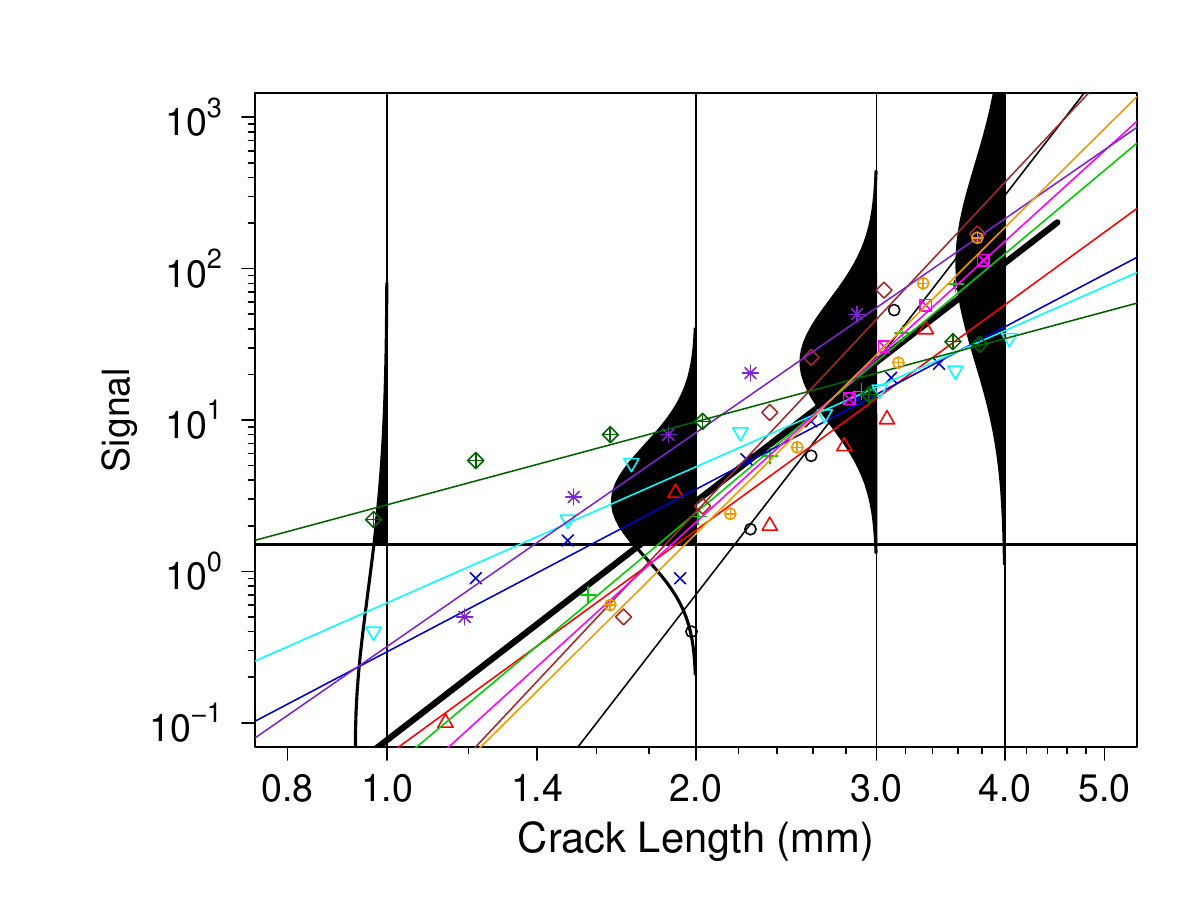}
\end{tabular}
\caption{A scatterplot matrix of the draws from the joint posterior
  distribution of the model parameters for the CVM data~(a) and a
  plot of the corresponding fitted RP model with POD indicated by
  the shaded area under the normal densities
  (bottom)~(b).}
\label{figure:random_effects_pair_response_CVM}
\end{figure}

Recall from Section~\ref{section:cvm.background.data}
that log transformations were used to
have an approximately linear relationship between the \DI{}
signal and crack
length for the CVM data. Those transformations are used here for the RP
model. The top plot in
Figure~\ref{figure:random_effects_pair_response_CVM}a, similar to the
previous examples, shows the scatterplot matrix of the draws from
the posterior distribution  and
Figure~\ref{figure:random_effects_pair_response_CVM}b is a fitted model plot
on log-log axes showing the fitted regression lines for
each crack/sensor pair and the estimates of the signal distribution
as a function of crack size with POD represented by the shaded in
areas of the estimated densities. 

\begin{figure}[tbp]
\centering
\includegraphics[width=0.7\columnwidth,trim=0 9.8pt 0 9.8pt,clip]{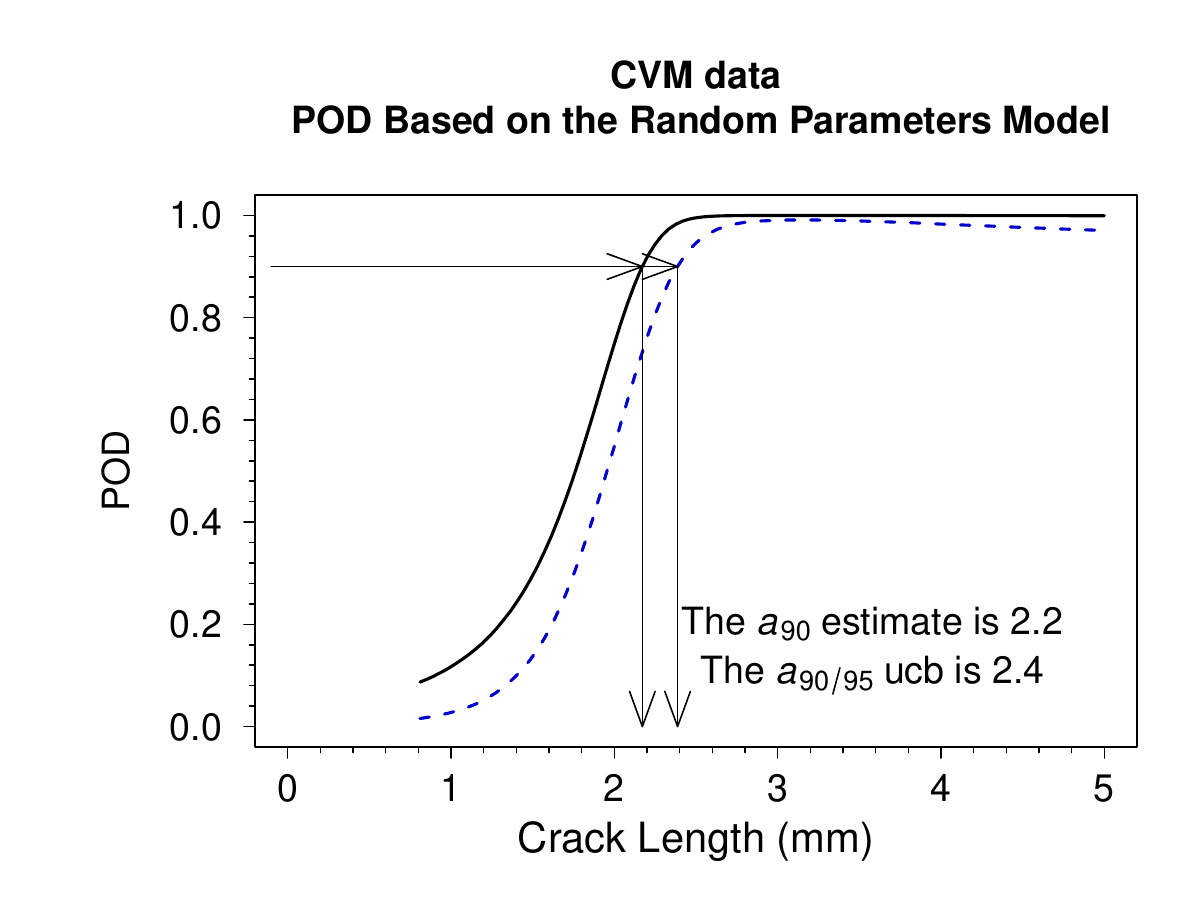}
\caption{A plot of the POD curve estimated from the draws from the
  joint distribution of the posterior distribution of the
  RP model parameters for the CVM data.}
\label{figure:pod.estimate.rp.CVM}
\end{figure}

Similar to the previous examples,
Figure~\ref{figure:pod.estimate.rp.CVM} shows the RP model POD
estimate and lower credible band, along with the $\hat{a}_{90}=2.2$
estimate and the $a_{90/95}=2.4$ upper credible bound.
The ratio $a_{90/95}/\hat{a}_{90}= 1.09$ quantifies estimation
precision and is considerably smaller than the corresponding ratios
of the LaD method $\hat{a}_{90}$
and $a_{90/95}$ values shown on the right-hand side of
Figure~\ref{figure:cvm_prob_pod_distributions}; those range from
1.25 to 1.73. For the PZT and CNT examples, the ratios for the RP
model (1.12 and 1.17, respectively) were also generally smaller than
those for the LaD models, although the differences were much less
pronounced. The stronger contrast for the CVM data is likely because
the LaD method does not use all of the information in the
signal-versus-crack-size data, and the resulting loss of information
is more consequential when the number of crack/sensor combinations
is small ($n = 10$) and there is substantial variability in the data,
as is the case in the CVM experiment.
In contrast, the hierarchical structure
of the RP model allows information to be shared across crack/sensor
combinations, making more efficient use of the limited data and
providing more precision than the LaD method for this
example.




Interestingly, and in contrast to the PZT and CNT examples, the
lower credible bound in Figure~\ref{figure:pod.estimate.rp.CVM}
reaches a maximum of 0.990 around crack size 3.1 mm and then begins
to decrease. The reason for this seemingly odd behavior is that in
the CVM example (again because of the number of crack/sensor
combinations, the weakly informative priors, and especially the
large amount of variability in the slopes) there is a nonnegligible
posterior probability (estimated to be 0.017) that the slope of the
signal versus crack size regression model is less than zero
(i.e., $\beta_{1} < 0$). Thus the POD function for some of the
posterior draws is decreasing.  For the PZT and CNT examples, where
the number of crack/sensor combinations was larger (13 and 19,
respectively, relative to 10 for the CVM experiment) and there was
less variability in the slopes, the corresponding posterior
probabilities are $5\times10^{-12}$ and $3\times10^{-84}$,
respectively.  Specifically,
\begin{align}
  \label{equation:pr.slope.negative}
  \Pr(\beta_{1} <0) = \Phi\left(-\frac{\mu_{\beta_1}}{\sigma_{\beta_1}}\right).
\end{align}
The estimates of $\Pr(\beta_{1} <0)$ were obtained by
evaluating~(\ref{equation:pr.slope.negative}) for each of the
posterior draws to obtain draws from the marginal posterior
distribution of~(\ref{equation:pr.slope.negative}). Then the point
estimates were obtained from the median of those draws.  Note, also,
that the point estimate of the random-slope standard deviation
$\sigmahat_{\beta_1}=2.48$ is large, relative to the point estimate of mean
$\muhat_{\beta_1}=5.23$.

%
%
%
%
%

\section{A Comparison of the size-of-damage-at-detection (SoDaD) and
  the Random-Parameters Model Methods of Estimating SHM POD}
\label{section:comparison.sodad.rp}
This section compares the SoDaD (which could also be called LaD in
our crack-detection examples) and the random-parameters (RP)
methods of estimating POD.
Table~\ref{table:pod_estimates_lad_rp} summarizes the numerical POD
estimates for the LaD distributions considered in the examples and
for the RP model.

\begin{table}[!b]
\centering
\caption{POD estimates for different LaD distributions and the RP
  method.}
\label{table:pod_estimates_lad_rp}
\scriptsize
\setlength{\tabcolsep}{3pt}
\renewcommand{\arraystretch}{1.08}
\resizebox{\textwidth}{!}{%
  \begin{tabular}{@{}ll lcc lcc lcc@{}}
\toprule
 & & \multicolumn{3}{c}{PZT}
   & \multicolumn{3}{c}{CNT}
   & \multicolumn{3}{c}{CVM} \\
\cmidrule(lr){3-5}
\cmidrule(lr){6-8}
\cmidrule(lr){9-11}
Method & Distribution
& Fit & $\hat{a}_{90}$ & $a_{90/95}$
& Fit & $\hat{a}_{90}$ & $a_{90/95}$
& Fit & $\hat{a}_{90}$ & $a_{90/95}$ \\
\midrule
LaD & Weibull
& excellent & 6.2 & 6.8
& good & 1.4 & 1.7
& good & 2.4 & 3.2 \\
LaD &  smallest extreme value
& excellent & 6.2 & 6.7
& excellent & 1.4 & 1.6
& poor & 2.4 & 3.0 \\
LaD & lognormal
& excellent & 6.3 & 7.1
& poor & 1.7 & 2.5
& excellent & 2.4 & 3.4 \\
LaD & normal
& excellent & 6.3 & 6.9
& good & 1.4 & 1.7
& good & 2.3 & 3.0 \\
LaD & Fr\'{e}chet
& fair & 6.7 & 8.2
& poor & 2.9 & 6.9
& excellent & 2.6 & 4.5 \\
LaD & largest extreme value
& fair & 6.5 & 7.6
& poor & 1.6 & 2.1
& good & 2.4 & 3.2 \\[1.0ex] 
RP & bivariate normal
& good & 5.9 & 6.6
& good & 1.2 & 1.4
& good & 2.2 & 2.4 \\
\bottomrule
\end{tabular}%
}
\end{table}

For the SoDaD
method, most fitted distributions provide good fits to the
data. However, the resulting POD estimates are not the same across
distributions. This difference is mainly due to the upper tail behavior of
the assumed LaD distribution, because $a_{90/95}$ is an upper
confidence bound on an upper-tail quantile
and is therefore more sensitive to the behavior in the upper tail
of the distribution
than the point estimate $\hat{a}_{90}$.
The RP method generally
gives smaller $a_{90/95}$ values than the SoDaD method because it
uses the repeated signal-response measurements rather than only the
first detection length. This results from the different definitions
of threshold crossing in the two methods.

\subsection{Characteristics and Advantages of the SoDaD Method}
\label{section:characteristics.advantages.sodad}
The SoDaD method has the following
characteristics and advantages.
\begin{enumerate}
\item Even though it ignores some information in the path data,
  it uses the most important information and is computationally and statistically simple.
    \item Because the SoDaD method is based on a single distribution with only
      two parameters, good precision can be expected in estimation,
      even with a small number of crack/sensor combinations. This
      has been seen in Monte Carlo simulations that have been
      conducted (details not reported here).
    \item The SoDaD method requires an assumption about the
      distribution of sizes of damage at detection (e.g., normal,
      lognormal, or other distributions used in the examples in this
      paper). This is a relatively simple assumption
      that is easy to perturb (relative to
      what is required in specifying the RP model). In some
      applications (e.g., the CNT data), the estimate of POD can be
      highly sensitive to the assumed distribution.
    \item Generally, there will be little information to
      discriminate among different distributions that might give
      vastly different $a_{90/95}$ values. It is, however, easy to
      do sensitivity analyses and choose a distribution that agrees
      with the data and is appropriately conservative.
    \item
      Statistical software to fit single distributions such as those
      used in our SoDaD/LaD examples is readily available, making it
      easy to implement the methods.
\end{enumerate}

\subsection{Characteristics and Limitations of the RP Method}
\label{section:characteristics.limitations.rp}
The repeated-measures random-parameters (RP) model/method has the
following characteristics and limitations:
\begin{enumerate}
    \item The model for the data (a bivariate distribution for the
      random slopes and intercepts and a separate distribution for
      the error term) is more complicated, requiring the estimation
      of five population model parameters.
    \item
      Although it could be argued that RP uses the available data
      more efficiently (SoDaD ignores information about the path),
      simulation studies indicate that the amount of information
      lost relative to SoDaD is small. Exceptions include situations
      like the CVM example, in which there is substantial
      specimen-to-specimen variability and the hierarchical
      structure of the RP model allows the sharing of information
      (sometimes called borrowing strength) across crack/sensor
      combinations. \citet{Gelman.et.al.2013} provide more information
      about the advantages of Bayesian hierarchical models.
\item Presently only a bivariate normal distribution is
      available as an RP model. It would be possible to generalize this to allow
      other bivariate distributions, but the task would be
      technically and computationally complicated. For example,
      numerical integration is needed for each evaluation of the
      likelihood function and also to compute each point on the POD
      curve. Without readily available software to do these
      computations, the kind of sensitivity
      analysis suggested for the SoDaD method is not yet available
      for the RP method.
    \item Use of the RP method requires consideration of the use of
      transformations to linearize the path (to correspond to the
      assumed linear regression model). This is similar to what is
      done in the MIL-HDBK 1823A method where there are options to
      transform both the experimental variable (e.g., crack length)
      and the signal response (e.g., \DI{}). Log and square
      root transformations are often used.
    \item Importantly, use of the RP method requires careful
      consideration of how much of the sample path information
      should be used in fitting the RP model. The most important
      information is in the data points close to the detection
      threshold. Especially when sample paths deviate from the linear model
      assumption, estimates can be biased and
      points far from the detection threshold can be
      highly influential.
    \item Having so many knobs to turn (e.g., in the previous three
      bullets) makes implementation challenging and, at the same
      time, makes the final results difficult to defend.
    \item More complicated computational algorithms are
      needed for the RP method. However, with modern computational
      tools that are
      available (hardware and software), this should not be a major
      impediment.
    \item
      A potential advantage of the RP method is that it provides a
      framework for better understanding and characterizing inspection
      variability and for implementing
      a version of MAPOD, as described in the next section.
\end{enumerate}

\FloatBarrier
\section{Application of Model-Assisted Probability of Detection
  (MAPOD) in SHM Applications}
\label{section:mapod}
The methodologies presented in the previous sections are primarily
developed for data-driven SHM-POD analysis. However, in many
practical applications, especially when experimental data are
limited, it is desirable to incorporate physics-based or simulation
models into the POD estimation process.

In this context, the statistical frameworks introduced in this paper
can be naturally extended to MAPOD
settings. This section briefly outlines how the proposed approaches
can be adapted and integrated with model-assisted methodologies.

\subsection{Background and related literature}
Because of the nature of SHM, it is difficult to conduct economical
POD studies like those suggested for traditional NDI in the main part of
\citet{milhdbk1823a2009}. For example, the handbook suggests
using 40 or more cracks; an SHM-POD study would therefore require 40 or
more sensor/crack combinations in which the cracks are grown over time.
\citet{aldrin2016best} outline general
considerations that will be needed to provide a valid
characterization of SHM capability.

MAPOD, in which physics-based inspection models
are used to help quantify POD for particular kinds of inspections,
has, for many years, been a subject of discussion and research
\citep[e.g.,][and the references in these papers]{thompson2008unified,li2014physical}. \citet{aldrin2011protocol} and
\citet{lindgren2011need}, among others, have outlined some of
the general needs for applying MAPOD concepts to quantify POD in SHM
applications. \citet{janapati2016damage} describe
some of the important steps that would be needed to provide
information for a MAPOD-based POD study for SHM.

\subsection{A framework for MAPOD in SHM applications}
The random-parameters model described in
Section~\ref{section:rp.model} provides a
framework for applying MAPOD in SHM applications. If the parameters
$\mu_{\beta_0},\; \mu_{\beta_1},\; \sigma_{\beta_0},\;
\sigma_{\beta_1},\; \rho, \;\text{and}\; \sigma_{\varepsilon}$ were
known, POD could be computed exactly using~(\ref{equation:rp.model.pod}) and
(\ref{eq:rp2}). The physical interpretations of these parameters are
given in Section~\ref{section:rp.model.basic.idea}
(e.g., the means and variances of
slopes and positions of the regression lines across the population
of possible crack-sensor combinations). In actual applications,
these parameters will never be completely known, but prior
information about the parameters is often available or can be obtained
in an economical manner. Bayesian estimation methods allow
combining such prior information about these parameters with data,
perhaps from experiments on a smaller sample of crack/sensor
combinations. For any given state of knowledge and available data,
one can use the methods described in
Sections~\ref{section:rp.model.pod} and~\ref{section:rp.model.bayesian} to compute
POD and lower credible bounds. When applying Bayesian methods, it is
possible to specify weakly informative (or diffuse) prior
distributions for some or all of the parameters. However, doing so
when there are limited data (i.e., experimental results on only a few
crack/sensor combinations) can result in considerable uncertainty in
estimates of POD, resulting in a lower confidence bound on POD that
is much less than the point estimate. Such a result would imply the
need for tests on more crack/sensor combinations or additional prior
information on one or more of the model parameters. Using the
Bayesian framework in this way will help determine how many
crack/sensor combinations will be needed to have the desired amount
of precision for estimating POD.

\subsection{Sources of information to inform the prior distributions}
The challenging part of implementing MAPOD is to obtain relevant
prior information about the model parameters in the form of
sensible, defensible prior probability distributions. Prior
information for MAPOD can be obtained from a variety of sources,
including combinations of transferable past experimental results,
mathematical models describing the physics of inspection, and data
from smaller, focused physical experiments that are less expensive
than testing additional crack/sensor combinations. One approach
frequently used in traditional NDI is to use a physics-based model
to generate pseudo POD data that is then analyzed by the usual
statistical methods for estimating POD,
as illustrated in~\citet{Buethe.etal2015path}. Such an approach could
be adapted to particular SHM-POD applications.

\FloatBarrier
\section{Concluding remarks and areas for future research}
\label{section:concluding.remarks}
Using in-situ SHM sensors, it is possible to remotely monitor the
integrity of specific parts of a structure during service and to
detect incipient damage before catastrophic failures occur. As
mentioned earlier, in POD studies for either NDI or SHM, it is
vitally important to recognize and properly capture and model all
important sources of variability. Statistical methods in
\citet{milhdbk1823a2009} assume that each target in a POD
study will be inspected only once. These methods need to be
generalized (similar to other generalizations that have been used in
the past for other applications, such as having explanatory
variables other than damage size) for SHM applications in which
there will be repeated measures on crack/sensor combinations.

This paper presents appropriate statistical methods for modeling data from
an SHM-POD study, allowing estimation of POD for a fixed sensor
detecting a crack that is propagating in a known direction near the
sensor. The methods presented here are applicable when there is a
scalar \DI{} or other response that will be used to make a
detection decision. The SoDaD method offers promise but ignores
useful information. The statistical generalization to the MIL-HDBK
1823A methods based on a random-parameters model can be used for
repeated measures SHM applications.  Further comments and areas for
future research include the following:
\begin{enumerate}
  \item
This paper illustrates the statistical methods and computations
that can be used to compute a POD function. The validity of such
computations depends on the source of the input data, driven by the
design of the POD experiment. For example, when SHM sensors are
mounted on flat plates in an experiment, the resulting POD estimate
will reflect the probability of finding cracks in flat plates and
any attempt to extend the results to other settings will require
justification.

  \item
Things change over time. SHM monitoring systems (e.g., those based
on guided waves) may be highly sensitive to changes in the
structural configuration as a function of time. Similarly, sensors
themselves can deteriorate over time, affecting system efficacy.

  \item
In traditional NDI applications, calibration is used to verify that
the inspection system is working as expected and to reduce
variability due to such factors as probe-to-probe variability and
possible changes in the inspection system over time. Relatedly,
corresponding methods will have to be developed to assure the
continuing efficacy of SHM monitoring over time.

  \item
The statistical methods presented in this paper, aimed at
traditional aerospace applications (as opposed to pipeline integrity
applications), NDI damage detection decisions are made on the basis
of current readings only. There is no attempt to compare current
readings with previous readings in applications where there are
multiple inspections. Statistically, under certain conditions, more
sensitive change detection methods can be obtained by combining
current readings with previous readings.
\citet[][Axiom II]{worden2007fundamental} states: ``The assessment of damage
requires a comparison between two system states.'' Of course, when a
change is detected, there needs to be a way to distinguish changes
due to a flaw versus changes due to sensor degradation or other
innocuous changes in or near to the SHM system.

  \item
The statistical models described in this paper are appropriate when
there is a scalar response (such as a \DI{}). A more
complicated model would be required for a vector signal response.

  \item
Traditionally, POD is usually given as a function of damage size
(e.g., crack length). In some SHM applications, it will be more
appropriate to have POD depend not only on damage size but also on
variables such as distance between the crack and the sensor(s) or
the point of initiation of the crack. The methods described here
could also be used (with straightforward extensions) to compute POD
from the results of a more complicated POD study that also varies
environmental variables such as temperature and loading. Using
well-established statistical methods for designing experiments would
allow the inclusion of such variables without having to
substantially increase the number of crack-sensor combinations in
the study.

  \item
The methods described in this paper were motivated and illustrated
by SHM methods for local or small zonal detection of cracks. Other
POD methods will have to be developed for SHM methods to detect
cracks or other damage (such a corrosion) over wider areas (sometime
referred to as global SHM).
\item
  In the application of the RP method to the CVM data in
  Section~\ref{section:cvm.rp.pod.estimates}, we saw that one of the
  potential problems with using the bivariate normal distribution to
  model the random intercepts and slopes is the non-zero probability
  of a negative slope. Also, in
  Section~\ref{section:characteristics.limitations.rp} we noted that
  it would be desirable to have alternative bivariate distributions
  to provide a more flexible modeling framework. A
  general family of bivariate distributions could be developed
  by choosing either location-scale or log-location-scale
  marginal distributions, separately for the slope and the
  intercept. Then those marginals could be linked by utilizing a
  copula function to describe the dependence structure. While
  computationally more complicated and demanding, the development of required
  software is feasible with modern computing capabilities.
\end{enumerate}

For our SHM examples, we provide a brief description of the SHM
systems in Section~\ref{considerations.shm.pod.study},
but it is important to note that we do
not attempt to compare the different SHM systems but merely uses the
available data from these systems to illustrate the applicability of
the statistical methods. Comparing different systems would require
that the systems be tested on the same kind of parts and with the
same kind of crack growth mechanisms. Also, and very importantly, in
order to compare POD curves, it is always necessary to establish
that the methods have the same probability of false alarms (e.g.,
concluding that damage exists when it does not). We have not done
this. For the examples shown in this paper, the respective
thresholds were provided by the experimenters.

\raggedbottom

\section*{Acknowledgments}
The writing of this paper was aided by inputs and comments received
from Shobbo Basu, Mark Davis, Tom Eason, Benjamin Eckstein, Dave
Forsyth, Rafik Hadjria, Christie Henry, Eric Lindgren, Holger
Speckmann, Floyd Spencer, and Paul Swindell. We appreciate their
helpful comments that helped improve this paper.

\section*{Funding Acknowledgments}
Qizheng Xia, William Meeker, and Qing Li received
no external funding for his work in the
preparation of this paper. The CNT crack gauge sensor research
presented in this paper was performed at the Metis Design
Corporation in Boston, MA, and the FAA Technical Center Atlantic
City, NJ, sponsored by the United States Air Force Research
Laboratories (AFRL) under SBIR Phase II topic AF14-065 contract
FA8650-15-C-2563 and by the FAA under CRADA ANG-TT-CRDA-0352. The
PZT and CVM sensor research presented in this paper was conducted at
Sandia National Laboratories and sponsored by the FAA William J.
Hughes Technical Center under the program lead of Paul
Swindell. Sandia National Laboratories is a multi-mission laboratory
managed and operated by National Technology and Engineering
Solutions of Sandia, LLC, a wholly-owned subsidiary of Honeywell
International, Inc., for the U.S. Department of Energy's National
Nuclear Security Administration under contract DE-NA0003525

\section*{Declaration of Conflicting Interests}
The authors declare that there are no conflicts of interest.

\section*{Data Availability Statement}
The data set used the current study are available in the
supplementary materials associated with this article.

\section*{Nomenclature}
\noindent\begin{tabular}{@{}>{\raggedright\arraybackslash}p{0.20\columnwidth}@{\hspace{0.03\columnwidth}}>{\raggedright\arraybackslash}p{0.73\columnwidth}@{}}
$a$ & Crack length or damage size in the $\hat{a}$ versus $a$ method\\
$\hat{a}$ & Signal response in the $\hat{a}$ versus $a$ method\\
$a_{90}$ & Crack length when POD is 0.90 \\
$a_{90/95}$ & 95\% upper confidence bound on $a_{90}$ \\
$\athresh$ & Detection threshold \\
CNT & Carbon nanotube \\
CVM & Comparative vacuum  monitoring\\
$i$ & Index for crack/sensor combinations \\
$j$ & Index for repeated measurements within a crack/sensor combination \\
LaD & Crack length at detection \\
$\lengthbar$ & Sample mean of observed crack lengths \\
$m_i$ & Number of repeated measurements for crack/sensor combination
$i$ \\
MAPOD & Model-assisted probability of detection \\
$n$ & Number of crack/sensor combinations or observations \\
NDI & Nondestructive inspection\\
PFA & Probability of a false alarm\\
POD & Probability of detection \\
PZT & Piezoelectric transducer\\
RP & Random parameter\\
$s$ & Sample standard deviation \\
SHM &Structural health monitoring\\
SoDaD & Size-of-damage-at-detection method\\
$x_i$ & Size-of-damage-at-detection (or crack length at detection)
observation, possibly transformed (e.g., by taking logs)\\
$\xbar$ & Sample mean \\
$\beta_0$ & Fixed-effect intercept in the simple signal-response model \\
$\beta_1$ & Fixed-effect slope in the simple signal-response model \\
$\beta_{0i}$ & Random intercept for crack/sensor combination $i$ \\
$\beta_{1i}$ & Random slope for crack/sensor combination $i$ \\
$\delta$ & Difference between crack length and mean crack length \\
$\varepsilon_i$ & Error term for observation $i$ \\
$\varepsilon_{ij}$ & Error term for measurement $j$ from crack/sensor combination $i$ \\
$\mu_{\beta_0}$ & Population mean of random intercepts \\
$\mu_{\beta_1}$ & Population mean of random slopes \\
$\rho$ & Correlation between random intercepts and slopes \\
$\sigma$ & Error standard deviation in the simple signal-response model \\
$\sigma_{\beta_0}$ & Standard deviation of random intercepts \\
$\sigma_{\beta_1}$ & Standard deviation of random slopes \\
$\sigma_{\varepsilon}$ & Within-crack/sensor response standard deviation \\
$\Phi_{\norm}(\cdot)$ & Standard normal cumulative distribution function \\
\end{tabular}

\FloatBarrier
\bibliographystyle{chicago}
\bibliography{references}

\end{document}